# An Overview of Plasma Confinement in Toroidal Systems

Fatemeh Dini, Reza Baghdadi, Reza Amrollahi
*Department of Physics and Nuclear Engineering, Amirkabir University of Technology, Tehran, Iran*

Sina Khorasani
*School of Electrical Engineering, Sharif University of Technology, Tehran, Iran*

Table of Contents

**Abstract:** This overview presents a tutorial introduction to the theory of magnetic plasma confinement in toroidal confinement systems with particular emphasis on axisymmetric equilibrium geometries, and tokamaks. The discussion covers three important aspects of plasma physics: Equilibrium, Stability, and Transport. The section on equilibrium will go through an introduction to ideal magnetohydrodynamics, curvilinear system of coordinates, flux coordinates, extensions to axisymmetric equilibrium, Grad-Shafranov Equation (GSE), Green's function formalism, as well as analytical and numerical solutions to GSE. The section on stability will address topics including Lyapunov Stability in nonlinear systems, energy principle, modal analysis, and simplifications for axisymmetric machines. The final section will consider transport in toroidal systems. We present the flux-surface-averaged system of equations describing classical and non-classical transport phenomena. Applications to the small-sized high-aspect-ratio Damavand tokamak will be described.

# I. Introduction

The increasing worldwide energy demand asks for new solutions and changes in the energy policy of the developed world, but the challenges are even greater for the emerging economies. Saving energy and using renewable energy sources will not be sufficient. Nuclear energy using fission is an important part of the worldwide energy mixture and has great potential, but there are concerns in many countries. A future possibility is the nuclear reaction of fusion, the source of solar energy. Though many scientific and technical issues have still to be resolved, controlled fusion is becoming more and more realistic. Two methods of nuclear reactions can be used to produce energy: fission – gaining energy through the break-up of heavy elements like uranium; and fusion – gaining energy by merging light elements such as deuterium and tritium. The fusion option is still far on the horizon, but international exploration has started in earnest these years. Nuclear fusion promises some welcome characteristics: an inexhaustible source of energy in light nucleus atoms; the inherent safety of a nuclear reaction that cannot be sustained in a non-controlled reaction; and few negative environmental implications. Research in controlled nuclear fusion has a self-sustainable burning plasma as its goal, and good progress has been made in recent years towards this objective by using both laser power and radiation to merge the light nuclei (inertial confinement) or using magnetic fields (magnetic confinement) to confine and merge deuterium and tritium. Large new facilities are currently under construction, the most prominent using magnetic confinement is ITER, which is seen as the international way towards the peaceful use of controlled nuclear fusion.

## I. 1. Energy Crisis

The daily increasing demand for energy in the world points out the growing need of the mankind to the various sources of energy. Renewable energies, despite their compatibility with the environment, are economical only in small scales of power delivery. On the other hand, reserves of fossil fuels are limited too, and also the obtained energy from burning fossil fuels causes the emission of carbon dioxide and particles, which in turn leads to the rise of the average temperature and air pollution. In Fig. I.1.1, it can be seen that the gap between the demand and delivery of crude oil is rapidly widening, as predicted over the next two decades.

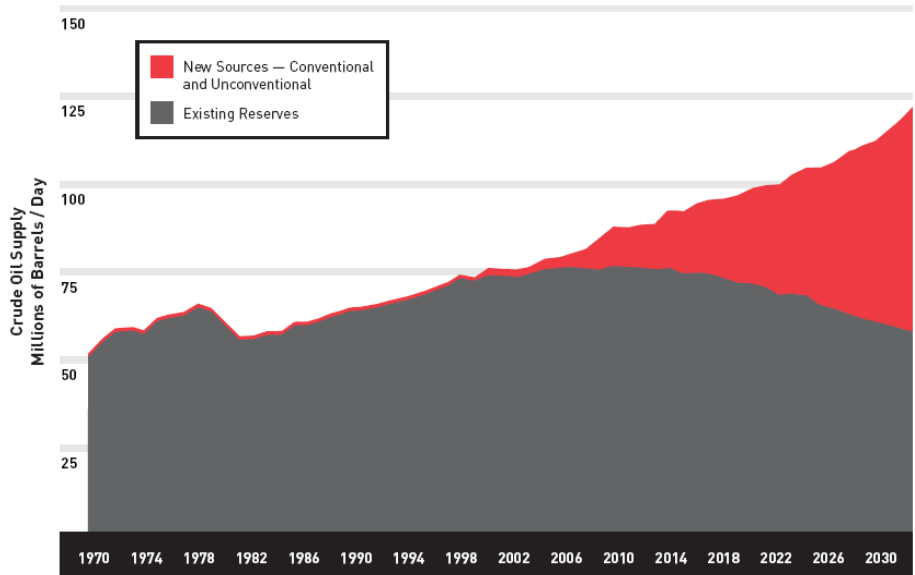

**Figure I.1.1**: The widening gap between oil delivery and demand (red) [1].

In the year 2030, the daily available access to the crude oil will be amounted to about 65 million barrels, and this is while there would be an extra 60 million barrels which should be replaced by other energy resources. Currently, more than 440 nuclear fission reactors around the globe produce 16% of the total spent energy by the mankind. The United States of America and France, respectively, with capacities of 98 and 63 giga Watts out of 104 and 59 nuclear reactors are the largest suppliers of nuclear electricity. With the completion of the Busheher nuclear reactors, Iran would join to the 33 countries in the world which are capable of producing nuclear electricity.

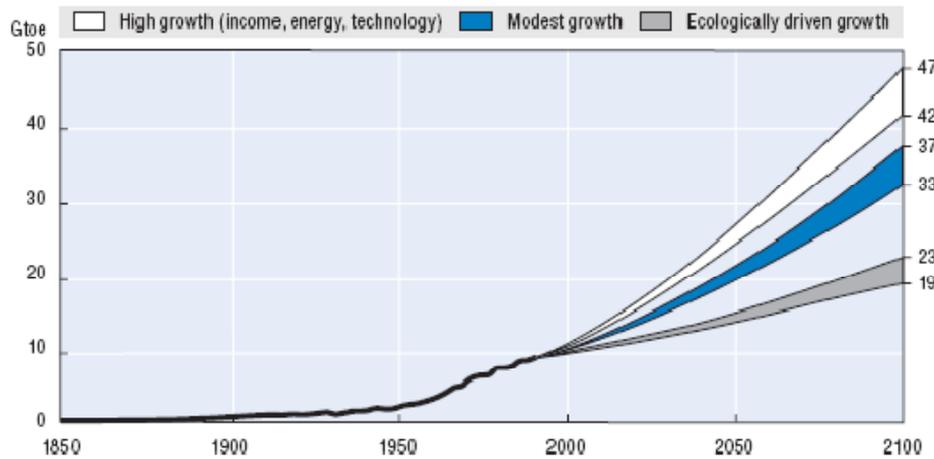

**Figure I.1.2:** Predicted energy demand till 2100 based on three different scenarios (billion tone crude oil equivalent) [1].

| Country or area | Operating | | Under construction | | Construction suspended | |
|---|---|---|---|---|---|---|
| | No | MW | No | MW | No | MW |
| Argentina | 2 | 935 | 1 | 692 | | |
| Armenia | 1 | 376 | | | | |
| Belgium | 7 | 5 728 | | | | |
| Brazil | 2 | 1 855 | | | | |
| Bulgaria | 6 | 3 538 | | | | |
| Canada | 14 | 9 998 | | | | |
| China | 3 | 2 167 | 8 | 6 370 | | |
| Czech Republic | 5 | 2 560 | 1 | 912 | | |
| Finland | 4 | 2 656 | | | | |
| France | 59 | 63 203 | | | | |
| Germany | 19 | 21 144 | | | | |
| Hungary | 4 | 1 755 | | | | |
| India | 14 | 2 548 | 2 | 900 | | |
| Iran | | | 2 | 1900 | | |
| Japan | 54 | 44 301 | 3 | 3 696 | | |
| Korea, S | 16 | 12 990 | 4 | 3 800 | | |
| Lithuania | 2 | 2 370 | | | | |
| Mexico | 2 | 1 364 | | | | |
| Netherlands | 1 | 452 | | | | |
| Pakistan | 2 | 425 | | | | |
| Romania | 1 | 655 | 1 | 655 | 3 | 1 965 |
| Russia | 30 | 20 793 | 3 | 2 625 | 7 | 6 628 |
| Slovakia | 6 | 2 472 | 2 | 840 | | |
| Slovenia | 1 | 679 | | | | |
| South Africa | 2 | 1 842 | | | | |
| Spain | 9 | 7 345 | | | | |
| Sweden | 11 | 9 460 | | | | |
| Switzerland | 5 | 3 170 | | | | |
| Taiwan | 6 | 4 884 | 1 | 1350 | | |
| UK | 33 | 12 528 | | | | |
| Ukraine | 13 | 11 195 | 2 | 1 900 | 3 | 2 859 |
| USA | 104 | 98 050 | | | 6 | 7 293 |
| Total | 438 | 353 425 | 30 | 25 640 | 19 | 18 745 |

**Table I.1.1:** Nuclear Reactors around the globe [2].

## I. 2. Nuclear Fission

In all of the nuclear reactors in the world, fission of heavy and unstable isotopes of Uranium makes the nuclear energy available, which is usually extract through a thermal cycle after first transforming into mechanical and subsequently electrical forms. The corresponding reactions are:

$$^{235}U + {}^{1}n \rightarrow \text{fission products} + \text{neturons} + \text{energy}\,(\sim 200 MeV) \qquad (I.1.1)$$

$$^{238}U + {}^{1}n \rightarrow {}^{239}U + \text{gamma rays} \qquad (I.1.2)$$

$$^{239}U \rightarrow {}^{239}Np \rightarrow {}^{239}Pu \quad (\text{a series of beta-decays}) \qquad (I.1.3)$$

In (I.1.1), the number of emitted neutrons and daughter nuclei might be different and range from 2 to 4. But the average number of neutrons is equal to 2.43. Neutrons may cause a chain reaction of (I.1.1) and transformation into $^{238}U$, or through continued reactions (I.1.2) and (I.1.3) produce the heavier element $^{239}Pu$. As we know, only 0.7% of the uranium in the nature is fissionable via thermal neutrons. The rest of existing uranium is in the form of $^{235}U$, which cannot be broken up by thermalized (slow) neutrons. Hence, the fuel used in the nuclear reactors is usually in the form of $U^2O$, with $^{235}U$ isotope enriched up to 4%. $^{239}Pu$ can be fissioned by neutrons, and $^{232}Th$ by absorption of one neutron transforms into $^{233}U$, which in turn is highly fissionable by thermal neutrons.

The fission of uranium causes a large energy density. The fission of only one gram of $^{235}U$ per day can generate an average power of 1MW. This is equivalent to burning of 3 tones of coal and more than 600 gallons of oil product, emitting 250 Kg of carbon dioxide. The released energy is carried by the kinetic energy of daughter nuclei, which is absorbed in the water pool of the reactor. In some designs such as pressurized water reactors (PWR), the thermal energy is exploited for evaporation of water in a separate cycle. But in boiling water reactors (BWR), the water in contact with the nuclear fuel is directly evaporated and used for driving turbines. Also, there exists the possibility of using fast neutrons in place of thermal neutrons with $^{239}Pu$ fuel. Therefore, $^{235}U$ reactors produce the necessary fuel for the former kind of reactors. Annually, about 100 tones of $^{239}Pu$ is obtained worldwide.

Considering the daily need to the production of nuclear electricity and applications of radioactive materials in various areas of energy, medicine, industry, agriculture, and research in countries, the use of nuclear energy is inevitable. Despite the advantages of using fission energy, many drawbacks are also associated with this nuclear technology as well, the problem of wastes being the most important. The transmutation of nuclear wastes containing or contaminated by radioactive materials is among the most important unsolved problems of this technology. It seems that simple methods are only explored for this purpose, and no acceptable plan for long time isolation or transmutation of nuclear wastes exists to date.

Until the early 1950s, dilution, air release, submerging in ocean floors, and concealing over deserts have been used. Since then other methods such as concealing in multilayer undergrounds and vacant mines are also proposed. But through the time, the production of nuclear wastes has raised so much that none of the mentioned methods would work in the long run. Five decades of exploiting nuclear reactors in the United States, only, has produced 50,000 million tones of spent nuclear fuel. It is anticipated that this trend would increase all the way to 20,000 million tones annually.

For achievement of a permanent solution, the Yukka mountain project with the capacity of 70,000 million tones has been under way, which clearly is insufficient (Fig. I.2.1). But even this project has been stopped due to its extraordinary cost of 6 billion dollars. The alternative proposed solution is irradiation of radioactive wastes by neutrons obtained from accelerated protons (Fig. I.2.2). In this method, nuclear wastes with long life times are converted into short-lived radioisotopes. Also, the generated heat from many of the burning isotopes such as $^{129}$I, $^{99}$Tc, $^{237}$Np, $^{90}$Sr, and $^{137}$Cs can be extracted by Pb and exploited for production of electricity needed to run the accelerators.

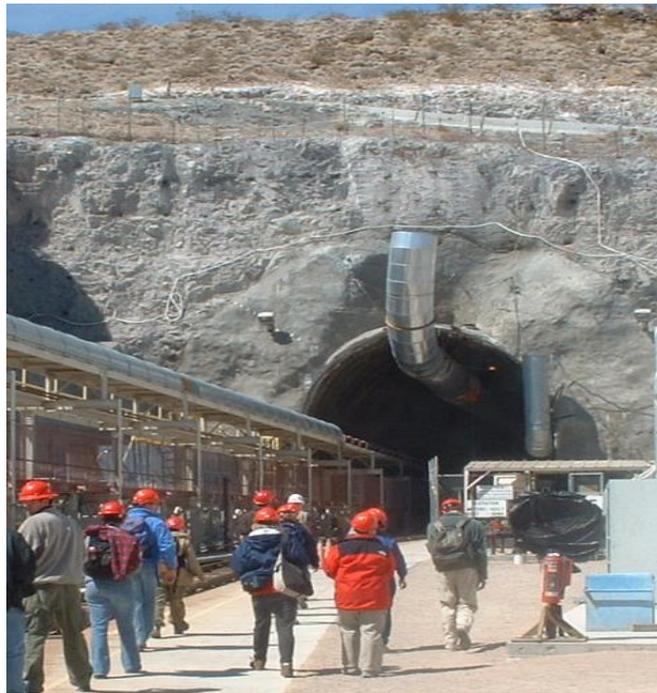

**Figure I.2.1:** The six billion dollar Yukka mountain project [6,7].

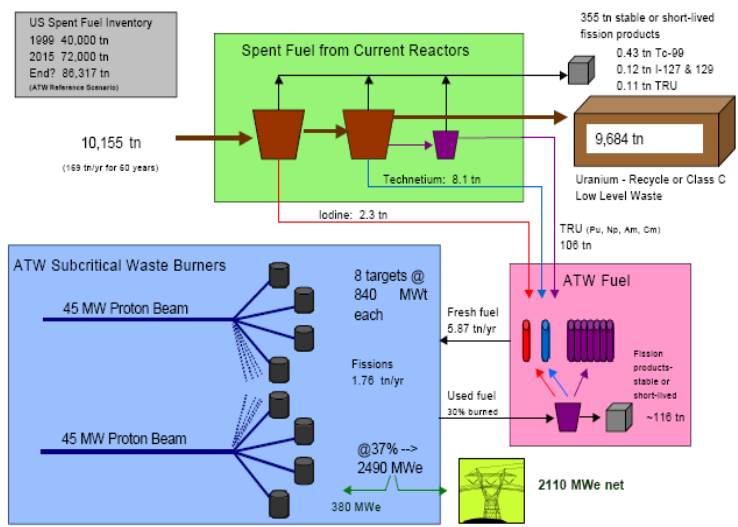

**Figure I.2.2:** Transmutation of nuclear wastes from fission reactors via proton accelerators [6,7].

Other replacements include Fission-Fusion hybrids and also using thermonuclear plasmas of tokamaks as neutron sources, both of which are based on the Fusion technology. Therefore, the nuclear fusion once completed can be used for energy production as well as transmutation of the nuclear waste from fission reactors. It should be mentioned, however, that still the cheapest energy in the world is not from nuclear, but rather coal resources (Fig. I.2.3).

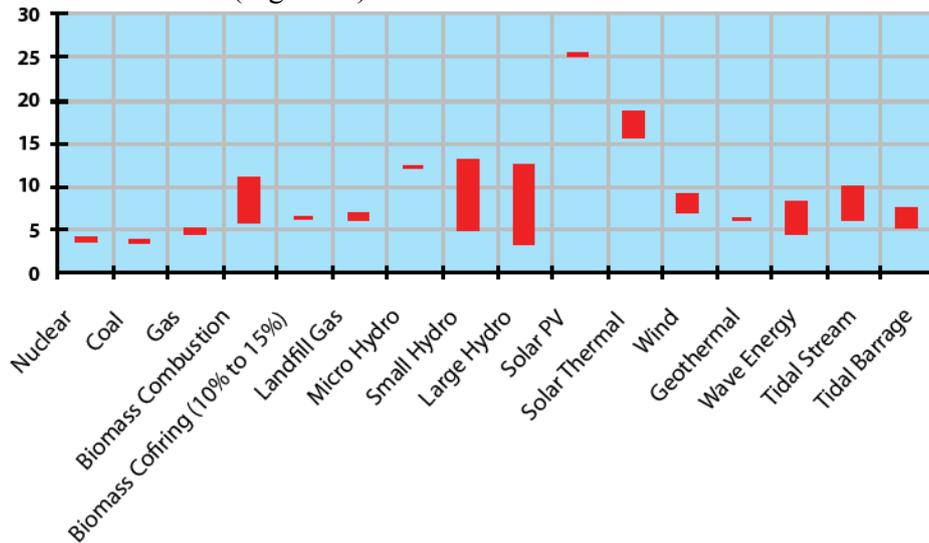

**Figure I.2.3:** Cost of electricity produced from different resources (cent per KWh).

## I. 3. Nuclear Fusion

As discussed above, the nuclear energy can be either obtained from the fission of heavy elements, or fusion of light elements. Generally speaking, whenever the heavier element has a lower potential energy compared to the sum of potential energies of two separate nuclei, the fusion reaction is plausible. The experiment reveals that iron with the atomic number 26 has the lowest level of potential energy, and therefore it would be the most stable nucleas. This shows that fusion of lighter elements than iron always generates energy, as the fission of heavier elements than iron does. But the released energy depends on the reaction cross section as well as the energy obtained from every individual reaction.

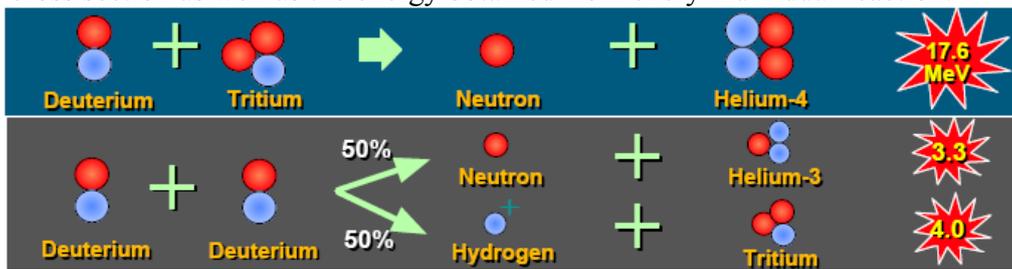

**Figure I.3.1:** First generation nuclear fusion reactions.

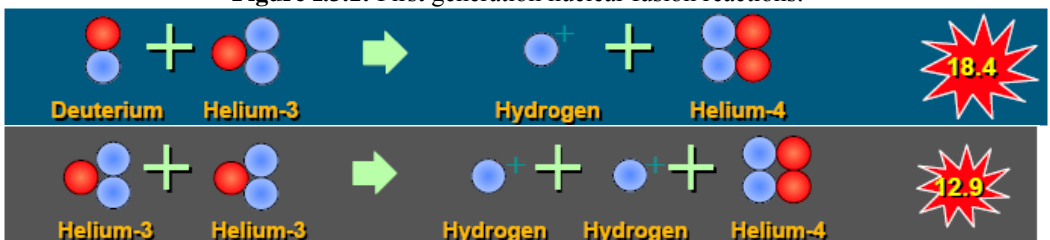

**Figure I.3.2:** Second (top) and third (bottom) generations of nuclear fusion reactions.

The nuclear fusion reactions take place in the universe in the center of stars among the nuclei of hydrogen and helium, and in white dwarfs among nuclei lighter than iron. The simplest nuclear fusion reactions which can be achieved on the earth are among the four lightest elements of the periodic table, and their isotopes. These include hydrogen (and its isotopes: deuterium and tritium), helium, lithium, and beryllium, each generating an enormous amount of energy. But the H-H reaction has a very small cross section and hence very small probability for taking place. In contrast, heavier elements than hydrogen or its isotopes can be used to obtain the four nuclear fusion reactions corresponding to three distinct generations.

The reactions belonging to the first generation occur between the isotopes of hydrogen, namely deuterium D=$^2$H, and tritium T=$^3$H. A significant amount of deuterium can be found on the surface of earth, and via industrial methods can be obtained from water in the form of heavy water D$_2$O. But tritium is the unstable and radioactive isotope of hydrogen with a life time of about 12 years, and therefore does not exits naturally. To produce tritium, reactions of fast neutrons with isotopes of lithium can be exploited as follows

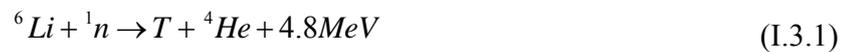
$$^6Li + {}^1n \rightarrow T + {}^4He + 4.8MeV \qquad (I.3.1)$$
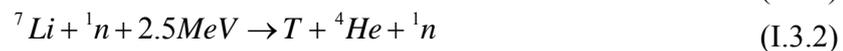
$$^7Li + {}^1n + 2.5MeV \rightarrow T + {}^4He + {}^1n \qquad (I.3.2)$$

The reactions belonging to the second generation does not produce neutrons, and therefore have the advantage that the collection of resulting energy, which in the first generation usually escapes in the form of kinetic energy of fast neutrons, would be much simpler. Among the reactions of first to third generations, the D-T reaction has the highest cross section, but this is maximized at the temperature of 100keV. However, experiments and theoretical calculations show that sustainable chain reactions could be achieved at much lower temperatures, being around 10keV. In other words, self-sustaining nuclear fusion reactions require a temperature of around 120×10$^6$K. At such elevated temperatures, matter could exist only in the form of plasma, and all atoms become fully ionized. Clearly, under such severe conditions, the problem of confinement and heating of thermonuclear plasmas forms the bottleneck of nuclear fusion technology; thermonuclear plasmas cannot be simply confined in a manner comparable to gases and liquids.

In stars, the thermonuclear plasma exists in the center and is inertially confined through the force of gravity (Fig. I.3.3). The strength of gravitational force and temperature is so high in the core, that nuclear fusion reactions take place on their own. When the fusion reactions among all of the light elements stop due to the termination of nuclear fuel, the star undergoes either a collapse or expansion depending on its mass. Heavy stars form white dwarfs with extremely high mass densities where nuclear fusion reactions continue until all of the fuel is transformed into iron, while lighter stars expand and continue to faintly radiate as a red giant.

On the earth, the time needed for confinement of thermonuclear plasma in order to achieve self-sustained reactions depends on the plasma density. For this reason, the plasma can be confined using ultra strong magnetic fields obtained by superconducting coils. The technique is referred to as the Magnetic Confinement Fusion (MCF). In the other approach, plasma is confined by pettawatt laser pulses having energies exceeding 10MJ, or accelerated particles, which uniformly irradiate a solidified spherical micro target. This technique is referred to as the Inertial Confinement Fusion (ICF). In MCF, the mean plasma density should be of the order of 10$^{19}$cm$^{-3}$, and its temperature peaks at 10keV. In ICF, the mean plasma density should exceed $10^{28} cm^{-3}$, which is at least four orders of magnitude, or 10,000 times, higher than the density of solids under standard conditions.

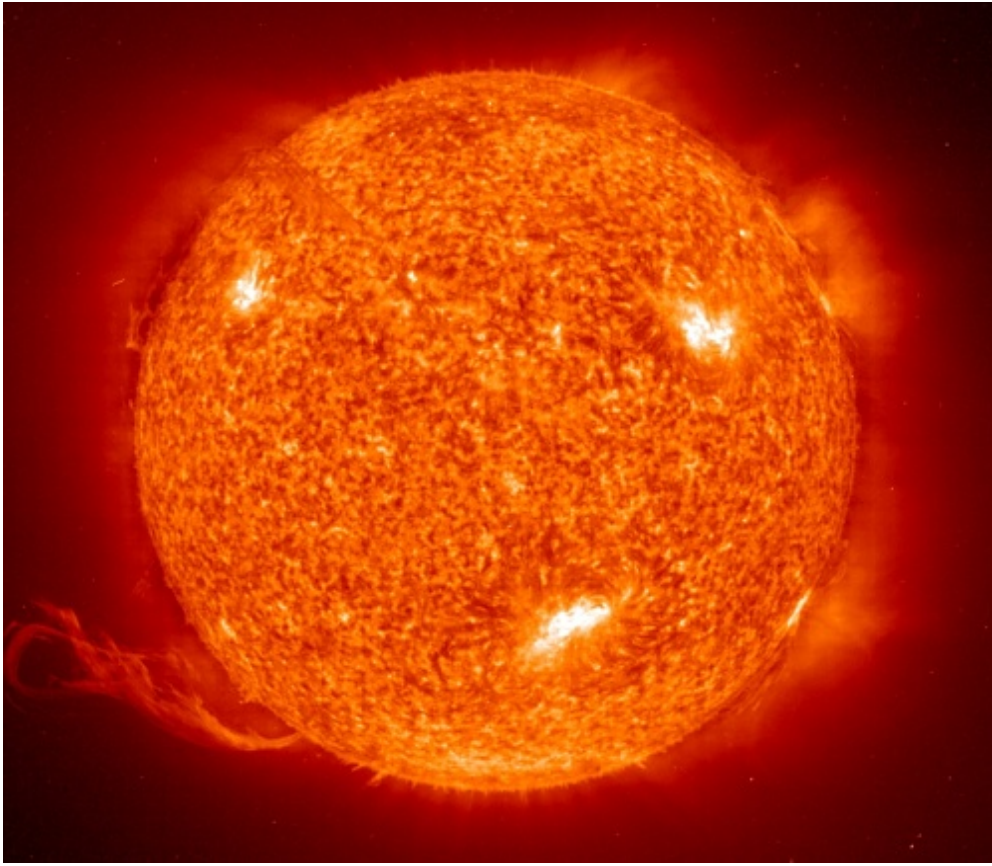

**Figure I.3.3:** Gravitational force of the sun makes the thermonuclear Fusion process gets rolling.

In this way, nuclear fusion reactions require the triple product of density, temperature, and confinement time to obey the following inequality, widely known as Lawson's criterion

$$\langle n \times T \times t \rangle > 10^{20} cm^{-3} \cdot keV \cdot s \qquad (I.3.3)$$

where $\langle \cdot \rangle$ sign represents the average. Therefore, the thermonuclear plasma in MCF should be kept at the temperature of several keVs for several seconds. Similarly, the confinement time in inertial fusion should be at least of the order of few nanoseconds.

In practice, the confinement of plasma for such time intervals is so difficult due to many instabilities, that the experimental thermonuclear plasmas have been heated only up to the ignition point. Under such circumstances, the ratio of output to input plasma power $Q$ is around 3, at which the heat generated by nuclear fusion reactions balances the plasma natural losses through plasma-wall interactions, radiations, and escape of energetic particles. But in order to obtain useful electrical power, this ratio should exceed 10.

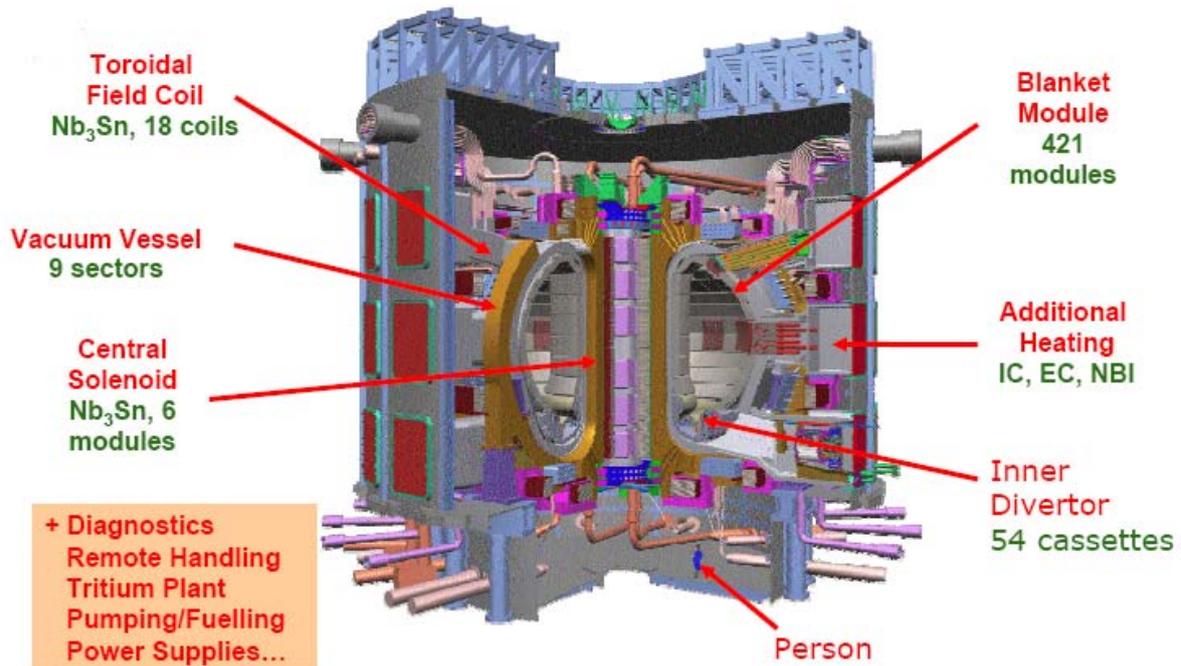

**Figure I.3.4:** The International thermonuclear Experimental Reactor (ITER) located in Caradache, France [3].

Currently, the largest existing project to achieve thermonuclear fusion is ITER (Fig. I.3.4), which is to be built in the city of Caradache, France. The multi-billion dollar ITER project is scheduled for operation by 2025, and is funded by many countries including the United States, Russia, European Union, China, South Korea, India, and Japan, and each country is responsible for fabrication of one or several parts of the project. ITER is based on a machine named Tokamak which benefits from the technology evolved from decades of research in MCF science and technology. In tokamak, plasma is confined in the form of a torus by very strong magnetic fields in a vacuum vessel. To attain the plasma stability conditions a large unidirectional toroidal electrical current should be maintained in the plasma. For the case of ITER tokamak, this toroidal current should be about 15MA. The plasma confinement time is designed to be at least 400sec. Calculations predict that the plasma passes the ignition point and $Q$ factor reaches 10. The total plasma volume in this giant machine amount to $840m^3$. The cross section of toroidal plasma in ITER tokamak has been shown in Fig. I.3.5, indicating the dimensions as well.

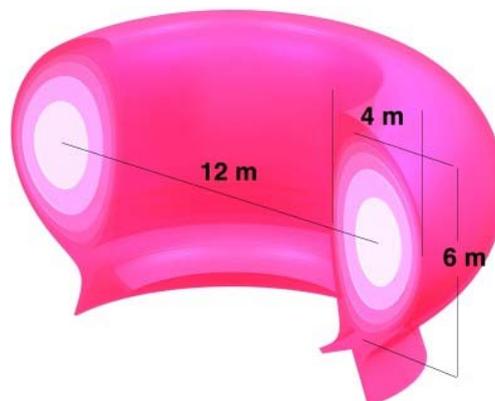

**Figure I.3.5:** Cross section of toroidal plasma in ITER tokamak [3].

## I. 4. Other Fusion Concepts

Tokamak is not the only proposed path to the controlled nuclear fusion. There are other designs, among them spherical tokamaks, stellarators, and laser fusion could be named out. Spherical tokamaks are much similar to tokamaks in the concepts of design and operation, with the main difference being their tight aspect ratio. Aspect ratio is defined as the ratio of plasma's major radius to its minor radius; for tokamaks this ratio is within the range 3-5, while for spherical tokamaks is typically less than 1.5. This allows operation at higher magnetic pressures, which results in better confinement properties. Currently, there exist two major spherical tokamak experiments in the world: the Mega-Ampere Spherical Tokamak (MAST) in Culham, United Kingdom, where the Joint European Torus (JET) tokamak resides, and the National Spherical Tokamak eXperiment (NSTX) at Princeton Plasma Physics Laboratory (PPPL) in New Jersey, the United States. Photographs of NSTX plasma and facility are seen in Fig. I.4.1.

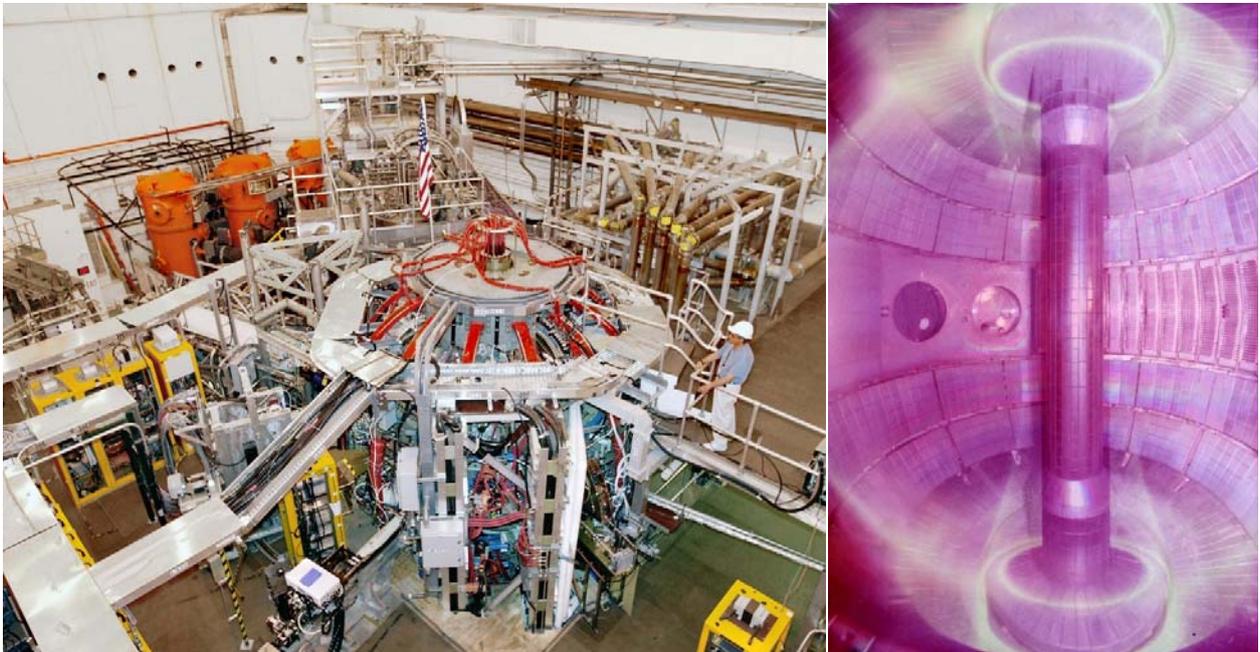

**Figure I.4.1:** The National Spherical Tokamak eXperiment (NSTX); left: facility; right: plasma in the vacuum vessel [4].

The stellarator concept is also similar to tokamaks, in the aspect that in both designs the plasma is produced and maintained in a toroidal vacuum vessel. However, while the stability of tokamak plasma is provided through the establishment of a DC toroidal plasma current, the stellarator plasma is stable without need to such a toroidal plasma current. The reason is that the stability is obtained by a complex topology of magnetic field windings which produce both the toroidal and poloidal magnetic fields. Stellarators are usually considered as too complex for realistic reactor designs, but they offer unlimited possibilities in plasma confinement. Wendelstein stellarator at Max Planck Institute of Plasma Physics (Fig. I.4.2), Germany, and National Compact Stellarator eXperiment (NCSX) (Fig. I.4.3) are examples of advanced stellarator configurations around the globe.

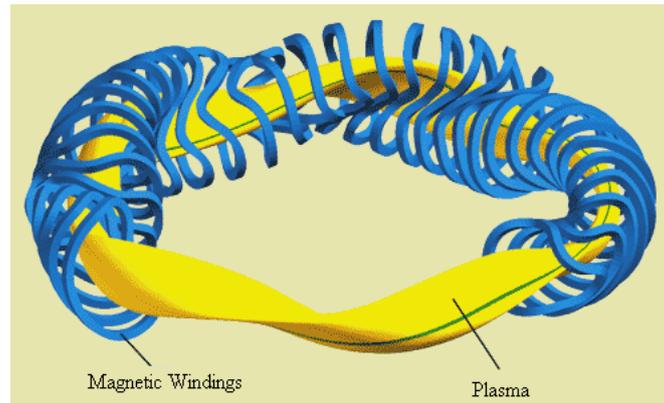

**Figure I.4.2:** Typical stellarator configuration.

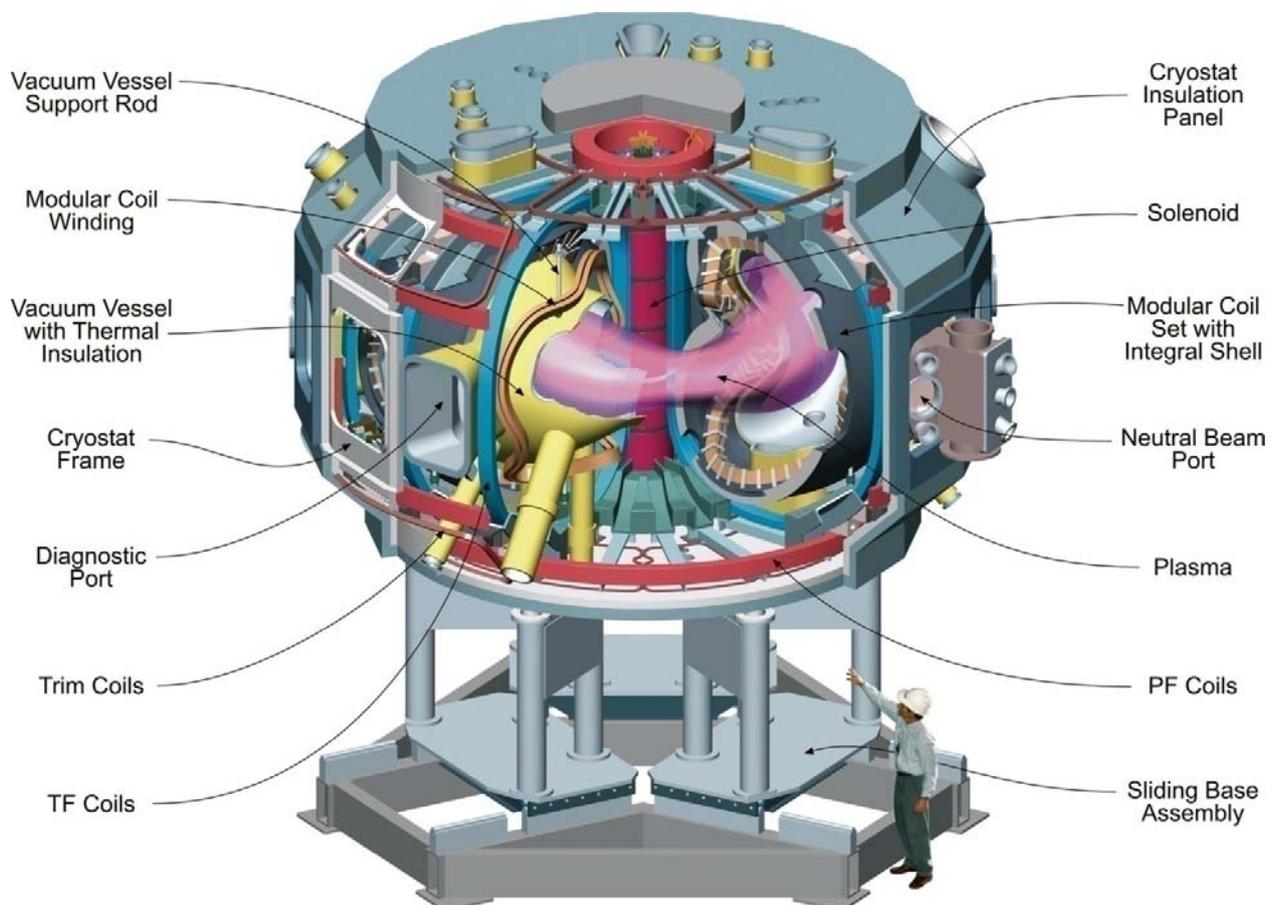

**Figure I.4.3:** National Compact Stellarator eXperiment at PPPL [4].

It is also worth mentioning about the modern fusion-fission hybrid concept [5], which connects the possibilities of both technologies, combining the benefits and eliminating the drawbacks. In this design, a fission reactor produces the output electrical power, which is also used to run a tokamak or z-pinch (another MCF concept) and a proton accelerator. Both of auxiliary systems produce fast neutrons to keep the fission energy yield as high as possible. Because of the energetic neutrons used, heavy elements such as uranium may break up into much smaller elements, releasing even more energy and much less radioactive waste. These designs [6,7] are nowadays bringing attractions up as the progress in controlled nuclear fusion has slightly slowed down.

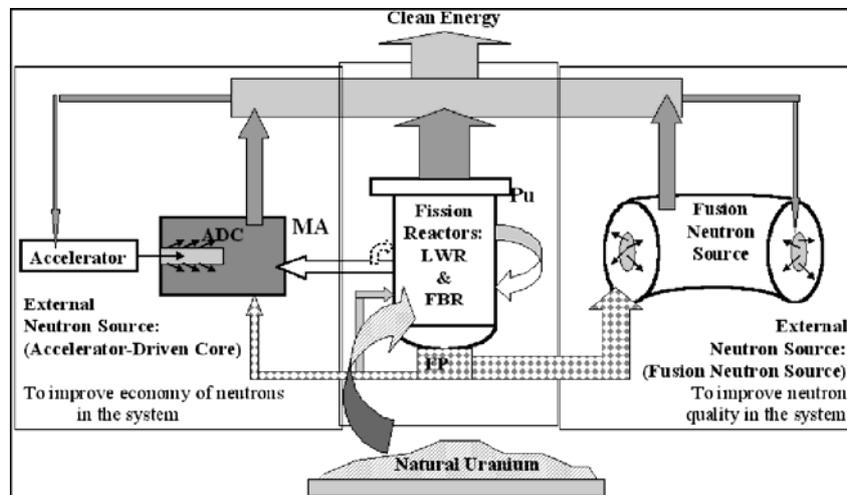

**Figure I.4.5:** Hybrid fission-fusion design to produce clean and efficient nuclear energy.

# II. Plasma Equilibrium

## II. 1. Ideal Magnetohydrodynamics (MHD)

Plasma is often misinterpreted as a "hot gas," but its conductivity and dynamic response to electricity and magnetism recognize it from a gas. The shape of the plasma and location of the plasma boundary deeply affect its stability. Since the electromagnetic fields control the movement of the plasma which itself induces electromagnetic fields, determining this shape may quickly lead to nonlinear equations. One simple way of studying magnetically confined plasmas with an emphasis on the shaping magnetic field topology is *magnetohydrodynamics* (MHD) model. MHD model first proposed by Hannes Olof Gösta Alfvén (Figure II.1.1). The word magnetohydrodynamic (MHD) is derived from *magneto-* meaning magnetic field, and *hydro-* meaning liquid, and *-dynamics* meaning movement.

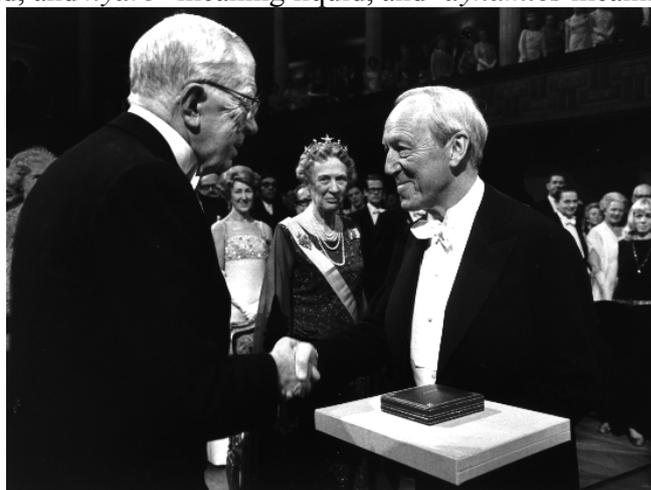

**Figure II.1.1:** Hannes Alfvén, the father of modern plasma science, receives Nobel Prize from the King of Sweden in 1970 [1].

MHD equations consist of the equation of fluid dynamics and Maxwell's equations that should be solved simultaneously. The MHD model is composed of the following relations

$$\frac{\partial \rho}{\partial t} + \nabla \cdot (\rho \mathbf{V}) = 0 \qquad \text{Continuity} \qquad (\text{II.1.1})$$

$$\rho \frac{\partial \mathbf{V}}{\partial t} + \nabla p = \mathbf{J} \times \mathbf{B} \qquad \text{Momentum} \qquad (\text{II.1.2})$$

$$\frac{1}{\sigma} \mathbf{J} = \mathbf{E} + \mathbf{V} \times \mathbf{B} \qquad \text{Ohm's Law} \qquad (\text{II.1.3})$$

$$\left.\begin{array}{l} \nabla \times \mathbf{E} = -\dfrac{\partial \mathbf{B}}{\partial t} \\[4pt] \dfrac{1}{\mu_0} \nabla \times \mathbf{B} = \mathbf{J} + \dfrac{\partial \mathbf{D}}{\partial t} \\[4pt] \nabla \cdot \mathbf{D} = \rho \\[4pt] \nabla \cdot \mathbf{B} = 0 \end{array}\right\} \qquad \text{Maxwell's Equations} \qquad (\text{II.1.4})$$

In the above equations, plasmas are described as magnetohydrodynamic fluids with mass density $\rho$, current density $\mathbf{J}$, mass flow velocity $\mathbf{V}$, and local electric $\mathbf{E}$ and magnetic $\mathbf{B}$ fields. As in a plasma we have both the ion and electron species, we should write MHD equations for both ions and electrons, separately, but charge neutrality of plasma enables us to approximate the plasma as a neutral fluid with zero local electric charge density. Furthermore, since the mass of ions is much larger than the mass of electrons (the ion-to-electron mass ratio is $m_i/m_e = 1836 A$, where A is the atomic weight of the ion) the contribution of ions govern the mass density of the plasma.

MHD establishes a relationship between the magnetic field $\mathbf{B}$ and plasma motion $\mathbf{V}$. Let us examine the relationship of these two parameters by applying curl operator to equation (II.1.3), which results in:

$$\nabla \times \left(\frac{\mathbf{J}}{\sigma}\right) = \nabla \times \mathbf{E} + \nabla \times (\mathbf{V} \times \mathbf{B}) \qquad (\text{II.1.5})$$

Now by using equation (II.1.4) we get:

$$\begin{aligned} \frac{\partial \mathbf{B}}{\partial t} &= \nabla \times \left(\mathbf{V} \times \mathbf{E} - \frac{\mathbf{J}}{\sigma}\right) = \nabla \times (\mathbf{V} \times \mathbf{B}) - \frac{1}{\sigma \mu_0} \nabla \times \mathbf{J} \\ &= \nabla \times (\mathbf{V} \times \mathbf{B}) - \frac{1}{\sigma \mu_0} \nabla \times (\nabla \times \mathbf{B}) \qquad (\text{II.1.6}) \\ &= \nabla \times (\mathbf{V} \times \mathbf{B}) - \frac{1}{\sigma \mu_0} \nabla^2 \mathbf{B} \end{aligned}$$

Equation (II.1.6) consists of two terms: the first term $\nabla \times (\mathbf{V} \times \mathbf{B})$, is the convection term and the second term proportional to $\nabla^2 \mathbf{B}$, represents the diffusion. Rate of change of the magnetic field is controlled by these two terms.

Assume that the velocity of plasma **V** is zero everywhere so that the plasma does not move, therefore the first term in equation (II.1.6) vanishes, and we get:

$$\frac{\partial \mathbf{B}}{\partial t} = \frac{1}{(\sigma \mu_0)} \nabla^2 \mathbf{B} = D_m \nabla^2 \mathbf{B} \qquad (\text{II}.1.7)$$

where $D_m$ is called the *diffusion coefficient* of plasma. If the resistivity is finite, the magnetic field diffuses into the plasma to remove local magnetic inhomogeneities, e.g., curves in the field, etc.

Ideal magnetohydrodynamics (MHD) describes the plasma as a single fluid with infinite conductivity. Hence by putting $\sigma = \infty$ in the Ohm's law (II.1.3), we obtain

$$\mathbf{E} + \mathbf{V} \times \mathbf{B} = \mathbf{0} \qquad (\text{II}.1.8)$$

In case of ideal MHD, $\sigma \to \infty$, the diffusion is very slow and the evolution of magnetic field **B** is solely determined by the plasma flow. For this reason the equation (II.1.7) recasts into the form

$$\frac{\partial \mathbf{B}}{\partial t} = \nabla \times (\mathbf{V} \times \mathbf{B}) \qquad (\text{II}.1.9)$$

The measure of the relative strengths of convection and diffusion is the *magnetic Reynolds number* $R_m$. Hence magnetic Reynolds number is a representation of combination of quantities that indicate the dynamic behavior of plasma. Reynolds number is the ratio of the first term to the second term on the right-hand-side of (II.1.6)

$$\frac{|\nabla \times (\mathbf{V} \times \mathbf{B})|}{\left|\frac{1}{\sigma \mu_0} \nabla^2 \mathbf{B}\right|} \approx \frac{\frac{VB}{L}}{\left(\frac{B}{L^2}\right)\left(\frac{1}{\sigma \mu_0}\right)} = \mu_0 V L \sigma \equiv R_m \qquad (\text{II}.1.10)$$

where $L$ is the typical plasma dimension. In (II.1.10) the magnetic Reynolds number is equal to the ratio of the magnetic diffusion time $\tau_R = \mu_0 L^2 \sigma$ to the Alfven transit time $\tau_H = L/V$, that is $R_m = \tau_R/\tau_{Hc}$. The magnetic field in a plasma changes according to a diffusion equation, when $R_m \ll 1$ while the lines of magnetic force are frozen in the plasma, when $R_m \gg 1$.

We can demonstrate frozen-In theorem in integral form as below:

$$\frac{d\Phi}{dt} = \frac{d}{dt} \iint_S \mathbf{B} \cdot d\mathbf{S} = 0 \qquad (\text{II}.1.11)$$

Hence, magnetic flux passing through any surface $S$ with the plasma motion is constant. When $R_m \to \infty, \sigma \to \infty$ the rate of change of the flux becomes zero. This means the magnetic flux is frozen in the plasma.

## II. 2. Curvilinear System of Coordinates

Using curvilinear system of coordinates in analytical and numerical computations of equilibrium, stability and transport of toroidal plasmas is vital. The purpose of this section is to review a few fundamental ideas about curvilinear system of coordinate in general. Flux, Boozer and Hamada coordinates are typical coordinate systems in study of magnetic fusion plasmas.

By definition, the position vector $\mathbf{r}$ in Cartesian coordinate system has three components $(x, y, z)$ along its basis vectors $(\hat{x}, \hat{y}, \hat{z})$, so that $\mathbf{r} = x\hat{x} + y\hat{y} + z\hat{z}$. We may represent the components of the position vector $x$ with $x^1$, $y$ with $x^2$ and $z$ with $x^3$, and similarly the basis vectors $\hat{x} = \nabla x$ with $\hat{x}_1 = \nabla x_1$, $\hat{y} = \nabla y$ with $\hat{x}_2 = \nabla x_2$ and $\hat{z} = \nabla z$ with $\hat{x}_3 = \nabla x_3$, to get $\mathbf{r} = x^1\hat{x}_1 + x^2\hat{x}_2 + x^3\hat{x}_3$, or simply $\mathbf{r} = x^i\hat{x}_i$ where the Einstein summation convention on repeated indices is adopted. The vectors $\hat{x}_j, j=1,2,3$ are called contravariant basis vectors, while $x_j, j=1,2,3$ are referred to as the covariant components. Similarly, the components $x^j, j=1,2,3$ are called contravariant components of position vector, and while $\hat{x}^j, j=1,2,3$ are referred to as the covariant bases. For the case of Cartesian coordinates, there is no distinction between contravariant $\mathbf{r} = x^i\hat{x}_i$ and covariant $\mathbf{r} = x_i\hat{x}^i$ representations, in the sense that $x_i = x^i$ and $\hat{x}_i = \hat{x}^i$.

A curvilinear system of coordinates $(\zeta^1, \zeta^2, \zeta^3)$ uniquely establishes a one-to-one correspondence to the Cartesian coordinates $(x^1, x^2, x^3)$ through the set of analytic relations

$$\zeta^j = \zeta^j(x^1, x^2, x^3), j = 1, 2, 3 \tag{II.2.1}$$

For instance, suppose that $(\xi^1, \xi^2, \xi^3)$ are components of the standard spherical coordinate system. Then

$$\begin{aligned}\xi^1 &= \sqrt{(x^1)^2 + (x^2)^2 + (x^3)^2} \\ \xi^2 &= -\tan^{-1}(x^2/x^1) \\ \xi^3 &= \cos^{-1}\left(x^3 / \sqrt{(x^1)^2 + (x^2)^2 + (x^3)^2}\right)\end{aligned} \tag{II.2.2}$$

In Cartesian coordinates, we represent any arbitrary vector quantity $\mathbf{A}$ with respect to its basis as

$$\mathbf{A} = A_1\hat{x}^1 + A_2\hat{x}^2 + A_3\hat{x}^3 = A_j\hat{x}^j \tag{II.2.3}$$

where the components $A_j, j = 1, 2, 3$ of the vector $\mathbf{A}$, are called covariant components of $\mathbf{A}$. In a curvilinear system of coordinates, we must use different basis vectors $\hat{\zeta}^j, j = 1, 2, 3$, given by

$$\hat{\zeta}^j = \nabla\zeta^j = \left(\frac{\partial}{\partial x^1}\hat{x}_1 + \frac{\partial}{\partial x^2}\hat{x}_2 + \frac{\partial}{\partial x^3}\hat{x}_3\right)\zeta^j = \frac{\partial \zeta^j}{\partial x^i}\hat{x}_i \tag{II.2.4}$$

Note that unlike the Cartesian coordinates, where the covariant bases $\hat{x}^j, j = 1, 2, 3$ are physically dimensionless, $\hat{\zeta}^j, j = 1, 2, 3$ may take on non-trivial physical dimensions. In general, these basis vectors need not to be unit vectors.

The condition for one-to-one correspondence of the curvilinear system of coordinates is that the basis vectors $\hat{\zeta}^j, j = 1, 2, 3$ construct a parallelepiped with non-vanishing volume. Mathematically, the Jacobian determinant defined as

$$J = \hat{\zeta}^1 \cdot \hat{\zeta}^2 \times \hat{\zeta}^3 = \nabla\zeta^1 \cdot \nabla\zeta^2 \times \nabla\zeta^3 = \begin{vmatrix} \frac{\partial \zeta^1}{\partial x^1} & \frac{\partial \zeta^1}{\partial x^2} & \frac{\partial \zeta^1}{\partial x^3} \\ \frac{\partial \zeta^2}{\partial x^1} & \frac{\partial \zeta^2}{\partial x^2} & \frac{\partial \zeta^2}{\partial x^3} \\ \frac{\partial \zeta^3}{\partial x^1} & \frac{\partial \zeta^3}{\partial x^2} & \frac{\partial \zeta^3}{\partial x^3} \end{vmatrix} \tag{II.2.5}$$

should not vanish. Furthermore, we suppose that the order of coordinates is chosen in such a way that the Jacobian $J$ is always positive, which corresponds to a right handed system. As examples, the values of Jacobian in spherical and cylindrical coordinate systems are $1/R\sin\theta$ and $1/r$, respectively.

Since $J > 0$, any vector such as $\mathbf{A}$ can be expanded in terms of the linearly independent bases as

$$\mathbf{A} = A_j\hat{\zeta}^j \tag{II.2.6}$$

where components are in covariant forms, and hence (II.2.6) is a covariant representation of $\mathbf{A}$. In order to find $A_1$ one may perform a dot product on both sides by $\hat{\zeta}^2 \times \hat{\zeta}^3$ to find:

$$\mathbf{A} \cdot \left(\hat{\zeta}^2 \times \hat{\zeta}^3\right) = A_1 J \tag{II.2.7}$$

By cyclic permutation of indices we get the relation

$$A_i = \frac{\varepsilon_{ijk}}{2J}\hat{\zeta}^j \times \hat{\zeta}^k \cdot \mathbf{A} \tag{II.2.8}$$

Here, $\varepsilon_{ijk}$ is Levi-Civita pseudo-tensor symbol and is given by:

$$\varepsilon_{ijk} = \hat{x}_i \cdot \hat{x}_j \times \hat{x}_k = \begin{cases} +1 & \text{If } i,j,k \text{ is an even permutation of } 1,2,3 \\ -1 & \text{If } i,j,k \text{ is an odd permutation of } 1,2,3 \\ 0 & \text{Otherwise} \end{cases} \quad \text{(II.2.9)}$$

so that only nonzero components are

$$\varepsilon_{123} = \varepsilon_{231} = \varepsilon_{312} = 1 \qquad \varepsilon_{132} = \varepsilon_{321} = \varepsilon_{213} = -1 \quad \text{(II.2.10)}$$

The factor 2 in the denominator of (II.2.8) comes from the fact that a summation convention is adopted on the right-hand-side because of the repeating indices.

An intelligent fellow could have chosen an alternative set of basis vectors derived from the covariant bases $\hat{\zeta}^j = \nabla \zeta^j, j = 1,2,3$, simply given by

$$\hat{\zeta}_i = \frac{\varepsilon_{ijk}}{2J} \hat{\zeta}^j \times \hat{\zeta}^k, i = 1,2,3 \quad \text{(II.2.11)}$$

It is easy to verify that the new contravariant bases $\hat{\zeta}_j, j = 1,2,3$ construct a parallelepiped with non-vanishing volume equal to $1/J = \hat{\zeta}_1 \cdot \hat{\zeta}_2 \times \hat{\zeta}_3$, and are hence linearly independent. It should be noted that covariant and contravariant bases usually have physically different dimensions, while in the Cartesian coordinates they coincide. Now (II.2.8) together with (II.2.11) gives the relation for covariant components of **A** as

$$A_j = \mathbf{A} \cdot \hat{\zeta}_j \quad \text{(II.2.12)}$$

Similarly, any vector such as **A** can be expanded in terms of the contravariant bases like (II.2.6) as

$$\mathbf{A} = A^j \hat{\zeta}_j = \frac{\varepsilon_{jkl} A^j}{2J} \hat{\zeta}^k \times \hat{\zeta}^l \quad \text{(II.2.13)}$$

Hence, (II.2.12) is a contravariant representation of **A**. In contrast to (II.2.12), the contravariant components $A^j, j = 1,2,3$ can now be easily found by performing an inner product with $\hat{\zeta}^j$ on both sides to give

$$A^i = A^i \delta_i^j = A^i \hat{\zeta}_i \cdot \hat{\zeta}^j = \mathbf{A} \cdot \hat{\zeta}^j \quad \text{(II.2.14)}$$

in which we have used the relation $\hat{\zeta}_i \cdot \hat{\zeta}^j = \delta_i^j$. Hence, we get the fairly easy relation for the contravariant components

$$A^j = \mathbf{A} \cdot \hat{\zeta}^j \qquad (\text{II}.2.15)$$

Figure II.2.1 shows the typical construction of contravariant and convariant bases in the two-dimensional plane and the curvilinear coordinates for which $\zeta^1 = \zeta^1(x^1, x^2)$, $\zeta^2 = \zeta^2(x^1, x^2)$, and $\zeta^3 = x^3$. In addition, suppose that we have positive Jacobian $J > 0$ everywhere.

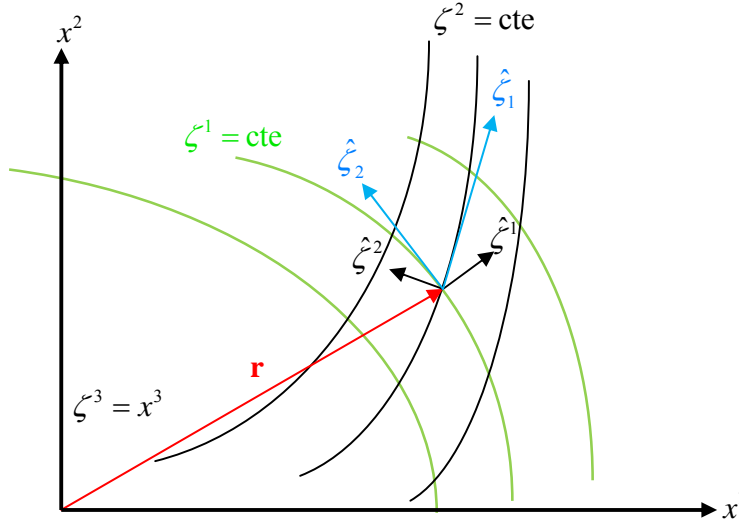

**Figure II.2.1:** Geometrical representation of contravariant and covariant bases in two-dimensional plane.

As it can be seen, the covariant bases $\hat{\zeta}^j$ are by their definition always normal to their respective constant contours given by $\zeta^j = \text{cte}$, while contravariant bases $(\hat{\zeta}_1, \hat{\zeta}_2) \parallel (\hat{\zeta}_2 \times \hat{\zeta}_3, \hat{\zeta}_3 \times \hat{\zeta}_1)$ are respectively tangent to the contours $(\zeta_2 = \text{cte}, \zeta_1 = \text{cte})$. Note that in this example, all vectors $(\hat{\zeta}_1, \hat{\zeta}_2)$ and $(\hat{\zeta}^1, \hat{\zeta}^2)$ lie on the two-dimensional plane $(x^1, x^2)$ because of the special choice of $\zeta^1 = \zeta^1(x^1, x^2)$, $\zeta^2 = \zeta^2(x^1, x^2)$, and $\zeta^3 = x^3$. Hence, the orthogonality relationship holds as $\hat{\zeta}^1 \cdot \hat{\zeta}_2 = \hat{\zeta}^2 \cdot \hat{\zeta}_1 = 0$. However, the normality of covariant bases and tangential property of contravariant bases to contours are quite universal rules, and applicable everywhere the Jacobian does not vanish. We stress again that in general these two sets of bases need to have neither similar physical dimensions nor identical directions.

The only system of coordinates for which both covariant and contravariant vectors share the same physical dimensions and directions is the rectangular Cartesian system of coordinates, and only for the square Cartesian coordinates two bases are equal. On the other hand, orthogonal systems of coordinates are marked with covariant and contravariant vectors pointing to the same directions, while having different physical dimensions. For this to happen, covariant vectors need to be mutually orthogonal. Examples of orthogonal coordinate systems include spherical and cylindrical coordinates. Hence, for an orthogonal coordinate system, we have two further orthogonality relationships given by $\hat{\zeta}_i \cdot \hat{\zeta}_j = 0$ and $\hat{\zeta}^i \cdot \hat{\zeta}^j = 0$, only if $i \neq j$. For these two coordinate systems, contravariant (and hence covariant) bases are always mutually normal, yet position dependent and changing direction from point to point. In contrast, for Cartesian coordinates the direction (as well as the length of) contravariant bases are fixed throughout the space.

### II.2.1 Transformation of Coordinates

We can transform components of an arbitrary vector from a given coordinate system $\left(\zeta^1, \zeta^2, \zeta^3\right)$ into another coordinate system $\left(\bar{\zeta}^1, \bar{\zeta}^2, \bar{\zeta}^3\right)$, related by

$$\begin{aligned} \bar{\zeta}^1\left(\zeta^1, \zeta^2, \zeta^3\right) \\ \bar{\zeta}^2\left(\zeta^1, \zeta^2, \zeta^3\right) \\ \bar{\zeta}^3\left(\zeta^1, \zeta^2, \zeta^3\right) \end{aligned} \tag{II.2.16}$$

By using the covariant representation of the vector **A**, we get

$$\mathbf{A} = A_j \nabla \zeta^j = \bar{A}_i \nabla \bar{\zeta}^i = A_j \frac{\partial \zeta^j}{\partial \bar{\zeta}^i} \nabla \bar{\zeta}^i \tag{II.2.17}$$

from which it can be concluded that

$$\bar{A}_i = A_j \frac{\partial \zeta^j}{\partial \bar{\zeta}^i} \tag{II.2.18}$$

It is easy to verify that transformation law for contravariant components is given by

$$\bar{A}^i = A^j \frac{\partial \bar{\zeta}^i}{\partial \zeta^j} \tag{II.2.19}$$

### II.2.2 Metric Tensor

In order to get contravariant components of vector **A** from its covariant components one should multiply the covariant components by the elements $g^{ij}$

$$A^j = A_i g^{ij} \tag{II.2.20}$$

where $g^{ij} = \hat{\zeta}^i \cdot \hat{\zeta}^j$ is the symmetric *metric tensor* and includes all necessary information about the coordinate system. In particular, if the coordinate system is orthogonal, the metric sensor will be diagonal. Determinant of metric tensor has a relation to Jacobian of coordinate system as

$$\left|g^{ij}\right| = J^2 \tag{II.2.21}$$

From equation (II.2.20) one can obtain the relation of covariant $A_i$ and contravariant components $A^i$ as

$$A_i = \left[g^{ij}\right]^{-1} A^j = g_{ij} A^j \tag{II.2.22}$$

where $g_{ij}$ is the covariant form of metric tensor. We furthermore define the determinant of the covariant metric tensor $g$ as

$$g \equiv |g_{ij}| = \frac{1}{J^2} \tag{II.2.23}$$

The relations (II.2.20) and (II.2.23) are frequently referred to as laws of raising and lowering indices, respectively. Together, these relations constitute the concept of 'index gymnastics'.

### II.2.3 Volume and Surface Elements
Volume element is defined as

$$dV = dx^1 dx^2 dx^3 = \frac{1}{J} d\zeta^1 d\zeta^2 d\zeta^3 = \sqrt{g}\, d\zeta^1 d\zeta^2 d\zeta^3 \tag{II.2.24}$$

Besides, the square line element for coordinate system $(\zeta_1, \zeta_2, \zeta_3)$ is

$$ds^2 = d\mathbf{r} \cdot d\mathbf{r} = \hat{\zeta}^i \cdot \hat{\zeta}^j d\zeta_i d\zeta_j = g^{ij} d\zeta_i d\zeta_j = g_{ij} d\zeta^i d\zeta^j \tag{II.2.25}$$

More often, metrics are represented by their respective line elements in the form (II.2.25), instead of expressing all independent components in matrix form.

### II.2.3 Dot and Cross Product
The inner (dot) product of two arbitrary vectors $\mathbf{A}$ and $\mathbf{B}$ may be easily found if one is represented in covariant and the other in contravariant forms

$$\mathbf{A} \cdot \mathbf{B} = \left(A^i \hat{\zeta}_i\right) \cdot \left(B_j \hat{\zeta}^j\right) = A^i B_j \hat{\zeta}_i \cdot \hat{\zeta}^j = A^i B_j \delta_i^j = A^i B_i = A_i B^i \tag{II.2.26}$$

In order to obtain the cross product, both vectors may be expressed in covariant form. Therefore

$$\begin{aligned} \mathbf{W} = \mathbf{A} \times \mathbf{B} &= \left(A_i \hat{\zeta}^i\right) \times \left(B_j \hat{\zeta}^j\right) \\ &= (A_2 B_3 - A_3 B_2) \hat{\zeta}^2 \times \hat{\zeta}^3 + (A_3 B_1 - A_1 B_3) \hat{\zeta}^3 \times \hat{\zeta}^1 + (A_1 B_2 - A_2 B_1) \hat{\zeta}^1 \times \hat{\zeta}^2 \\ &= J(A_2 B_3 - A_3 B_2) \hat{\zeta}_1 + J(A_3 B_1 - A_1 B_3) \hat{\zeta}_2 + J(A_1 B_2 - A_2 B_1) \hat{\zeta}_3 \end{aligned} \tag{II.2.27}$$

Through comparison to $\mathbf{W} = W^i \hat{\zeta}_i$, the contravariant components of $\mathbf{W}$ can be thus found as

$$W^i = \mathbf{W} \cdot \hat{\zeta}^i = \varepsilon^{ijk} A_j B_k J \tag{II.2.28}$$

where we here define the contravariant Levi-Civita symbol as $\varepsilon^{ijk} = \varepsilon_{ijk}$. The covariant components of $\mathbf{W} = W_i \hat{\zeta}^i$ may be directly obtained from (II.2.28) as

$$W_i = g_{ij} W^j = g_{ij} \varepsilon^{jkl} A_k B_l J = g \varepsilon_i^{kl} A_k B_l J = \frac{1}{J} \varepsilon_i^{jk} A_j B_k \tag{II.2.29}$$

Again, we adopt the definition $\varepsilon_i^{jk} = \varepsilon_{ijk}$, that is the Levi-Civita pseudo-tensor does not transform according to the transformation laws of index gymnastics.

**II.2.4 Gradient, Divergence and Curl Operator**
The gradient operator is by definition given by

$$\nabla = \nabla \zeta^i \frac{\partial}{\partial \zeta^i} = \hat{\zeta}^i \frac{\partial}{\partial \zeta^i} \tag{II.2.30}$$

By applying the gradient operator to a scalar function $f(\zeta^1, \zeta^2, \zeta^3)$ we get a vector field as

$$\mathbf{S} = \nabla f(\zeta^1, \zeta^2, \zeta^3) = \hat{\zeta}^i \frac{\partial f}{\partial \zeta^i} \tag{II.2.31}$$

Comparing to (II.2.27), we directly obtain the covariant components of $\mathbf{S}$ as

$$S_i = \frac{\partial f}{\partial \zeta^i} \tag{II.2.32}$$

Now, using (II.2.22) one can find the contravariant components of $\mathbf{S}$ as

$$S^i = g^{ij} \frac{\partial f}{\partial \zeta^j} \tag{II.2.33}$$

One can take the directional derivative of any vector $\mathbf{A}$ along curvilinear coordinates

$$\frac{\partial}{\partial \zeta^j} \mathbf{A} = \frac{\partial}{\partial \zeta^j} \left( A_i \hat{\zeta}^i \right) = \frac{\partial A_i}{\partial \zeta^j} \hat{\zeta}^i + A_i \frac{\partial \hat{\zeta}^i}{\partial \zeta^j} \tag{II.2.34}$$

Here, the second term expresses the dependence of basis vectors on coordinates, and identically vanishes for Cartesian coordinates. Hence, we can write the covariant components of directional derivative of vector $\mathbf{A}$, or $A_{i,j}$ as

$$A_{i,j} = \left(\frac{\partial}{\partial \zeta^j} \mathbf{A}\right)_i = \frac{\partial}{\partial \zeta^j} \mathbf{A} \cdot \hat{\zeta}_i$$
$$= \frac{\partial A_i}{\partial \zeta^j} + \Gamma^k_{ji} A_k \tag{II.2.35}$$

On the other hand, the contravariant components of directional derivative of vector $\mathbf{A}$, $A^i_{,j}$ can be obtained as

$$A^i_{,j} = \left(\frac{\partial}{\partial \zeta^j} \mathbf{A}\right)^i = \left[\frac{\partial}{\partial \zeta^j}\left(A^k \hat{\zeta}_k\right)\right] \cdot \hat{\zeta}^i = \frac{\partial A^i}{\partial \zeta^j} + A^k \frac{\partial \hat{\zeta}_k}{\partial \zeta^j} \cdot \hat{\zeta}^i$$
$$= \frac{\partial A^j}{\partial \zeta^j} - \Gamma^i_{jk} A^k \tag{II.2.36}$$

In the latter relations, $\Gamma$ is referred to as the *Christoffel Symbol* and is defined as

$$\Gamma^k_{ji} \triangleq \frac{\partial \hat{\zeta}^k}{\partial \zeta^j} \cdot \hat{\zeta}_i = -\frac{\partial \hat{\zeta}_i}{\partial \zeta^j} \cdot \hat{\zeta}^k \tag{II.2.37}$$

The Christoffel Symbol can be presented in terms of the metric tensor and more convenient form

$$\Gamma^i_{jk} = -\frac{1}{2} g^{im} \left(\frac{\partial g_{mi}}{\partial \zeta^j} + \frac{\partial g_{mj}}{\partial \zeta^i} - \frac{\partial g_{jk}}{\partial \zeta^m}\right) \tag{II.2.38}$$

The divergence operator $\nabla \cdot$ acts on a vector field and is defined as

$$\nabla \cdot \mathbf{A} = \hat{\zeta}^i \frac{\partial}{\partial \zeta^i} \cdot \left(A^j \hat{\zeta}_j\right) = \frac{\partial A^j}{\partial \zeta^i} \hat{\zeta}_j \cdot \hat{\zeta}^i + A^j \frac{\partial \hat{\zeta}_j}{\partial \zeta^i} \cdot \hat{\zeta}^i = \frac{\partial A^i}{\partial \zeta^i} + A^j \frac{\partial}{\partial \zeta^i} \left(\frac{\varepsilon_{jkl}}{2J} \hat{\zeta}^k \times \hat{\zeta}^l\right) \cdot \hat{\zeta}^i$$
$$= \frac{\partial A^i}{\partial \zeta^i} + A^j \frac{\varepsilon_{jkl}}{2J} \frac{\partial}{\partial \zeta^i} \left(\hat{\zeta}^k \times \hat{\zeta}^l\right) \cdot \hat{\zeta}^i + A^j \varepsilon_{jkl} \left(\hat{\zeta}^k \times \hat{\zeta}^l\right) \cdot \hat{\zeta}^i \frac{\partial}{\partial \zeta^i}\left(\frac{1}{2J}\right) \tag{II.2.39}$$
$$= \frac{\partial A^i}{\partial \zeta^i} + A^j \varepsilon_{jkl} \varepsilon_{ikl} J \frac{\partial}{\partial \zeta^i}\left(\frac{1}{2J}\right) = \frac{\partial A^i}{\partial \zeta^i} + A^j 2J \frac{\partial}{\partial \zeta^j}\left(\frac{1}{2J}\right)$$

in which we have made use of the identity $\partial \zeta^j / \partial \zeta^i = \delta^{ij}$ and thus $\partial \hat{\zeta}^j / \partial \zeta^i = \mathbf{0}$, as well as $\varepsilon_{jkl} \varepsilon_{ikl} = 2\delta_{ij}$. Finally the relation of divergence of a vector field in curvilinear coordinates simplifies to the compact form

$$\nabla \cdot \mathbf{A} = J \frac{\partial}{\partial \zeta^i}\left(\frac{1}{J} A^j\right) \tag{II.2.40}$$

Similarly, it is possible to express the rotation or curl of a vector field $\mathbf{A}$, which is the vector product of the operator $\nabla$ and the vector $\mathbf{A}$ given by

$$\nabla \times \mathbf{A} = \hat{\zeta}^i \frac{\partial}{\partial \zeta^i} \times \mathbf{A} \tag{II.2.41}$$

The contravariant components of (II.2.41) are

$$(\nabla \times \mathbf{A})^i = \hat{\zeta}^i \cdot \hat{\zeta}^j \times \frac{\partial}{\partial \zeta^j}\left(A_k \hat{\zeta}^k\right) = \hat{\zeta}^i \cdot \hat{\zeta}^j \times \hat{\zeta}^k \frac{\partial A_k}{\partial \zeta^j} + A_k \hat{\zeta}^i \cdot \hat{\zeta}^j \times \frac{\partial \hat{\zeta}^k}{\partial \zeta^j}$$
$$= J\varepsilon_{ijk} \frac{\partial}{\partial \zeta^j} A_k \tag{II.2.42}$$

Here, the identity $\partial \hat{\zeta}^j / \partial \zeta^i = \mathbf{0}$ is exploited again.

## II. 3. Flux Coordinates

Many operators take on simple forms in flat coordinate systems such as Cartesian coordinates and are easy to remember and evaluate. However, when problems deal with toroidally symmetric systems it is often helpful to use coordinate systems, which exploit the toroidal symmetry and in particular nested toroidal shape of closed magnetic surfaces. In order to study toroidal devices such as tokamaks, we have to choose the handiest coordinate system so that the equations of equilibrium as well as major plasma parameters become straightforward. At first, we consider the *Primitive Toroidal Coordinates*, in which an arbitrary point in 3-space can be uniquely identified by a set of one radial and two angle coordinates. Then we proceed to study the *Flux Coordinates*, in which poloidal cross section of magnetic surfaces look as concentric circles. We also mention Boozer and Hamada systems as particular cases of flux coordinates.

### II. 3.1 Primitive Toroidal Coordinates

Figure II.3.1 shows primitive toroidal coordinates $(r_0, \theta_0, \zeta_0)$ in which $r_0$ is the distance measured from the plasma major axis, $\theta_0$ is the poloidal angle and $\zeta_0$ is the toroidal angle; axisymmetric equilibrium rules out any dependence on the toroidal angle $\zeta_0$.

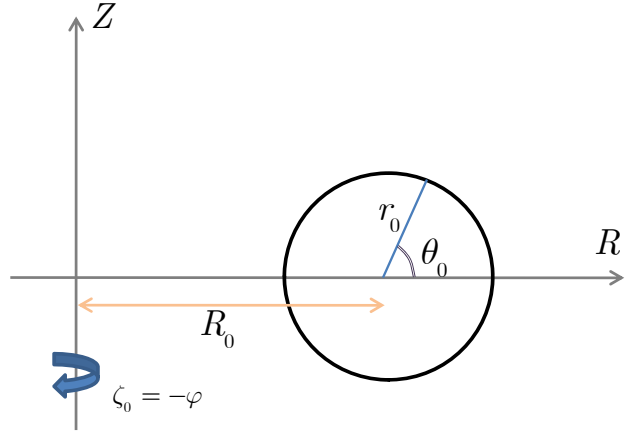

**Figure II.3.1:** Primitive toroidal coordinates.

The values assumed by these coordinates are physically limited according to $0 \leq r_0 < \infty$, $0 \leq \theta_0 < 2\pi$ and $0 \leq \zeta_0 < 2\pi$. Relations between primitive toroidal and cylindrical coordinates $(R, \varphi, Z)$ are

$$\begin{aligned} r_0 &= \sqrt{(R-R_0)^2 + Z^2} \\ \theta_0 &= \tan^{-1}\left(\frac{Z}{R-R_0}\right) \\ \zeta_0 &= -\varphi \end{aligned} \qquad \text{(II.3.1)}$$

where $R_0$ is major radius of plasma, and the minus sign in the third equation is for maintaining right-handedness of the system. This primitive toroidal coordinates is evidently orthogonal and its metric tensor is therefore diagonal. Consequently the squared differential length is

$$\begin{aligned} ds^2 &= dR^2 + R^2 d\varphi^2 + dz^2 \\ &= dr_0^2 + r_0^2 d\theta_0^2 + (R_0 + r_0 \cos\theta_0)^2 d\zeta_0^2 \end{aligned} \qquad \text{(II.3.2)}$$

Hence the metric coefficients of primitive toroidal coordinates are given by $g_{rr} = 1$, $g_{\theta\theta} = r_0$ and $g_{\zeta\zeta} = R = R_0 + r_0 \cos\theta_0$; all other metric coefficients are zero. Using equation (II.2.5), the Jacobian determinant of this system is found to be $1/rR_0$.

The gradient of a scalar field $f$ in primitive coordinates is simply

$$\nabla f = \frac{\partial f}{\partial r_0}\hat{r}_0 + \frac{1}{r_0}\frac{\partial f}{\partial \theta_0}\hat{\theta}_0 + \frac{1}{R_0 + r_0 \cos\theta_0}\frac{\partial f}{\partial \zeta_0}\hat{\zeta}_0 \qquad \text{(II.3.3)}$$

The divergence and curl of a vector field **A** are

$$\nabla \cdot \mathbf{A} = \frac{1}{R_0 + r_0 \cos\theta_0} \left\{ \frac{1}{r_0} \frac{\partial}{\partial r_0} \left[ r_0 \left( R_0 + r_0 \cos\theta_0 \right) A_{r_0} \right] \right\}$$
$$+ \frac{\partial A_{\zeta_0}}{\partial \zeta_0} + \frac{1}{r_0} \frac{\partial}{\partial \theta_0} \left[ \left( R_0 + r_0 \cos\theta_0 \right) A_{\theta_0} \right] \quad \text{(II.3.4)}$$

$$\nabla \times \mathbf{A} = \frac{1}{\left( R_0 + r_0 \cos\theta_0 \right)} \left\{ \frac{\partial A_{\theta_0}}{\partial \zeta_0} - \frac{1}{r_0} \frac{\partial}{\partial \theta_0} \left[ \left( R_0 + r_0 \cos\theta_0 \right) A_{\zeta_0} \right] \right\} \hat{r}_0$$
$$+ \frac{1}{\left( R_0 + r_0 \cos\theta_0 \right)} \left\{ \frac{\partial}{\partial r_0} \left[ \left( R_0 + r_0 \cos\theta_0 \right) A_{\zeta_0} \right] - \frac{\partial A_{r_0}}{\partial \zeta_0} \right\} \hat{\theta}_0 \quad \text{(II.3.5)}$$
$$+ \frac{1}{r_0} \left\{ \frac{\partial A_{r_0}}{\partial \theta_0} - \frac{\partial}{\partial r_0} \left( r_0 A_{\theta_0} \right) \right\} \hat{\zeta}_0$$

It can be shown that the non-vanishing Christoffel symbols in primitive toroidal coordinates are

$$\Gamma^2_{21} = -\Gamma^2_{12} = \frac{\cos\theta_0}{R_0 + r_0 \cos\theta_0}$$
$$\Gamma^2_{23} = -\Gamma^2_{32} = \frac{-\sin\theta_0}{R_0 + r_0 \cos\theta_0} \quad \text{(II.3.6)}$$
$$\Gamma^3_{31} = -\Gamma^3_{13} = \frac{1}{r_0}$$

## II. 3.2 Flux Coordinates

In confined toroidal plasmas, magnetic field lines define closed magnetic surfaces due to a famous 'hairy ball' theorem proven by Poincaré, which implies that field lines of a non-zero magnetic field must cover a toroidal surface, as shown in Fig. II.3.2. These define surfaces, to which the particles are approximately constrained, known as flux surfaces. The surfaces are mathematically expressed as constant poloidal flux surfaces, denoted by $\psi = \text{cte}$. Magnetic surfaces for equilibrium plasmas with no external current drive coincide with isobar, i.e. constant pressure surfaces. Figure II.3.3 shows magnetic field lines as well as current density lines, which lie on these nested isobaric flux surfaces.

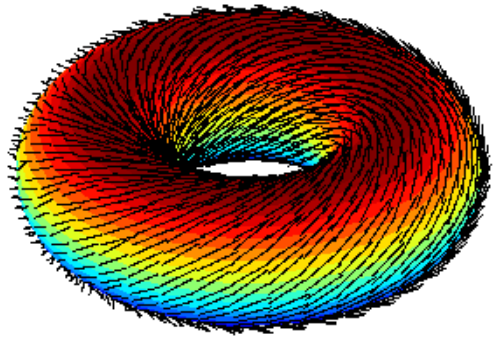

**Figure II.3.2:** A hairy doughnot.

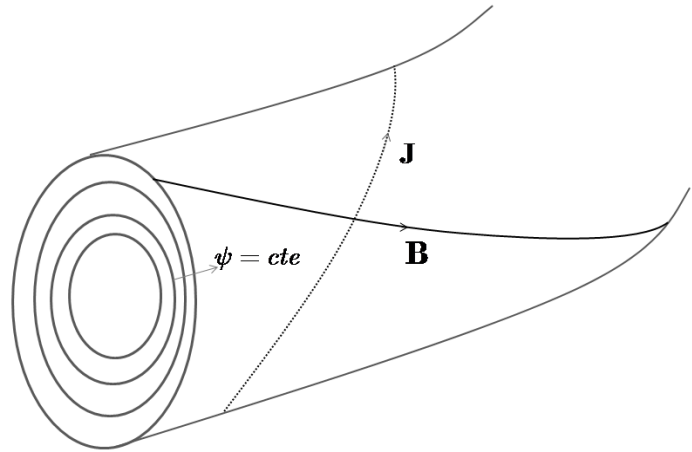

**Figure II.3.3:** Both magnetic field and current density lines lie on nested magnetic surfaces.

The flux coordinates $(\psi, \chi, \zeta)$ shown in Figure II.3.4, represent the true complicated physical shape of magnetic surfaces, and are here expressed as functions of the primitive toroidal coordinates $(r_0, \theta_0, \zeta_0)$. $\psi$ denotes the poloidal magnetic flux (or any monotonic function of), and $\chi$ and $\zeta$ are respectively referred to as poloidal and toroidal angles. The latter two coordinates are not true angles although they have the dimension of radians. Therefore we have

$$\begin{aligned} \psi &= \psi(r_0, \theta_0, \zeta_0) \\ \chi &= \chi(r_0, \theta_0, \zeta_0) \\ \zeta &= \zeta(r_0, \theta_0, \zeta_0) \end{aligned} \qquad \text{(II.3.7)}$$

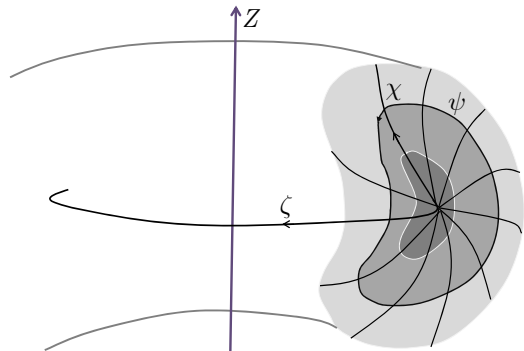

**Figure II.3.4:** Flux coordinates

Since the coordinates $\chi$ and $\zeta$ are similar to angles, then any physical quantity such as $A = A(\psi, \chi, \zeta)$ in flux coordinates should be periodic as $A = A(\psi, \chi + 2m\pi, \zeta + 2n\pi)$. This necessitates the transformation (II.3.7) to be of the form

$$\psi = \psi_0(r) + \sum_{\substack{m,n \\ (m,n)\neq(0,0)}} \psi_{mn}(r) e^{i(m\theta+n\varphi)}$$

$$\chi = \theta + \sum_{m,n} \theta_{mn}(r) e^{i(m\theta+n\varphi)} \qquad \text{(II.3.8)}$$

$$\zeta = -\varphi + \sum_{m,n} \varphi_{mn}(r) e^{i(m\theta+n\varphi)}$$

From (II.2.24) the volume element in Flux coordinate is defined as

$$dV = \sqrt{g}\, d\zeta^1 d\zeta^2 d\zeta^3 = J^{-1} d\psi d\chi d\zeta \qquad \text{(II.3.9)}$$

where the Jacobian is

$$J = \begin{vmatrix} \psi_x & \psi_y & \psi_z \\ \chi_x & \chi_y & \chi_z \\ \zeta_x & \zeta_y & \zeta_z \end{vmatrix} = \hat{\psi}\cdot\hat{\chi}\times\hat{\zeta} > 0 \qquad \text{(II.3.10)}$$

Here, $\hat{\psi} = \nabla\psi$, $\hat{\chi} = \nabla\chi$, and $\hat{\zeta} = \nabla\zeta$ are the covariant basis vectors. Hence, the corresponding contravariant bases in flux coordinate system are $\hat{\chi}\times\hat{\zeta}$, $\hat{\zeta}\times\hat{\psi}$, and $\hat{\psi}\times\hat{\chi}$, respectively. Hence, the contravariant representations of magnetic field as well as current density in flux coordinates are

$$\mathbf{B} = B^\psi(\psi,\chi,\zeta)\frac{\hat{\chi}\times\hat{\zeta}}{J} + B^\chi(\psi,\chi,\zeta)\frac{\hat{\zeta}\times\hat{\psi}}{J} + B^\zeta(\psi,\chi,\zeta)\frac{\hat{\psi}\times\hat{\chi}}{J} \qquad \text{(II.3.11)}$$

$$\mathbf{J} = J^\psi(\psi,\chi,\zeta)\frac{\hat{\chi}\times\hat{\zeta}}{J} + J^\chi(\psi,\chi,\zeta)\frac{\hat{\zeta}\times\hat{\psi}}{J} + J^\zeta(\psi,\chi,\zeta)\frac{\hat{\psi}\times\hat{\chi}}{J} \qquad \text{(II.3.12)}$$

where

$$\begin{aligned} B^\psi &= \mathbf{B}\cdot\hat{\psi} & J^\psi &= \mathbf{J}\cdot\hat{\psi} \\ B^\chi &= \mathbf{B}\cdot\hat{\chi} & J^\chi &= \mathbf{J}\cdot\hat{\chi} \\ B^\zeta &= \mathbf{B}\cdot\hat{\zeta} & J^\zeta &= \mathbf{J}\cdot\hat{\zeta} \end{aligned} \qquad \text{(II.3.13)}$$

As both magnetic field lines and current density lines lie on magnetic surfaces, we must have

$$\begin{aligned} B^\psi &= \mathbf{B}\cdot\hat{\psi} = 0 \\ J^\psi &= \mathbf{J}\cdot\hat{\psi} = 0 \end{aligned} \qquad \text{(II.3.14)}$$

since contravariant field components are tangent to the surfaces and have no component along the corresponding covariant bases. On the other from Maxwell's equation we know that $\nabla \cdot \mathbf{B} = 0$, and therefore by applying divergence operator we obtain

$$\nabla \cdot \mathbf{B} = J\left[\frac{\partial}{\partial \zeta}\left(\frac{B^\zeta}{J}\right) + \frac{\partial}{\partial \chi}\left(\frac{B^\chi}{J}\right)\right] = 0 \quad \text{(II.3.15)}$$

which results in

$$\frac{\partial}{\partial \zeta}\left(\frac{B^\zeta}{J}\right) = -\frac{\partial}{\partial \chi}\left(\frac{B^\chi}{J}\right) \quad \text{(II.3.16)}$$

The continuity equation for electric charge also reads

$$\nabla \cdot \mathbf{J} = -\frac{\partial \rho}{\partial t} \quad \text{(II.3.17)}$$

But the time derivative $\partial/\partial t$ vanishes under steady-state assumption, therefore $\nabla \cdot \mathbf{J} = 0$, and similar to (II.3.16) we get

$$\frac{\partial}{\partial \zeta}\left(\frac{J^\zeta}{J}\right) = -\frac{\partial}{\partial \chi}\left(\frac{J^\chi}{J}\right) \quad \text{(II.3.18)}$$

Now we adopt the notations $b^\zeta = B^\zeta/J$, $b^\chi = B^\chi/J$, and $j^\zeta = J^\zeta/J$, $j^\chi = J^\chi/J$, which by (II.3.16) allows us to write down

$$\begin{aligned} b^\chi(\psi,\chi,\zeta) &= \overline{b}^\chi(\psi) - \widehat{b}(\psi)\chi - \frac{\partial \widetilde{b}(\psi,\chi,\zeta)}{\partial \zeta} \\ b^\zeta(\psi,\chi,\zeta) &= \overline{b}^\zeta(\psi) + \widehat{b}(\psi)\zeta + \frac{\partial \widetilde{b}(\psi,\chi,\zeta)}{\partial \chi} \end{aligned} \quad \text{(II.3.19)}$$

$$\begin{aligned} j^\chi(\psi,\chi,\zeta) &= \overline{j}^\chi(\psi) - \widehat{j}(\psi)\chi - \frac{\partial \widetilde{j}(\psi,\chi,\zeta)}{\partial \zeta} \\ j^\zeta(\psi,\chi,\zeta) &= \overline{j}^\zeta(\psi) + \widehat{j}(\psi)\zeta + \frac{\partial \widetilde{j}(\psi,\chi,\zeta)}{\partial \chi} \end{aligned} \quad \text{(II.3.20)}$$

Because of periodicity with respect to the angular coordinates $\chi$ and $\zeta$, we need $\widehat{b}(\psi) = \widehat{j}(\psi) = 0$. On the other hand the auxiliary functions need to obey

$$\tilde{b}(\psi,\chi,\zeta) = \tilde{b}(\psi,\chi+2m\pi,\zeta+2n\pi)$$
$$\tilde{j}(\psi,\chi,\zeta) = \tilde{j}(\psi,\chi+2m\pi,\zeta+2n\pi)$$
(II.3.21)

Hence, (II.3.11) and (II.3.12) can be rewritten as

$$\mathbf{B} = b^\chi(\psi,\chi,\zeta)\hat{\zeta}\times\hat{\psi} + b^\zeta(\psi,\chi,\zeta)\hat{\psi}\times\hat{\chi}$$
$$= \left[\overline{b}^\chi(\psi) - \frac{\partial \tilde{b}(\psi,\chi,\zeta)}{\partial \zeta}\right]\hat{\zeta}\times\hat{\psi} + \left[\overline{b}^\zeta(\psi) + \frac{\partial \tilde{b}(\psi,\chi,\zeta)}{\partial \chi}\right]\hat{\psi}\times\hat{\chi}$$
(II.3.21)

$$\mathbf{J} = j^\chi(\psi,\chi,\zeta)\hat{\zeta}\times\hat{\psi} + j^\zeta(\psi,\chi,\zeta)\hat{\psi}\times\hat{\chi}$$
$$= \left[\overline{j}^\chi(\psi) - \frac{\partial \tilde{j}(\psi,\chi,\zeta)}{\partial \zeta}\right]\hat{\zeta}\times\hat{\psi} + \left[\overline{j}^\zeta(\psi) + \frac{\partial \tilde{j}(\psi,\chi,\zeta)}{\partial \chi}\right]\hat{\psi}\times\hat{\chi}$$
(II.3.22)

But, the current density **J** and magnetic field **B** are further interrelated by the Ampere's law $\nabla\times\mathbf{B} = \mu_0\mathbf{J}$, and thus noting (II.3.22) we get the alternative form

$$\mathbf{J} = \frac{1}{\mu_0}\nabla\times\mathbf{B} = \nabla\times\left[-\tilde{c}(\psi,\chi,\zeta)\hat{\psi} + c_\chi(\psi)\hat{\chi} + c_\zeta(\psi)\hat{\zeta}\right]$$
(II.3.23)

for which

$$\overline{j}^\chi(\psi) = +\frac{dc_\zeta(\psi)}{d\psi}$$
$$\overline{j}^\zeta(\psi) = -\frac{dc_\chi(\psi)}{d\psi}$$
(II.3.24)

$$\tilde{c}(\psi,\chi,\zeta) = \tilde{c}(\psi,\chi+2m\pi,\zeta+2n\pi)$$
(II.3.25)

must hold according to (II.2.42) and (II.3.8). The magnetic field can be thus derived from

$$\mathbf{B} = \mu_0\left[-\tilde{c}(\psi,\chi,\zeta)\hat{\psi} + b_\chi(\psi)\hat{\chi} + b_\zeta(\psi)\hat{\zeta} + \nabla g(\psi,\chi,\zeta)\right]$$
(II.3.26)

where $g(\psi,\chi,\zeta)$ is an arbitrary function given by

$$g(\psi,\chi,\zeta) = \tilde{g}(\psi,\chi,\zeta) + a_\chi\chi + a_\zeta\zeta + a$$
$$\tilde{g}(\psi,\chi,\zeta) = \tilde{g}(\psi,\chi+2m\pi,\zeta+2n\pi)$$
(II.3.27)

in which $a$, $a_\chi$ and $a_\zeta$ are constants. It is obvious that the latter two constants are trivial and may be respectively absorbed in the functions $b_\chi(\psi)$ and $b_\zeta(\psi)$, and thus may be safely ignored.

### II. 3.3. Boozer Coordinates

Allen Boozer showed that by a proper transformation of coordinates, one could even get rid of $\tilde{g}(\psi,\chi,\zeta)$. This transformation leads us to the so-called Boozer coordinate system. The required transformation is

$$\begin{aligned}
\psi' &= \psi \\
\chi' &= \chi + \overline{A}_\chi(\psi)\tilde{G}(\psi,\chi,\zeta) + \overline{C}_\chi(\psi)\tilde{F}(\psi,\chi,\zeta) \\
\zeta' &= \zeta + \overline{A}_\zeta(\psi)\tilde{G}(\psi,\chi,\zeta) + \overline{C}_\zeta(\psi)\tilde{F}(\psi,\chi,\zeta)
\end{aligned} \qquad \text{(II.3.28)}$$

with $\overline{A}_\chi$, $\overline{A}_\zeta$, $\overline{C}_\chi$, $\overline{C}_\zeta$, and $\tilde{G}$ and $\tilde{F}$ being arbitrary functions satisfying $\langle\tilde{G}\rangle = \langle\tilde{F}\rangle = 0$; here, $\langle\cdot\rangle$ stands for angular average. Appropriate candidate for these functions are

$$\begin{aligned}
\psi' &= \psi \\
\chi' &= \chi + \frac{1}{b^\zeta c_\zeta + b^\chi c_\chi}\left(b^\chi \tilde{g} + \tilde{b}c_\zeta\right) \\
\zeta' &= \zeta + \frac{1}{b^\zeta c_\zeta + b^\chi c_\chi}\left(b^\zeta \tilde{g} - \tilde{b}c_\chi\right)
\end{aligned} \qquad \text{(II.3.29)}$$

Finally, one can find covariant and contravariant forms of the magnetic field as follows

$$\begin{aligned}
\mathbf{B} &= b^\chi \hat{\zeta} \times \hat{\psi} + b^\zeta \hat{\psi} \times \hat{\chi} \\
&= \tilde{c}\hat{\psi} + B_\chi \hat{\chi} + B_\zeta \hat{\zeta}
\end{aligned} \qquad \text{(II.3.30)}$$

It may be shown that

$$\tilde{c}(\psi,\chi,\zeta) = \left(b^\chi \tilde{g} + B_\zeta \tilde{b}\right)\frac{dB_\chi}{d\psi} + \left(b^\zeta \tilde{g} - B_\chi \tilde{b}\right)\frac{dB_\zeta}{d\psi} \qquad \text{(II.3.31)}$$

$$\begin{aligned}
b^\zeta(\psi) &= q(\psi) \\
b^\chi(\psi) &= 1 \\
B_\chi(\psi) &= \mu_0 I_t(\psi)/2\pi \triangleq i_t \\
B_\zeta(\psi) &= \mu_0\left[I_p^{\text{coil}}(\psi) - I_p^{\text{plasma}}(\psi)\right]/2\pi \triangleq i_p(\psi)
\end{aligned} \qquad \text{(II.3.32)}$$

Here, $q(\psi)$ is the safety factor of plasma, which is number of turns the helical magnetic field lines in a tokamak makes round the major circumference per each turn of the minor circumference, $I_t(\psi)$ is the toroidal current within the magnetic surface $\psi$, and $i_p(\psi)$ is the poloidal current difference between poloidal coils and plasma within the magnetic surface $\psi$.

Upon substituting (II.3.31) and (II.3.32) into (II.3.30) we obtain the fairly simple forms of contravariant and covariant representations of the magnetic field as

$$\begin{aligned}\mathbf{B} &= q(\psi)\hat{\psi}\times\hat{\chi}+\hat{\zeta}\times\hat{\psi} \\ &= i_t(\psi)\hat{\chi}+i_p(\psi)\hat{\zeta}+\tilde{c}\hat{\psi}\end{aligned} \quad\quad (\text{II}.3.33)$$

This shows that the covariant and contravariant components of the magnetic field are given as

$$\begin{array}{ll} B^\psi = 0 & B_\psi = \tilde{c} \\ B^\chi = J & B_\chi = i_t(\psi) \\ B^\zeta = Jq(\psi) & B_\zeta = i_p(\psi) \end{array} \quad\quad (\text{II}.3.34)$$

Equation (II.3.34) displays both covariant and contravariant components of magnetic field in flux coordinates. The squared magnitude of magnetic field $B^2 = \mathbf{B}\cdot\mathbf{B}$, may be readily found from $B^2 = B_i B^i$ as

$$\begin{aligned}\mathbf{B}\cdot\mathbf{B} &= \left[\tilde{k}(\psi,\chi,\zeta)\hat{\psi}+i_t(\psi)\hat{\chi}+i_p(\psi)\hat{\zeta}\right]\cdot\left[q(\psi)\hat{\psi}\times\hat{\chi}+\hat{\zeta}\times\hat{\psi}\right] \\ &= Jqi_p(\psi)+Ji_t(\psi)\end{aligned} \quad\quad (\text{II}.3.35)$$

Jacobian can thus be determined as

$$J = \frac{B^2}{q(\psi)i_p(\psi)+i_t(\psi)} = \frac{B_p^2+B_t^2}{q(\psi)i_p(\psi)+i_t(\psi)} \quad\quad (\text{II}.3.36)$$

### II. 3.4. Hamada Coordinates

In general, the Jacobian vary as a function of coordinates like $J = J(\psi,\chi,\zeta)$. Hamada coordinates $(\psi_H,\chi_H,\zeta_H)$ are chosen in such a way that the Jacobian $J$ is made a flux label; a scalar flux label function has the characteristic that its gradient is always parallel to $\hat{\psi}$ and hence a function of only $\psi$. For the particular choice of $\psi_H = V(\psi)/(2\pi)^2$, $\chi_H = \chi$, and $\zeta_H = \zeta$ it can be shown that $J = 1$, where $V(\psi)$ is the volume of magnetic flux tube bounded by the poloidal flux $\psi$. By virtue of Hamada coordinates, (II.3.34), (II.3.12) and (II.3.14) we also readily obtain

$$B^\psi = 0 \qquad\qquad J^\psi = 0$$
$$B^\chi = B^\chi(\psi) = 1 \qquad\qquad J^\chi = J^\chi(\psi) \qquad\qquad \text{(II.3.35)}$$
$$B^\zeta = B^\zeta(\psi) = q(\psi) \qquad\qquad J^\zeta = J^\zeta(\psi)$$

In other words, all contravariant components of fields become flux functions. Use of Hamada coordinates also implies many other attractive features, some of which will be discussed in the following. It can be furthermore shown that for a toroidal plasma equilibrium Hamada coordinates exists either in absence of pressure anisotropy or under axisymmetry. The former condition is automatically met in most practical situations where no external heating mechanism is used.

## II. 4. Extensions to Axisymmetric Equilibrium

In order to have steady state fusion energy, the hot plasma of Tokamak or other promising toroidal devices such as stellarator should be kept away at equilibrium from the first wall. Without use of strong magnetic field, the confinement of this hot plasma is out of reach. Tokamaks are axisymmetric machines which make their analysis much easier. Although recent progress in this field has resulted in some novel equilibrium configurations [29-32], however, we limit the discussion to the well-established cases.

### II. 4.1 MHD Equilibrium

From MHD momentum balance equation (II.1.2) we have:

$$\rho \frac{d\mathbf{V}}{dt} = -\nabla p + \mathbf{J} \times \mathbf{B} \qquad\qquad \text{(II.4.1)}$$

By using Ampere's law, we can eliminate the current density $\mathbf{J}$ from the $\mathbf{J} \times \mathbf{B}$ force term to get

$$\mathbf{J} \times \mathbf{B} = \frac{1}{\mu_0}(\nabla \times \mathbf{B}) \times \mathbf{B} \qquad\qquad \text{(II.4.2)}$$

Now by means of vector identity, $\nabla(\mathbf{A} \cdot \mathbf{B}) = (\mathbf{A} \cdot \nabla)\mathbf{B} + \mathbf{A} \times (\nabla \times \mathbf{B}) + (\mathbf{B} \cdot \nabla)\mathbf{A} + \mathbf{B} \times (\nabla \times \mathbf{A})$, one can rewrite (II.4.2) as

$$\mathbf{J} \times \mathbf{B} = \frac{1}{\mu_0}(\mathbf{B} \cdot \nabla)\mathbf{B} - \nabla\left(\frac{B^2}{2\mu_0}\right) \qquad\qquad \text{(II.4.3)}$$

The left-hand-side of (II.4.1) under equilibrium vanishes and therefore by substituting (II.4.3) in (II.4.1) we have

$$\nabla p = \mathbf{J} \times \mathbf{B} = \frac{1}{\mu_0}(\mathbf{B} \cdot \nabla)\mathbf{B} - \nabla\left(\frac{B^2}{2\mu_0}\right) \qquad\qquad \text{(II.4.4)}$$

This is equilibrium equation which states that under equilibrium, the pressure gradient is balanced by forces due to magnetic field curvature and pressure gradient. The next thing that may be inferred from (II.4.4) is that the current and magnetic field lie on isobaric surfaces. We accepted this fact without proof, but this consequence arises from the fact $\mathbf{J} \cdot \nabla p = \mathbf{B} \cdot \nabla p = 0$ while $\nabla p$ is normal to isobar surfaces.

Now rewriting (II.4.4) gives

$$\nabla \left( p + \frac{B^2}{2\mu_0} \right) = \frac{1}{\mu_0} (\mathbf{B} \cdot \nabla) \mathbf{B} \tag{II.4.5}$$

When the field lines are straight and parallel (with no curvature), the right-hand-side of (II.4.5) vanishes and it reduces to a simple form that the total pressure is constant everywhere within a plasma

$$p + \frac{B^2}{2\mu_0} = \frac{B^2}{2\mu_0}(1+\beta) = \text{cte} \tag{II.4.6}$$

where $\beta$ is here defined as

$$\beta \triangleq \frac{p}{B^2/2\mu_0} \tag{II.4.7}$$

According to (II.4.7), $\beta$ is the ratio of plasma pressure to magnetic field pressure and is one of the figures of merit for magnetic confinement devices. It should be mentioned that for practical confinement geometries (II.4.7) is not applicable and an averaged definition for $\beta$ is needed. In view of the fact that fusion reactivity increases with plasma pressure, a high value of beta is a sign of good performance. The highest value of beta achieved in a large tokamak is about 13%, though higher values are theoretically possible at lower aspect ratio. There is a theoretical limit on the maximum $\beta$ that can be achieved in a magnetic plasma and is due to deterioration in the confinement. The *Troyon $\beta$ limit* which states that for a stable plasma operation $\beta$ cannot be greater than $gI/aB$ where $g$ is Troyon coefficient and has a value of about 3.5 for conventional tokamaks, $I$ is the plasma current in Mega Amperes, $a$ is the minor radius in meters and $B$ is the toroidal field in Tesla.

### II. 4.2 Z-pinch Equilibrium

As an example, we are going to evaluate equilibrium of an ideal Z-pinch, a conceptual one-dimensional magnetic confinement device, which confines the plasma in cylindrical geometry by using an axial current and poloidal magnetic windings. In cylindrical coordinates $(r, \varphi, z)$ the equilibrium equation (II.4.4), for Z-pinch takes the form

$$\frac{\partial p}{\partial r} = -J_z B_\varphi \tag{II.4.8}$$

Using Ampere's law,

$$J_z = \frac{1}{\mu_0}\frac{\partial}{\partial r}\left(rB_\varphi\right) \tag{II.4.9}$$

Substituting (II.4.9) into (II.4.8) gives

$$\frac{\partial p}{\partial r} = -\frac{1}{\mu_0}\left(\frac{B_\varphi^2}{r} + \frac{\partial}{\partial r}\frac{B_\varphi^2}{2}\right) \tag{II.4.10}$$

Assuming that a uniform current distribution $J_z = \text{const}$ flows along the z-axis for $r \leq a$, and by integrating (II.4.9) one can obtain

$$B_\varphi(r) = \begin{cases} \dfrac{\mu_0}{2}J_z r, & r \leq a \\ \dfrac{\mu_0}{2}J_z \dfrac{a^2}{r}, & r > a \end{cases} \tag{II.4.11}$$

Now (II.4.11) can be integrated to give equilibrium pressure distribution for $r \leq a$ as follows

$$p(r) = \frac{1}{4}\mu_0 J_z^2\left(a^2 - r^2\right) \tag{II.4.12}$$

Magnetic field and pressure Profiles of Z-pinch for uniform current density are depicted in Figure II.4.1.

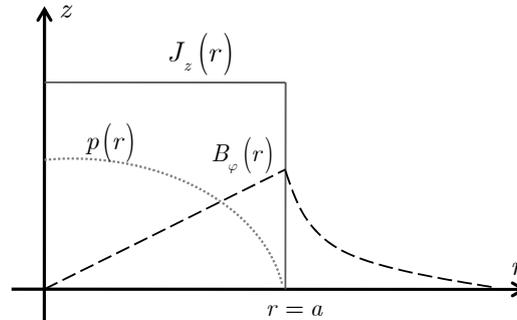

**Figure II.4.1:** Z-pinch profiles

## II. 4.3 $\theta$-pinch Equilibrium

Another conceptual one-dimensional magnetic confinement device, which confines the plasma in cylindrical geometry is $\theta$-pinch. Due to the fact that in a $\theta$-pinch, the current is azimuthal and the magnetic field is axial, the equilibrium Equation (II.4.4) for a $\theta$-pinch becomes

$$\frac{\partial p}{\partial r} = J_\varphi B_z \qquad (\text{II}.4.13)$$

We can eliminate $J_\varphi$ in (I.4.13) by using Ampere's Law

$$J_\varphi = -\frac{1}{\mu_0}\frac{\partial B_z}{\partial r} \qquad (\text{II}.4.14)$$

Substituting (II.4.14) in (II.4.13) yields

$$\frac{\partial p}{\partial r} = -\frac{B_z}{\mu_0}\frac{\partial}{\partial r}B_z \qquad (\text{II}.4.15)$$

Rewriting (II.4.15) gives

$$\frac{\partial}{\partial r}\left(\frac{B_z^2}{2\mu_0} + p\right) = 0 \qquad (\text{II}.4.16)$$

Hence the solution is

$$\frac{B_z^2}{2\mu_0} + p = \frac{B_{\text{ext}}^2}{2\mu_0} = \text{const} \qquad (\text{II}.4.17)$$

where the first term is magnetic pressure, the second term is plasma pressure, and $B_{\text{ext}}$ represents the external magnetic field on the plasma edge. So the total pressure for a $\theta$-pinch is constant. $\theta$-pinch profiles are illustrated in Figure II.4.2.

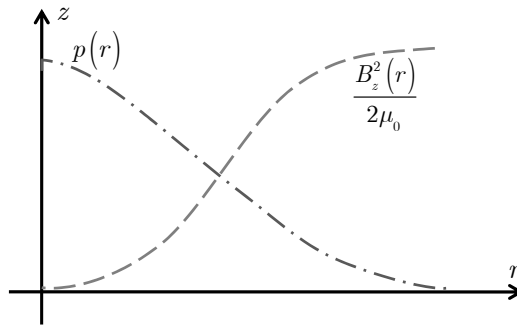

**Figure II.4.2:** $\theta$-pinch profiles

## II. 4.4 Screw Pinch Equilibrium
Here, we present the case which the cylindrical plasma column contains both axial and azimuthal current density distributions which leads us to axial as well as azimuthal magnetic fields. This configuration is known as *Screw Pinch*, or also sometimes known as *straight tokamak*. For a screw pinch which the magnetic field lines are helical, from the fourth of Maxwell's Equations (II.1.4), we get:

$$\frac{1}{r}\frac{\partial B_\varphi}{\partial \varphi} + \frac{\partial B_z}{\partial z} = 0 \qquad (II.4.18)$$

From Ampere's law one has

$$\mu_0 \left( J_\varphi \hat{\varphi} + J_z \hat{z} \right) = -\frac{\partial B_z}{\partial r}\hat{\varphi} + \frac{1}{r}\frac{\partial}{\partial r}\left( rB_\varphi \right)\hat{z} \qquad (II.4.19)$$

Using equilibrium equation gives

$$\frac{\partial p}{\partial r} = J_\varphi B_z - J_z B_\varphi = -\frac{\partial}{\partial r}\left(\frac{B_z^2}{2\mu_0}\right) - \frac{\partial}{\partial r}\left(\frac{B_\varphi^2}{2\mu_0}\right) - \frac{B_\varphi^2}{\mu_0 r} \qquad (II.4.20)$$

Rewriting the above equation gives rise to the governing equation for a screw pinch as

$$\frac{\partial}{\partial r}\left( p + \frac{B_z^2}{2\mu_0} + \frac{B_\varphi^2}{2\mu_0} \right) = \frac{B_\varphi^2}{\mu_0 r} \qquad (II.4.21)$$

This result reveals that knowledge of equilibrium configuration of a screw pinch requires the solution of an ordinary differential equation with three unknowns. Hence, we normally require information about the profiles of two parameters at least, which might be extracted from transport equations or experimental measurements. The same situation applies to tokamaks and will be discussed later.

### II. 4.5 Force Free Equilibrium

An equilibrium is called to be *force free*, if **J** and **B** are parallel; as a result, plasma pressure should have zero gradient. In cases where $\beta \ll 1$ and one with good approximation could ignore $\nabla p$, the force free equilibrium can be achieved. In this situation, current flows along field lines, so we have:

$$\mathbf{J} = k(r)\mathbf{B} \qquad (II.4.22)$$

in which $k(r)$ is constant along field lines. Taking divergence from the above gives

$$\nabla \cdot \mathbf{J} = \nabla \cdot [k(r)\mathbf{B}] = k(r)\nabla \cdot \mathbf{B} + (\mathbf{B} \cdot \nabla)k(r) = 0 \qquad (II.4.23)$$

Since from Maxwell's equation $\nabla \cdot \mathbf{B} = 0$, one can conclude that:

$$(\mathbf{B} \cdot \nabla)k(r) = 0 \qquad (II.4.24)$$

Now, we substitute (II.4.22) into Ampere's law $\nabla \times \mathbf{B} = \mu_0 \mathbf{J}$, which gives

$$\nabla \times \mathbf{B} = \mu_0 k(r) \mathbf{B} \tag{II.4.25}$$

Applying the curl operator on (II.4.25) results in

$$\nabla \times (\nabla \times \mathbf{B}) = \nabla \times [\mu_0 k(r) \mathbf{B}] \tag{II.4.26}$$

By using vector identities $\nabla \times (k\mathbf{B}) = k\nabla \times \mathbf{B} + \nabla k \times \mathbf{B}$ and $\nabla \times (\nabla \times \mathbf{B}) = \nabla(\nabla \cdot \mathbf{B}) - \nabla^2 \mathbf{B}$, (II.4.25) recasts into

$$\nabla^2 \mathbf{B} + [\mu_0 k(r)]^2 \mathbf{B} = -\mu_0 \nabla k(r) \times \mathbf{B} \tag{II.4.27}$$

The above differential equation may be easily solved if one neglects the radial dependence of $k(r)$. Expanding the above in terms of axial $B_z$ and azimuthal $B_\varphi$ fields gives the set of linear differential equations

$$r^2 \frac{d^2 B_\varphi}{dr^2} + r \frac{dB_\varphi}{dr} + (K^2 r^2 - 1) B_\varphi = 0 \tag{II.4.28}$$

$$r^2 \frac{d^2 B_z}{dr^2} + r \frac{dB_z}{dr} + K^2 r^2 B_z = 0 \tag{II.4.29}$$

Here, $K = \mu_0 k$. Solutions of (II.4.28) and (II.4.29) are simply given by Bessel's functions of the first kind and integer order as

$$B_\varphi(r) = B_0 J_1(\mu_0 k r) \tag{II.4.30}$$
$$B_z(r) = B_0 J_0(\mu_0 k r) \tag{II.4.31}$$

in which $B_0$ is the maximum axial field on the plasma axis.

## II. 5. Grad-Shafranov Equation (GSE)

The ideal MHD of axisymmetric toroidal plasma in tokamaks is described by *Grad-Shafranov Equation* (GSE) that was first proposed by *H. Grad and H. Rubin* (1958) and *Shafranov* (1966) for poloidal flux function. Here, we derive the GSE in flux coordinate system.

In tokamaks it is convenient to express magnetic field in mixed covariant-contravariant representation. As current lines lie on constant magnetic flux surfaces we have

$$\mathbf{J} \cdot \nabla \psi = J^\psi = \frac{1}{\mu_0} (\nabla \times \mathbf{B})^\psi = 0 \tag{II.5.1}$$

Or equivalently

$$J\left(\frac{\partial B_\zeta}{\partial \chi} - \frac{\partial B_\chi}{\partial \zeta}\right) = 0 \tag{II.5.2}$$

Axisymmetry requires that $\partial/\partial\zeta = 0$, and furthermore $J > 0$. Thus

$$\frac{\partial B_\chi}{\partial \zeta} = \frac{\partial B_\zeta}{\partial \chi} = 0 \tag{II.5.3}$$

It is clear that $B_\zeta$ is only function of $\psi$. More often in the context of tokamaks, the notation $B_\zeta(\psi) = I(\psi)$ is adopted. We also notice that $B_\zeta$ is not the same as the magnitude of toroidal magnetic field $B_t$; these two parameters even do not share the same physical dimensions, while they point to the same direction, i.e. $\mathbf{B}_t \parallel B_\zeta \hat{\zeta}$. Anyway, the magnetic field of tokamak may be decomposed into its toroidal and poloidal field components as

$$\mathbf{B} = \mathbf{B}_t + \mathbf{B}_p \tag{II.5.4}$$

where $\mathbf{B}_t = B_\zeta \hat{\psi} = I(\psi)\hat{\psi}$ and $\mathbf{B}_p = \hat{\zeta} \times \hat{\psi}$. Hence rewriting (II.5.4) gives the mixed covariant-contravariant representation of magnetic field in tokamaks as

$$\mathbf{B} = \mathbf{B}_t + \mathbf{B}_p = I(\psi)\hat{\zeta} + \hat{\zeta} \times \hat{\psi} \tag{II.5.5}$$

The magnitude of toroidal magnetic field is hence

$$B_t = |\mathbf{B}_t| = I(\psi)|\hat{\zeta}| = \frac{I(\psi)}{R} \tag{II.5.6}$$

Here, we have noticed that $\zeta$ is simply the angular coordinate of cylindrical system due to axisymmetry. Therefore, $|\hat{\zeta}| = \sqrt{g_{\zeta\zeta}} = 1/R$, and

$$I(\psi) = B_\zeta(\psi) = RB_t \tag{II.5.7}$$

This shows that the magnitude of toroidal field is *not* a flux function, while the covariant component $B_\zeta(\psi)$ is. Similarly, the magnitude of poloidal magnetic field is given as

$$B_p = |\mathbf{B}_p| = |\hat{\zeta} \times \hat{\psi}| = |\hat{\zeta}||\hat{\psi}|\sin(\hat{\zeta},\hat{\psi}) \tag{II.5.8}$$

Here, $(\hat{\zeta}, \hat{\psi})$ is the angle made by the basis vectors $\hat{\zeta}$ and $\hat{\psi}$. Axisymmetry of tokamaks excludes dependence on $\zeta$ coordinate, and hence this would be a right angle. This would mean that four elements of the metric tensor should vanish, that is $g_{\psi\zeta} = g_{\chi\zeta} = g_{\zeta\psi} = g_{\zeta\chi} = 0$, and $g^{\psi\zeta} = g^{\chi\zeta} = g^{\zeta\psi} = g^{\zeta\chi} = 0$. Another conclusion is that $g_{\zeta\zeta} g^{\zeta\zeta} = 1$. Therefore

$$B_p = \left|\hat{\zeta}\right|\left|\hat{\psi}\right| = \frac{1}{R}\left|\hat{\psi}\right| \qquad (\text{II.5.9})$$

In order to derive Grad-Shafranov Equation we begin with equilibrium equation (II.4.4). Upon substitution of **J** by the rotation of magnetic field we get

$$\begin{aligned}\mu_0 \mathbf{J} &= \nabla \times \left[I(\psi)\hat{\zeta} + \hat{\zeta} \times \hat{\psi}\right] \\ &= \nabla I(\psi) \times \hat{\zeta} + I(\psi)\nabla \times \hat{\zeta} + \nabla \times \left(\hat{\zeta} \times \hat{\psi}\right) \\ &= \frac{\partial I(\psi)}{\partial \psi}\hat{\psi} \times \hat{\zeta} + I(\psi)\nabla \times \hat{\zeta} + \nabla \times \left(\hat{\zeta} \times \hat{\psi}\right)\end{aligned} \qquad (\text{II.5.10})$$

where $\hat{\psi}$ in the cylindrical coordinates $(R, \varphi, Z)$ is given through the definition of covariant bases as

$$\begin{aligned}\hat{\psi} &= \nabla\psi = \left(\frac{\partial}{\partial R}\hat{r} - \frac{1}{R}\frac{\partial}{\partial\varphi}\hat{\varphi} + \frac{\partial}{\partial Z}\hat{z}\right)\psi \\ &= \frac{\partial\psi}{\partial R}\hat{r} + \frac{\partial\psi}{\partial Z}\hat{z}\end{aligned} \qquad (\text{II.5.11})$$

Here, we have taken the fact into account that from axisymmetry we have $\partial/\partial\varphi = 0$. Since $\zeta$ in axisymmetric flux coordinates is the same as $-\varphi$ in cylindrical coordinates, we can write

$$\hat{\zeta} = \nabla\zeta = -\frac{1}{R}\hat{\varphi} \qquad (\text{II.5.12})$$

By substitution of (II.5.12) and (II.5.11) we get the expression for contravariant component of the magnetic field as

$$\hat{\zeta} \times \hat{\psi} = \nabla\psi \times \frac{1}{R}\hat{\varphi} = \frac{\partial\psi}{\partial R}\hat{z} - \frac{\partial\psi}{\partial Z}\hat{r} \qquad (\text{II.5.13})$$

Now we are able to simplify (II.5.10) as follows

$$\mu_0 \mathbf{J} = \frac{\partial I(\psi)}{\partial \psi} \hat{\psi} \times \hat{\zeta} - \left( \frac{\partial}{\partial R} \hat{r} + \frac{\partial}{\partial Z} \hat{z} \right) \times \left( -\frac{1}{R} \frac{\partial \psi}{\partial R} \hat{z} + \frac{1}{R} \frac{\partial \psi}{\partial Z} \hat{r} \right)$$

$$= \frac{\partial I(\psi)}{\partial \psi} \hat{\psi} \times \hat{\zeta} - \frac{\partial}{\partial R} \left( -\frac{1}{R} \frac{\partial \psi}{\partial R} \right) \hat{r} \times \hat{z} + \frac{\partial}{\partial Z} \left( \frac{1}{R} \frac{\partial \psi}{\partial Z} \right) \hat{z} \times \hat{r}$$

$$= \frac{\partial I(\psi)}{\partial \psi} \hat{\psi} \times \hat{\zeta} - \hat{\varphi} \left[ \frac{\partial}{\partial R} \left( \frac{1}{R} \frac{\partial}{\partial R} \right) + \frac{\partial}{\partial Z} \left( \frac{1}{R} \frac{\partial}{\partial Z} \right) \right] \psi \qquad \text{(II.5.14)}$$

$$= \frac{\partial I(\psi)}{\partial \psi} \hat{\psi} \times \hat{\zeta} - \frac{\hat{\varphi}}{R} \left\{ R \frac{\partial}{\partial R} \left( \frac{1}{R} \frac{\partial}{\partial R} \right) + \frac{\partial^2}{\partial Z^2} \right\} \psi$$

Since the first term on the right-hand-side of (II.5.14) is not in toroidal direction, we can write the toroidal current density as

$$\mathbf{J}_t = J_t \hat{\varphi} = -\frac{1}{\mu_0} \frac{\hat{\varphi}}{R} \Delta^* \psi \qquad \text{(II.5.15)}$$

where

$$\Delta^* = R \frac{\partial}{\partial R} \left( \frac{1}{R} \frac{\partial}{\partial R} \right) + \frac{\partial^2}{\partial Z^2} = \frac{\partial^2}{\partial R^2} - \frac{1}{R} \frac{\partial}{\partial R} + \frac{\partial^2}{\partial Z^2} \qquad \text{(II.5.16)}$$

is so-called Grad-Shafranov operator. Hence, the GSE reads

$$\frac{1}{R} \Delta^* \psi = -\mu_0 J_t \qquad \text{(II.5.17)}$$

It is now very instructive to go back to (II.4.4) to find out

$$\mathbf{J} \times \mathbf{B} = \nabla p = \frac{dp}{d\psi} \hat{\psi}$$

$$= \frac{1}{\mu_0} \left\{ -\left[ I'(\psi) \hat{\zeta} \times \hat{\psi} + \frac{\Delta^* \psi}{r} \hat{\varphi} \right] \times \left( I(\psi) \hat{\zeta} + \hat{\zeta} \times \hat{\psi} \right) \right\}$$

$$= \frac{1}{\mu_0} \left\{ \frac{I(\psi) I'(\psi)}{r^2} (\hat{\psi} \times \hat{\varphi}) \times \hat{\varphi} + \frac{I'(\psi)}{r^2} (\hat{\psi} \times \hat{\varphi}) \times (\hat{\varphi} \times \hat{\psi}) \right.$$

$$\left. + \frac{\Delta^* \psi}{r^2} \hat{\varphi} \times I(\psi) \hat{\varphi} + \frac{\Delta^* \psi}{r^2} \hat{\varphi} \times (\hat{\varphi} \times \hat{\psi}) \right\} \qquad \text{(II.5.18)}$$

$$= \frac{1}{\mu_0} \left[ \frac{I(\psi) I'(\psi)}{R^2} (\hat{\psi} \times \hat{\varphi}) \times \hat{\varphi} + \frac{\Delta^* \psi}{R^2} \hat{\varphi} \times (\hat{\varphi} \times \hat{\psi}) \right]$$

$$= \frac{1}{\mu_0} \left[ -\frac{I(\psi) I'(\psi)}{R^2} \hat{\psi} + \frac{\Delta^* \psi}{R^2} \hat{\psi} \right]$$

Or equivalently

$$\frac{dp}{d\psi} = \frac{1}{\mu_0}\left\{-\frac{I(\psi)I'(\psi)}{R^2} + \frac{\Delta^*\psi}{R^2}\right\} \quad \text{(II.5.19)}$$

By rearranging (II.5.19) we arrive at the alternative form of GSE given by

$$\frac{1}{R}\Delta^*\psi = \mu_0 R \frac{dp(\psi)}{d\psi} + \frac{I(\psi)}{R}\frac{dI(\psi)}{d\psi} \quad \text{(II.5.20)}$$

This form of GSE has the advantage that the right-hand-side is also given in terms of the poloidal flux $\psi$. This however makes the above equation nonlinear in terms of the poloidal flux. More often, profiles of pressure $p(\psi)$ and toroidal field function $I(\psi) = RB_t$ are either known from experiment, or self-consistent solution of transport equations. Alternatively, prescribed polynomial forms are assumed for these two functions. The nature of the axisymmetric equilibrium in a tokamak is thus to a large extent determined by the choice of the free functions $p(\psi)$, $I(\psi)$ and the boundary conditions.

We may notice the right-hand-side of (II.5.20) is because of (II.5.17) actually the toroidal current density $J_t$ and hence a flux function. If the toroidal current density is assumed to have a linear dependence on flux as $J_t(\psi) \approx J_0 + J_1\psi$, then (II.5.20) allows exact solutions in terms of Bessel's functions. But numerical solution is inevitable for more complicated profiles.

A straightforward solution of (II.5.20) with the special choice of $p(\psi) = \psi$ and $I(\psi) = I_0$, known as Solov'ev solution has been shown to exist by Vitali Shafranov, which is given by

$$\psi(R,Z) = \frac{2}{a^2}R^2 - \frac{1}{a^4}R^4 - \frac{4b^2}{a^4}R^2Z^2 \quad \text{(II.5.21)}$$

where $a$ and $b$ are constants which determine the final equilibrium configuration. This solution is noticeably useful in description of a wide range of plasma equilibria in tokamaks.

## II. 6. Green's Function Formalism

In previous section we derived GSE from MHD equilibrium equation of a toroidal plasma. By solving Grad-Shafranov equation one can find flux distribution of magnetic flux. Once the flux distribution is known, it is easy to reconstruct the plasma boundary and the shape of nested magnetic surfaces. There are lots of different methods which have been proposed to solve GSE, which are categorized in Fig. II.6.1. In this section we are going to study Green's Function Method as an analytical solution to GSE.

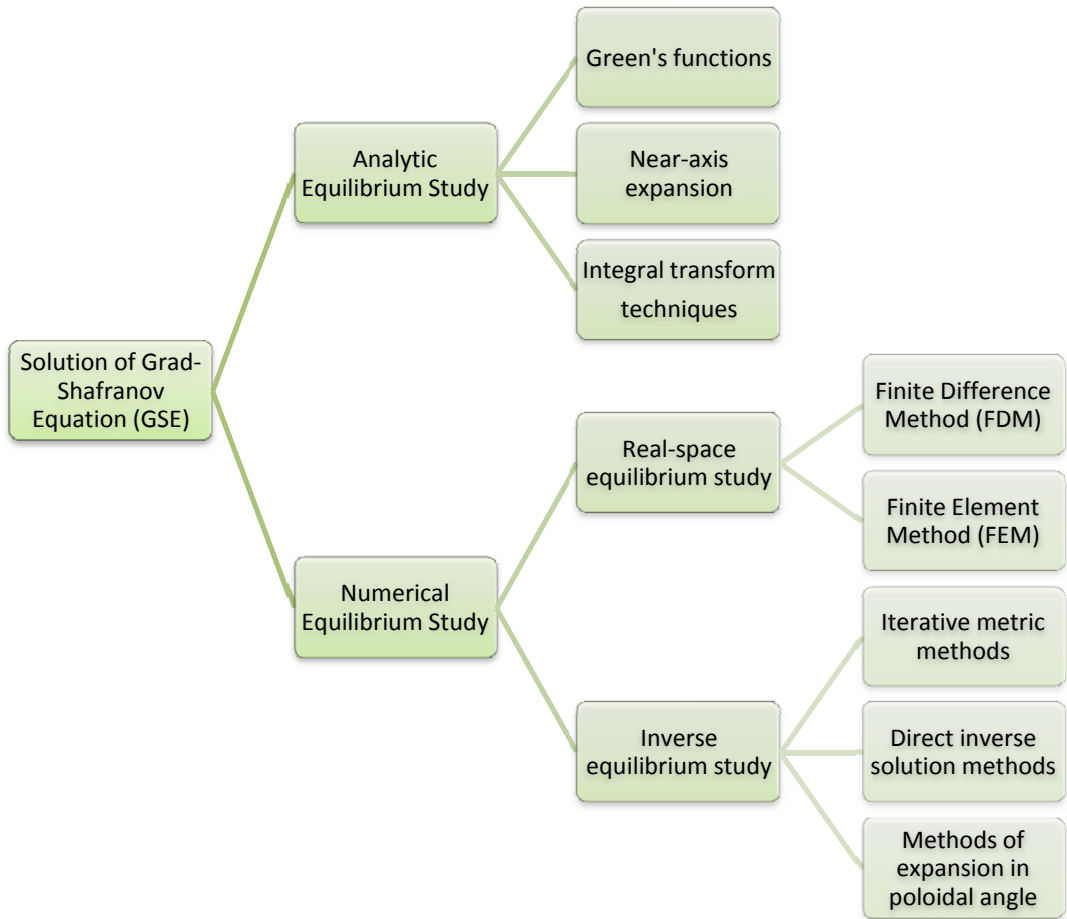

**Figure II.6.1:** Different solutions to GSE

## II. 6.1  Green's Function for GSE

The axisymmetric magnetostatics in cylindrical coordinates is described by the GSE equation:

$$\Delta^*\psi = -\mu_0 r J_t \qquad (\text{II.6.1})$$

in which $\psi = rA_\varphi$ is the magnetic poloidal flux, and where $A_\varphi$ is the toroidal component of the magnetic vector potential. Also, $J_t$ is the toroidal current density and $\Delta^*$ is the elliptic Grad-Shafranov operator which is already defined in (II.5.16). This concept is shown graphically in Fig. II.6.2. From now on, the cylindrical coordinates are represented by the set of coordinates $(r,\varphi,z)$.

In (II.6.1), we disregard the inherent dependence of $J_t$ on the poloidal flux $\psi$, making the GSE a linear differential equation. From a system engineering point of view, GSE represents a Linear Time Invariant (LTI) system, whose impulse response is given by its associated Green' function. Here in order to find flux function, we seek solutions of the form

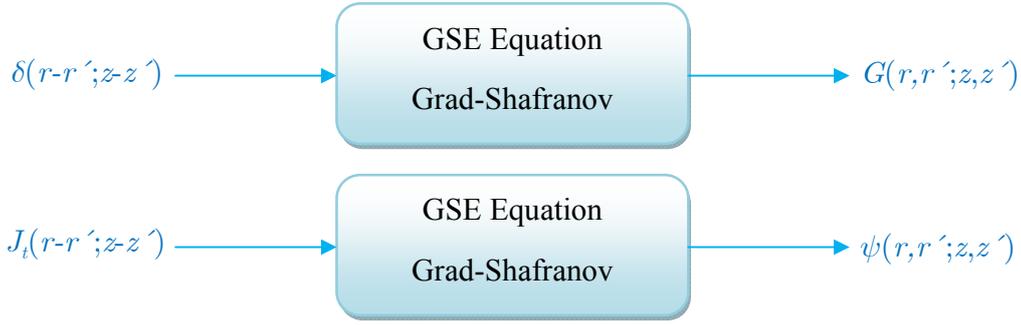

**Figure II.6.2:** Green's Function Method

$$\psi(r,z) = \int_{-\infty}^{\infty}\int_{0}^{\infty} G(r,r',z,z')J_t(r',z')\,dr'dz' \tag{II.6.2}$$

or equivalently

$$\psi(r,z) = \int_{-\infty}^{\infty}\int_{0}^{\infty} G(\mathbf{r},\mathbf{r}')J_t(\mathbf{r}')\,dr'dz' \tag{II.6.3}$$

where $\mathbf{r} = (r,z)$ is the two-dimensional position vector on the constant $\varphi$-plane, and $G(\mathbf{r},\mathbf{r}')$ is referred to as the Green's function, obtained through the solution of the following equation:

$$\Delta^*\psi = \mu_0\delta(\mathbf{r}-\mathbf{r}') = \mu_0\delta(r-r')\delta(z-z') \tag{II.6.4}$$

with $\delta(\cdot)$ being the Dirac's delta function.

At first, we first examine the solution to the homogeneous Grad–Shafranov equation,

$$\Delta^*\psi = 0 \tag{II.6.5}$$

and then proceed to construct the Green's function. Now let $\psi(r,z) = R(r)Z(z)$ and using separation of variables we get

$$\Delta^*[R(r)Z(z)] = \left(\frac{\partial^2}{\partial r^2} - \frac{1}{r}\frac{\partial}{\partial r} + \frac{\partial^2}{\partial z^2}\right)R(r)Z(z) = 0 \tag{II.6.6}$$

For the sake of simplicity, let $\partial/\partial r = (\cdot)'$ and $\partial/\partial z = (\dot{\cdot})$, and rewrite (II.6.6) as

$$ZR'' - Z\frac{1}{r}R' + R\ddot{Z} = 0 \tag{II.6.7}$$

Dividing both sides by $Z(z)R(r)$ yields

$$\frac{R''}{R} - \frac{1}{r}\frac{R'}{R} + \frac{\ddot{Z}}{Z} = 0 \tag{II.6.8}$$

As $R$ and $Z$ are only function of coordinates $r$ and $z$, respectively, one can write

$$\frac{R''}{R} - \frac{1}{r}\frac{R'}{R} = -\frac{\ddot{Z}}{Z} = -k^2 \tag{II.6.9}$$

where $k^2$ is a real-valued constant. Therefore we should have

$$\frac{\ddot{Z}}{Z} = k^2 \tag{II.6.10}$$

and

$$\frac{R''}{R} - \frac{1}{r}\frac{R'}{R} = -k^2 \tag{II.6.11}$$

Bounded solutions of (II.6.10) require that $k^2 < 0$, and are in the form

$$Z(z) = a_k \cos(kz) + b_k \sin(kz) \tag{II.6.12}$$

Letting $R(r) = rA(r)$ in (II.6.11) gives the Modified Bessel Function with the general solution

$$R(r) = r\left[c_k K_1(kr) + d_k I_1(kr)\right] \tag{II.6.13}$$

in which $a_k$, $b_k$, $c_k$, and $d_k$ are constants, and $K_1(.)$ and $I_1(.)$ are the first order modified Bessel functions. Hence, the proposed eigen-solution $\psi(r,z) = Z(z)R(r)$ becomes

$$\psi_k(r,z) = r\left[a_k \cos(kz) + b_k \sin(kz)\right]\left[c_k K_1(kr) + d_k I_1(kr)\right] \tag{II.6.14}$$

Now since for the modified Bessel functions we have

$$\lim_{r \to 0} I_1(kr) = \infty \tag{II.6.15}$$

$d_k$ should be zero, and hence

$$\psi_k(r,z) = r\left[a_k \cos(kz) + b_k \sin(kz)\right] K_1(kr) \tag{II.6.16}$$

Green's function may be constituted from a proper superposition of eigen-functions (II.6.16) using integration on all $k$, given by

$$\psi(r,z) = \int_0^\infty \psi_k(r,z)\, dk = \int_0^\infty r\left[a_k \cos(kz) + b_k \sin(kz)\right] K_1(kr)\, dk \tag{II.6.17}$$

The *reciprocity property* of the Green's function as understood from linearity of the GSE system requires that

$$G(\mathbf{r} - \mathbf{r}') \equiv G(\mathbf{r}' - \mathbf{r}) \tag{II.6.18}$$

Consequently, symmetry with regard to the change of arguments reduces Green's function to

$$G(\mathbf{r} - \mathbf{r}') = \int_0^\infty r a_k \cos\left[k(z-z')\right] K_1\left[k(r-r')\right] dk \tag{II.6.19}$$

Substituting (II.6.19) in (II.6.4) gives

$$\Delta^* G(\mathbf{r},\mathbf{r}') = \Delta^* \int_0^\infty a_k(r,r') \cos\left[k(z-z')\right] dk = -r\mu_0 \delta(\mathbf{r}-\mathbf{r}') \tag{II.6.20}$$

where $a_k(r,r') = r a_k K_1\left[k(r-r')\right]$. Applying the expanded form of Grad-Shafranov operator on (II.6.20) yields

$$\Delta^* G(\mathbf{r},\mathbf{r}') = \int_0^\infty \left(\frac{\partial^2}{\partial r^2} - \frac{1}{r}\frac{\partial}{\partial r}\right) a_k(r,r') \cos\left[k(z-z')\right] + a_k(r,r') \frac{\partial^2}{\partial z^2} \cos\left[k(z-z')\right] dk \tag{II.6.21}$$

which may be rewritten in the form

$$\Delta^* G(\mathbf{r},\mathbf{r}') = \int_0^\infty \left[\left(\frac{\partial^2}{\partial r^2} - \frac{1}{r}\frac{\partial}{\partial r} - k^2\right) a_k(r,r')\right] \cos\left[k(z-z')\right] dk \tag{II.6.22}$$

Now we adopt the definition

$$\Delta_k^* = \frac{\partial^2}{\partial r^2} - \frac{1}{r}\frac{\partial}{\partial r} - k^2 \tag{II.6.23}$$

and rewrite (I.6.22) as

$$\Delta^* G(\mathbf{r},\mathbf{r}') = \int_0^\infty \left[\Delta_k^* a_k(r,r')\right] \cos\left[k(z-z')\right] dk$$
$$= -r\mu_0 \delta(\mathbf{r}-\mathbf{r}') = r\mu_0 \delta(r-r')\delta(z-z') \tag{II.6.24}$$

But Dirac's delta function may be defined as

$$2\pi\delta(z) \triangleq \int_{-\infty}^\infty e^{jkz} dk = \int_{-\infty}^\infty \cos(kz) + j\sin(kz)\, dk \tag{II.6.25}$$

Since the right-hand-side of (II.6.24) is real, (II.6.25) simplifies as

$$\int_0^\infty \cos(kz) = \pi\delta(z) \tag{II.6.26}$$

Comparing (II.6.24) and (II.6.26) results in the linear ordinary differential equation

$$\Delta_k^* a_k(r,r') = \frac{\partial^2}{\partial r^2} a_k(r,r') - \frac{1}{r}\frac{\partial}{\partial r} a_k(r,r') - k^2 a_k(r,r')$$
$$= -\frac{r}{\pi}\mu_0 \delta(r-r') \tag{II.6.27}$$

which has the solution

$$a_k = \begin{cases} \dfrac{rr'}{\pi A} K_1(kr) I_1(kr') & r > r' \\ \dfrac{rr'}{\pi A} K_1(kr') I_1(kr) & r < r' \end{cases} \tag{II.6.28}$$

This can be written in the more compact and convenient form

$$a_k(r,r') = \frac{r_> r_<}{\pi A} I_1(kr_<) K_1(kr_>) \tag{II.6.29}$$

in which $A$ is the Wronskian of functions $r'I_1(kr')$ and $r'K_1(kr')$; furthermore, $r_< = \min(r,r')$ and $r_> = \max(r,r')$. Since the unknown coefficients $a_k(r,r')$ are determined, one can obtain the integral form of Green's Function as

$$G(\mathbf{r},\mathbf{r}') = \mu_0 \frac{rr'}{\pi} \int_0^\infty I_1(kr_<)K_1(kr_>)\cos[k(z-z')]dk \tag{II.6.30}$$

Surprisingly, (II.6.30) allows a very simple closed form integral given by

$$\begin{aligned} G(\mathbf{r},\mathbf{r}') &= \mu_0 \frac{\sqrt{rr'}}{2\pi} Q_{\frac{1}{2}}\left[\frac{r^2 + r'^2 + (z-z')^2}{2rr'}\right] \\ &= \mu_0 \frac{\sqrt{rr'}}{2\pi} Q_{\frac{1}{2}}\left(\frac{|\mathbf{r}-\mathbf{r}'|^2}{2rr'} + 1\right) \end{aligned} \tag{II.6.31}$$

where $Q_{1/2}(\cdot)$ is the Legendre function of the second kind, satisfying

$$(1-x^2)y''(x) - 2xy'(x) + \nu(\nu+1)y(x) = 0 \tag{II.6.32}$$

with $\nu = 1/2$. It is noticeable that the latter result justifies the requirement for the reciprocity property of the Green's function as stated above. This completes our assertion.

The asymptotic behavior of Legendre functions $Q_\nu(x)$ at large $x$ may be studied by the integral

$$\begin{aligned} Q_\nu(x) &= \int_0^\infty \frac{d\theta}{(1+\cosh\theta)^{\nu+1}} x^{-(1+\nu)} \qquad \nu > -1 \\ &\approx x^{-(1+\nu)} \frac{\sqrt{\pi}}{2^{1+\nu}} \frac{\Gamma(\nu+1)}{\Gamma\left(\nu+\frac{3}{2}\right)} \end{aligned} \tag{II.6.33}$$

Hence for $\nu = 1/2$ the asymptotic expansion is $Q_{1/2} \approx 32^{-1/2}\pi x^{-3/2}$. Therefore

$$G(\mathbf{r},\mathbf{r}') \approx \mu_0 \frac{r^2 r'^2}{4[r^2 + r'^2(z-z')^2]^{\frac{3}{2}}} \tag{II.6.34}$$

which satisfies all of the requirements on the boundaries as

$$\lim_{r \to 0} G(\mathbf{r} - \mathbf{r}') = 0$$
$$\lim_{r \to \infty} G(\mathbf{r} - \mathbf{r}') = 0 \qquad \text{(II.6.35)}$$
$$\lim_{\mathbf{r} \to \mathbf{r}'} G(\mathbf{r} - \mathbf{r}') = \infty$$

A contour plot of the Green's function is illustrated in Fig. II.6.3.

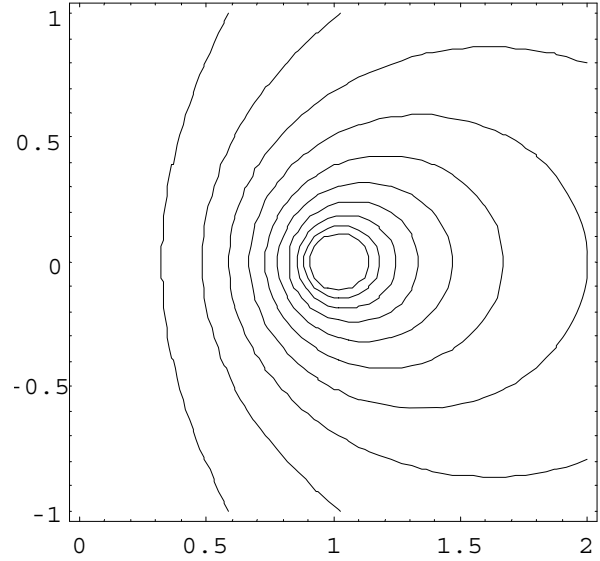

**Figure II.6.3:** Contour plot of the Green's function as given by (15).

### I. 6.2  Application of Green's Function to Different Current Density Distributions

As it was mentioned the poloidal flux function $\psi$ can be accurately obtained by through the Green's function formalism once the toroidal current density profile is known, following (II.6.2). As examples, we discuss the resultant magnetic fields due to a current loop and a solenoid.

*a)   Current loop*

In order to find poloidal flux of a current loop placed at the equatorial plane $z = 0$, we may immediately use the Green's function with the solution $\psi(r,z) = G(r,a;z,0)$, in which $a$ is the radius of loop. But for illustration purposes, we use the asymptotic expression of Green's function. This results in the following for $\psi(r,z)$

$$\psi(r,z) \approx \mu_0 \int_{-\infty}^{\infty} \int_0^{\infty} \frac{r^2 r'^2}{4\left[r^2 + r'^2(z-z')^2\right]^{\frac{3}{2}}} J_t(\mathbf{r}') dr' dz' \qquad \text{(II.6.36)}$$

For the current loop, the corresponding toroidal current density is

$$J_t(\mathbf{r}') = I_0 \delta(r' - a) \delta(z') \qquad \text{(II.6.37)}$$

where $I_0$ is the current passing throught the loop. As a result, the magnetic poloidal flux will be

$$\psi(r,z) \approx \mu_0 \int_{-\infty}^{\infty}\int_0^{\infty} \frac{r^2 r'^2}{4[r^2 + r'^2(z-z')^2]^{\frac{3}{2}}} I_0 \delta(r'-a)\delta(z')dr'dz'$$

$$= \mu_0 I_0 \frac{r^2 a^2}{4(r^2 + a^2 + z^2)^{\frac{3}{2}}}$$

(II.6.38)

Since $\psi(r,z) = rA_\varphi$ we have:

$$\mathbf{B} = -\frac{1}{r}\frac{\partial \psi(r,z)}{\partial z}\hat{r} + \frac{1}{r}\frac{\partial \psi(r,z)}{\partial r}\hat{z}$$

(II.6.39)

or

$$\mathbf{B} = \frac{\mu_0}{4} I_0 \left\{ \frac{3ra^2 z}{(r^2+a^2+z^2)^{\frac{5}{2}}}\hat{r} + \frac{a^2\left[2(r^2+a^2+z^2)-3\right]}{(r^2+a^2+z^2)^{\frac{5}{2}}}\hat{z} \right\}$$

(II.6.40)

*b)    Solenoid with toroidal current density*

In this case, we consider the toroidal current distribution of a cylinder with radius $a$, which carries uniform current density $J_0$ in the poloidal direction. We furthermore postulate that the cylinder's axis coincides with the $z$-axis. Hence, the corresponding current density is

$$\mathbf{J}(r,z) = J_t(r,z)\hat{\varphi} = J_0 \delta(r-a)\hat{\varphi}$$

(II.6.41)

Using (II.6.36) yields:

$$\psi(r,z) \approx \mu_0 \int_{-\infty}^{\infty}\int_0^{\infty} \frac{r^2 r'^2}{4[r^2+r'^2+(z-r')^2]^{\frac{3}{2}}} J_0 \delta(r'-a)dr'dz'$$

$$= \int_{-\infty}^{\infty} \frac{J_0 \mu_0 r^2 a^2}{4[r^2+a^2+(z-z')^2]^{\frac{3}{2}}} dz'$$

$$= \frac{J_0 \mu_0 r^2 a^2}{2(r^2+a^2)}$$

(II.6.42)

Using (II.6.39) for determining the magnetic field results in

$$\mathbf{B} = -\frac{1}{r}\frac{\partial \psi(r,z)}{\partial z}\hat{r} + \frac{1}{r}\frac{\partial \psi(r,z)}{\partial r}\hat{z} = \frac{J_0 \mu_0 a^4}{(r^2 + a^2)^2}\hat{z} \quad \text{(II.6.43)}$$

Finally, the on-axis magnetic field is given by the well-known expression

$$\mathbf{B} = \hat{z}\frac{\mu_0 J_0 b^2}{2}\lim_{r \to 0^+}\frac{2b^2}{(r^2 + b^2)^2} = \hat{z}\mu_0 J_0 \quad \text{(II.6.44)}$$

## II. 7. Analytical and Numerical Solutions to GSE

### II. 7.1 Analytical Solution

Grad-Shafranov equation (I.6.1) is normally expressed in cylindrical coordinates. This equation can be written in primitive toroidal coordinates by applying the transformations

$$\begin{aligned} r &= R_0 + r_0 \cos\theta_0 \\ z &= r_0 \sin\theta_0 \\ \varphi &= -\zeta_0 \end{aligned} \quad \text{(II.7.1)}$$

Therefore we get

$$\left[\frac{1}{r_0}\frac{\partial}{\partial r_0}\left(r_0 \frac{\partial}{\partial r_0}\right) + \frac{1}{r_0^2}\frac{\partial^2}{\partial \theta_0^2}\right]\psi - \frac{1}{R_0 + r_0 \cos\theta_0}\left(\cos\theta_0 \frac{\partial}{\partial r_0} - \frac{\sin\theta_0}{r_0}\frac{\partial}{\partial \theta_0}\right)\psi$$
$$= -\mu_0 \left(R_0 + r_0 \cos\theta_0\right)^2 \frac{\partial p(\psi)}{\partial \psi} - \mu_0^2 I(\psi)\frac{\partial I(\psi)}{\partial \psi} \quad \text{(II.7.2)}$$

Now, we assume that the poloidal flux is composed of circular and non-circular contributions as

$$\psi(r_0, \theta_0) = \psi_0(r_0) + \psi_1(r_0, \theta_0) \quad \text{(II.7.3)}$$

Here, $\psi_1(r_0, \theta_0)$ represents the deviation from the concentric nested circular magnetic surfaces in tokamaks, and hence is responsible for characteristics such as *Shafranov Shift*, *Triangularity* and *Elongation*. It is evident that the dominant term in (II.7.3) is due to $\psi(r_0)$, which plays a significant role in construction of magnetic flux surfaces.

For large aspect ratio tokamaks, $\psi_1(r_0, \theta_0)$ may be treated as a perturbation function satisfying $\psi_0(r_0) \gg \psi_1(r_0, \theta_0)$, so that (II.7.2) for $\psi_0(r_0)$ could be written as

$$\frac{1}{r}\frac{\partial}{\partial r}\left(r\frac{\partial}{\partial r}\right)\psi_o(r) = -\mu_0 R_0^2 \frac{\partial p(\psi_o)}{\partial \psi_o} - \mu_0^2 I(\psi_o)\frac{\partial I(\psi_o)}{\partial \psi_o} \qquad (II.7.4)$$

Similarly, (II.7.2) for the perturbation term $\psi_1(r_0,\theta_0)$ may be approximated as

$$\left[\frac{1}{r}\frac{\partial}{\partial r}\left(r\frac{\partial}{\partial r}\right) + \frac{1}{r^2}\frac{\partial^2}{\partial \theta^2}\right]\psi_1(r,\theta) \approx \frac{\cos\theta}{R_0}\frac{\partial}{\partial r}\psi_0 - 2\mu_0 R_0 r \cos\theta \frac{\partial p(\psi_0)}{\partial \psi_0} -$$
$$\frac{1}{\left(\frac{\partial \psi_0}{\partial r}\right)}\frac{\partial}{\partial r}\left[\mu_0 R_0^2 \frac{\partial p(\psi)}{\partial \psi_0} - \mu_0^2 I(\psi_0)\frac{\partial I(\psi_0)}{\partial \psi_0}\right]\psi_1(r,\theta) \qquad (II.7.5)$$

Hereinafter, we drop the subscript "0" denoting the primitive toroidal coordinates for the sake of convenience.

Input pressure and toroidal field profiles given by $\partial p(\psi_0)/\partial \psi_0$ and $I(\psi_0)\partial I(\psi_0)/\partial \psi_0$ on the right-hand-side of (II.5.20) can be expanded as Taylor series

$$\frac{\partial p(\psi_0)}{\partial \psi_0} = \sum_{n=0}^{\infty} P_n \psi_0^n = P_0 + P_1 \psi_0 + P_2 \psi_0^2 + \cdots + P_n \psi_0^n + \cdots$$
$$I(\psi_0)\frac{\partial I(\psi_0)}{\partial \psi_0} = \sum_{n=0}^{\infty} I_n \psi_0^n = I_0 + I_1 \psi_0 + I_2 \psi_0^2 + \cdots I_n \psi_0^n + \cdots \qquad (II.7.6)$$

Substituting (II.7.6) into (II.7.4) and (II.7.5) results in

$$\frac{1}{r}\frac{\partial}{\partial r}\left(r\frac{\partial}{\partial r}\right)\psi_o(r) = A_0 + A_1 \psi_0 + A_2 \psi_0^2 + \cdots \qquad (II.7.7)$$

and

$$\left[\frac{1}{r}\frac{\partial}{\partial r}\left(r\frac{\partial}{\partial r}\right) + \frac{1}{r^2}\frac{\partial^2}{\partial \theta^2}\right]\psi_1(r,\theta) \approx \frac{\cos\theta}{R_0}\frac{\partial}{\partial r}\psi_0$$
$$- (A_0 + 2A_1 \psi_0 + \ldots)\psi_1 - r\cos\theta(B_0 + B_1 \psi_0 + B_2 \psi_0^2 + \ldots) \qquad (II.7.8)$$

In order to solve GSE analytically, one should discard all those terms that makes GSE nonlinear. Hence after retaining linear terms (II.7.7) reduces to

$$\frac{1}{r}\frac{\partial}{\partial r}\left(r\frac{\partial}{\partial r}\right)\psi_0(r) = -A_0 - A_1\psi_0 \tag{II.7.9}$$

or equivalently

$$\frac{\partial^2\psi_0}{\partial r^2} + \frac{1}{r}\frac{\partial\psi_0}{\partial r} + A_1\psi_0 = -A_0 \tag{II.7.10}$$

Since the poloidal flux $\psi_0$ is a potential function, whose gradient is physically important giving rise to magnetic field, we may freely set $A_0 = 0$ for the moment. Now, letting $A_1 = k^2$ and $A_0 = 0$ gives

$$\frac{\partial^2\psi_0}{\partial r^2} + \frac{1}{r}\frac{\partial\psi_0}{\partial r} + k^2\psi_0 = 0 \tag{II.7.11}$$

The solution of homogenous equation is

$$\psi_{0h} = \psi_c J_0(kr) \tag{II.7.12}$$

No need to mention that $J_0(kr)$ is dimensionless, and hence $\psi_c$ appearing in (II.7.12) is a constant with the dimension of Weber. Now we are ready to add up the particular solution of (II.7.10) in the form of $\psi_{0p} = A$, which by putting in (II.7.12) and solving equation for $A$, yields

$$\psi_{0p} = A = \frac{-A_0}{k^2} \tag{II.7.13}$$

Now the general solution is

$$\psi_0 = \psi_{0h} + \psi_{0p} = \psi_c J_0(kr) - \frac{A_0}{k^2} \tag{II.7.14}$$

There are three unknowns $\psi_c, k$ and $A_0$ which can be determined by imposing plasma constraints. The (arbitrary) choice of $\psi_0(r=0) = 0$, gives

$$\psi_c = \frac{A_0}{k^2} \tag{II.7.15}$$

Now by rewriting (II.7.14) we obtain the solution given by

$$\psi_0 = \psi_c[J_0(kr) - 1] \tag{II.7.16}$$

Now we turn to the perturbation function $\psi_1(r_0,\theta_0)$; by inserting (II.7.16) into (II.7.8) and neglecting terms $A_n, n \geq 1$ we get

$$\left[\frac{1}{r}\frac{\partial}{\partial r}\left(r\frac{\partial}{\partial r}\right) + \frac{1}{r^2}\frac{\partial^2}{\partial \theta^2} + k^2\right]\psi_1(r,\theta) = f(r,\theta) \tag{II.7.17}$$

in which

$$f(r,\theta) = \frac{\cos\theta}{R_0}\frac{\partial \psi_o}{\partial r} - r\cos\theta(B_0 + B_1\psi_0) \tag{II.7.18}$$

The equation (II.7.17) can be analytically solved by means of Green's function technique. For this purpose we first define the Helmholtz operator in two-dimensional polar coordinates as

$$\mathbb{L} = \frac{1}{r}\frac{\partial}{\partial r}\left(r\frac{\partial}{\partial r}\right) + \frac{1}{r^2}\frac{\partial^2}{\partial \theta^2} + k^2 \tag{II.7.19}$$

The appropriate Green's function for solution of (II.7.17) hence obeys

$$\mathbb{L}_k G(r,r';\theta,\theta') = -\delta(\mathbf{r},\mathbf{r}') = -\frac{1}{r}\delta(r,r')\delta(\theta,\theta') \tag{II.7.20}$$

in which $\mathbf{r} = r\cos\theta\hat{x} + r\sin\theta\hat{y}$ is the two-dimensional position vector. The Green's function is well-known to be

$$G(\mathbf{r}-\mathbf{r}') = \text{Re}\left[\frac{j}{4}H_0^{(1)}(k|\mathbf{r}-\mathbf{r}'|)\right] \tag{II.7.21}$$

where $H_0^{(1)}(\cdot)$ is *Hankel's function of the first kind and zeroth order*. Now $\psi_1(r,\theta)$ can be readily determined by the convolution integral

$$\psi_1(r,\theta) = \int_0^{2\pi}\int_0^{\infty} G(\mathbf{r},\mathbf{r}')f(\mathbf{r}')r'dr'd\theta' \tag{II.7.22}$$

where

$$f(\mathbf{r}) = \left\{\frac{k\psi_c}{R_0}J_1(kr) + r\left[(B_0 - B_1\psi_c) + B_1\psi_c J_0(kr)\right]\right\}\cos\theta \tag{II.7.23}$$

As it will be discussed later, the sawtooth instability causes the safety factor on the plasma axis to be fixed to unity, that is $q(0) = 1$. This may be used to obtain the other constraint to find the remaining unknown coefficient. Hence, we first develop an expression for safety factor. Starting with the toroidal flux $\phi(r)$ for approximately circular cross section we have

$$\phi(r) = \int_0^r \int_0^{2\pi} B_t(\rho,\theta)\rho d\theta d\rho \tag{II.7.24}$$

Here, $B_t(r,\theta)$ is toroidal magnetic field across the poloidal cross section of plasma. Solov'ev equilibrium allows us to make the very good approximation

$$B_t(r,\theta) \approx \frac{B_{t0}}{1 + \dfrac{r}{R_0}\cos\theta} \tag{II.7.25}$$

Substituting (II.7.25) in (II.7.24) yields

$$\phi(r) = \int_0^r \int_0^{2\pi} \frac{B_{t0}}{1 + \dfrac{r}{R_0}\cos\theta} \rho d\theta d\rho \tag{II.7.26}$$

Since the term $r\cos\theta/R_0$ in denominator is always less than unity, one can use the binomial expansion theorem to obtain

$$\begin{aligned}\phi(r) &= \int_0^r \int_0^{2\pi} B_{t0} \sum_{n=0}^{\infty} \left(\frac{-\rho}{R_0}\cos\theta\right)^n \rho d\theta d\rho \\ &= \sum_{n=0}^{\infty} \frac{-1}{R_0} B_{t0} \int_0^r \int_0^{2\pi} \left(\frac{-\rho}{R_0}\cos\theta\right)^n \rho d\theta d\rho\end{aligned} \tag{II.7.27}$$

Using the identity

$$\int_0^{2\pi} (\cos\theta)^n d\theta = \begin{cases} \dfrac{2\sqrt{\pi}}{n!}\Gamma\left(n + \dfrac{1}{2}\right) & n = 0,2,4,... \\ 0 & n = 1,3,5,... \end{cases} \tag{II.7.28}$$

we reach at the expression for toroidal flux

$$\phi(r) = 2\sqrt{\pi} B_{t0} \sum_{n=0}^{\infty} \frac{\Gamma\left(n + \frac{1}{2}\right)}{n!} \int_0^r \left(\frac{\rho}{R_0}\right)^{2n} \rho \, d\rho$$

$$= \sqrt{\pi} B_{t0} r^2 \sum_{n=0}^{\infty} \frac{\Gamma\left(n + \frac{1}{2}\right)}{(n+1)!} \left(\frac{r}{R_0}\right)^{2n} \quad \text{(II.7.29)}$$

$$= \pi r^2 B_{t0} \left[1 + \frac{1}{4}\left(\frac{r}{R_0}\right)^2 + \frac{3}{24}\left(\frac{r}{R_0}\right)^4 + \cdots\right]$$

As it can be seen here, within zeroth-order approximation we have $\phi(r) \approx \pi r^2 B_{t0}$, which shows that the toroidal magnetic flux is approximately equal to the product of cross-sectional area of the outermost magnetic surface and toroidal magnetic field on the plasma axis.

Now, the safety factor is defined as

$$q = \frac{1}{\iota} = -\frac{\phi'}{\psi'} \quad \text{(II.7.30)}$$

in which $\iota$ is called the rotational transform. Hence, we get

$$q(r) = \frac{2\sqrt{\pi} B_{t0}}{k \psi_c J_1(kr)} \sum_{n=0}^{\infty} \frac{\Gamma\left(n + \frac{1}{2}\right)}{n!} \frac{r^{2n+1}}{R_o^{2n}} \quad \text{(II.7.31)}$$

Here, $B_{t0}$ and $R_0$ are design parameters of the tokamak machine. It is instructive if we take a look at safety factor on the plasma axis in the limit of $r \to 0$. Under these assumptions, we have $J_1(kr) \sim \frac{1}{2} kr$ by the corresponding asymptotic expansion near origin. Together with the condition imposed by sawtooth instability $q(0) = 1$, we obtain one missing equation to determine the unknown coefficients

$$q(0) = \frac{4\pi B_{t0}}{k^2 \psi_c} = 1 \quad \text{(II.7.32)}$$

and finally

$$k^2 = A_1 = \frac{4\pi B_{t0}}{\psi_c} \quad \text{(II.7.33)}$$

Other unknown parameters can be found by having the toroidal current density function, integration of which gives the total plasma current. By substitution of magnetic flux (II.7.16) into GSE we obtain the following

$$J_t(r) = \frac{-1}{\mu_0 R_0}\left(\frac{1}{r}\frac{d}{dr} + \frac{d^2}{dr^2}\right)\psi_0(r)$$
$$= \frac{-\psi_c}{\mu_0 R_0}\left[\frac{1}{r}J_0'(kr) + J_0''(kr)\right] \quad \text{(II.7.34)}$$

After some manipulation we get the fairly convenient form

$$J_t(r) = \frac{\psi_c}{2r\mu_0 R_0}\left\{2J_1(kr) + kr\left[J_0(kr) - J_2(kr)\right]\right\} \quad \text{(II.7.35)}$$

It can be readily seen that the maximum plasma current occurs on the plasma axis and is given by $J_t(0) = k^2\psi_c/\mu_0 R_0$. Now the plasma current can be computed as

$$I_p \approx \int_0^a J_t(r)2\pi r\,dr = \frac{2\pi k\psi_c a J_1(ka)}{\mu_0 R_0} \quad \text{(I.7.1)}$$

As $I_p$ is one of the design parameters of tokamaks, (II.7.33) can be simultaneously solved with (II.7.36) to determine the equilibrium.

For example, the unknowns $k$ and $\psi_0$ for Damavand Tokamak with the main parameters listed in Table II.7.1 are found as

$$\begin{aligned} k &= 41.614\ m^{-1} \\ \psi_c &= 7.98223\times 10^{-3}\ Wb \end{aligned} \quad \text{(II.7.37)}$$

Figure II.7.1 shows the plasma configuration together with poloidal and toroidal coils in the large-aspect-ratio Damavand tokamak.

**Table I.7.1:** Main parameters of Damavand tokamak.

| Parameter | Value |
|---|---|
| Major Radius | 37cm |
| Minor Radius | 7cm |
| Aspect Ratio | 5.1 |
| Toroidal Magnetic Field | 1.2T |
| Elongation | 1.2 |
| Peak Plasma Current | 40kA |
| Peak Plasma Density | $10^{19}cm^{-3}$ |
| Peak Electron Temperature | 300eV |
| Peak Ion Temperature | 150eV |
| Number of Toroidal Field Coils | 20 |
| Discharge Duration | 25ms |

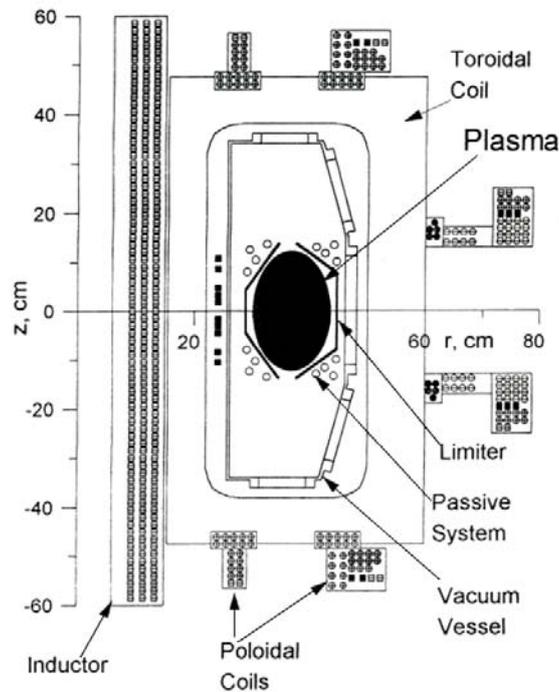

**Figure II.7.1.** Cross section of Damavand tokamak facility.

In Fig. II.7.2, variations of poloidal flux versus minor radius is demonstrated. As it is normally expected, the poloidal flux on the plasma axis is zero. This is due to the convention used here for definition of poloidal flux; one could equivalently use any other reference for poloidal flux $\psi$, as only derivatives of this function are important for determination of magnetic fields which are real physical quantities.

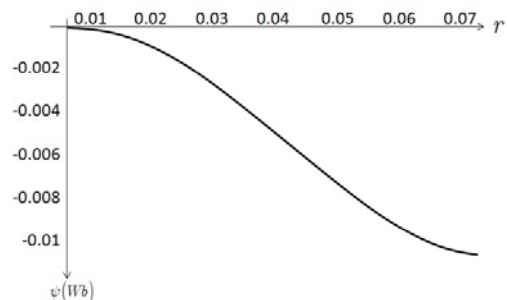

**Figure II.7.2:** Poloidal magnetic flux versus minor radius in Damavand.

Figure II.7.3 shows the variations of safety factor versus plasma minor axis. As the boundary conditions for safety factor on the plasma axis and edge require, safety factor is a monotonic increasing function of plasma minor radius and reaches from the minimum of 1 on the axis to a maximum of 4 on the boundary. Figure II.7.4 illustrates the toroidal current density function versus minor radius. As it is expected, the toroidal current density reaches its maximum on the plasma axis.

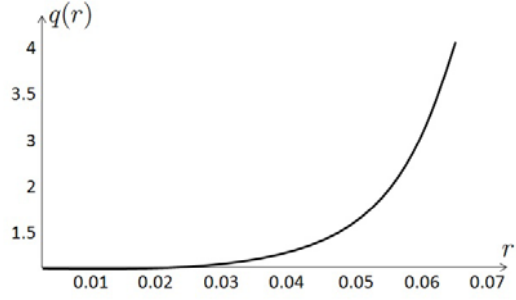

**Figure II.7.3:** Safety factor versus minor radius in Damavand.

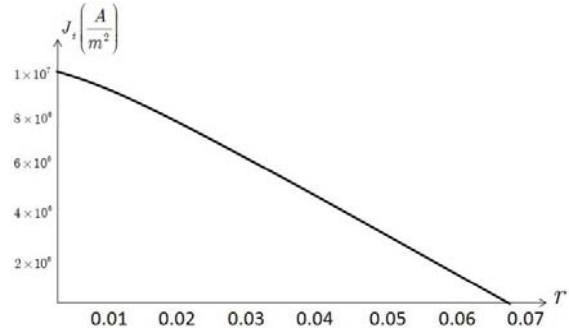

**Figure II.7.4:** Toroidal current density versus minor radius in Damavand.

## II. 7.2 Numerical Solution

As GSE intrinsically is a non-linear partial differential equation (PDE), the use of numerical solution is inevitable for description of axisymmetric plasma equilibria. Various numerical methods have been proposed to solve GSE, which could be found in the literature. The *Finite Element method* (FEM) is the most popular general purpose technique for computing accurate solutions to PDEs, which we hereby exploit to solve GSE. The family of FEMs may be divided into Galerkin and variational approaches, in both of which the solution is expanded on a set of eigenfunctions. In this section, the variational formulation of FEM, based on first-order triangular elements is presented.

The GSE (II.6.1) is here redisplayed for the sake of convenience

$$\Delta^*\psi = \left(\frac{\partial^2}{\partial r^2} - \frac{1}{r}\frac{\partial}{\partial r} + \frac{\partial^2}{\partial z^2}\right)\psi = -\mu_0 r J_t \qquad (II.7.38)$$

It can be shown that (II.7.38) may be regarded as an Euler-Ostogradskii equation of the functional

$$\Pi(\psi) = \iint \left(\frac{1}{2r}|\nabla\psi|^2 - \mu_0 J_t \psi\right) dr\, dz \qquad (II.7.39)$$

where the integration is taken over a domain $\Omega$ in the two-dimensional $(r,z)$ plane, illustrated in Fig. II.7.5, and $\nabla = \partial/\partial r\, \hat{r} + \partial/\partial z\, \hat{z}$ is the two-dimensional gradient.

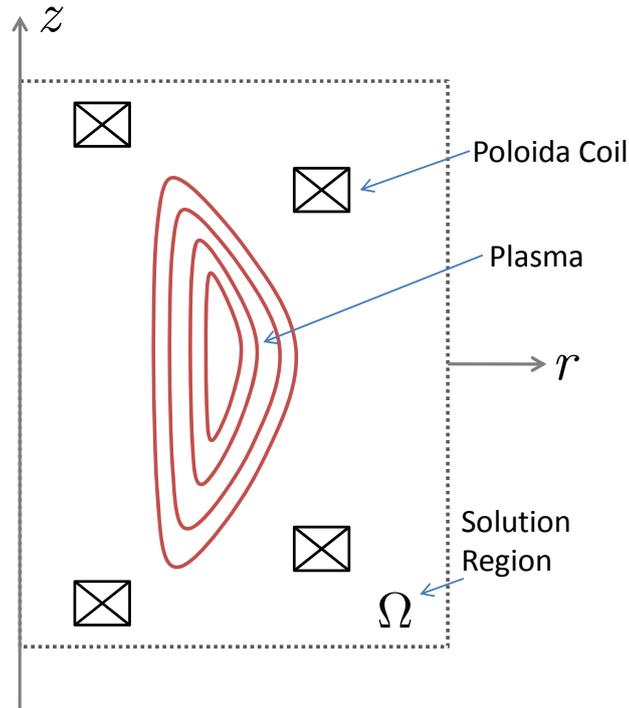

**Figure I.7.5:** Typical solution region for numerical methods.

The basic idea of the FEM is to make a piecewise approximation, that is the solution of a problem is achieved by dividing the region of interest into small regions called elements, and approximating the solution over each element by simple function with prescribed forms. The functions used to represent the behavior of the solution within an element are called interpolation functions; the simplest choice is linear dependence to coordinates referring to first-order elements. For example, the simplex element in two dimensions is a triangle with three nodes (corners). Nodes are usually shared by more than one element and it is desirable to find the nodal values of unknown functions through a set of algebraic operations which simultaneously extremize (II.7.39).

The choice of simplex triangle elements, allows us to express the variations of discretized function over the element with index $e$ as

$$\psi^e(r,z) = a^e + b^e r + c^e z \qquad (\text{II.7.40})$$

where superscript $e$ refer to element $e$, and unknown constants $a$, $b$ and $c$ are easily determined from:

$$\begin{bmatrix} a^e \\ b^e \\ c^e \end{bmatrix} = \begin{bmatrix} 1 & r_i^e & z_i^e \\ 1 & r_j^e & z_j^e \\ 1 & r_k^e & z_k^e \end{bmatrix}^{-1} \begin{bmatrix} \psi_i \\ \psi_j \\ \psi_k \end{bmatrix} = \mathbf{D}^e \Psi^e \qquad (\text{II.7.41})$$

Here, $i$, $j$, and $k$ refer to indices of nodes of element $e$. Furthermore, $r_l^e$ and $z_l^e$ correspond to radial and longitudinal coordinates of node $l$, belonging to element $e$ with $l$ standing either of $i$, $j$, or $k$. It is also customary to define the shape functions $N_l^e, l = i, j, k$ for the element $e$ as

$$\mathbf{N}^e(r,z) \equiv \begin{bmatrix} N_i^e(r,z) \\ N_j^e(r,z) \\ N_k^e(r,z) \end{bmatrix} \equiv \mathbf{D}^{e^T} \begin{bmatrix} 1 \\ r \\ z \end{bmatrix} \tag{II.7.42}$$

Therefore we have:

$$\psi^e(r,z) = \mathbf{N}^{e^T}(r,z) \Psi^e \tag{II.7.43}$$

Gradient of $\Psi$ is needed in (II.7.39), so one can approximate the gradient of unknown function over the element $e$ as

$$\nabla \psi^e = \nabla \mathbf{N}^{e^T} \Psi^e = \begin{bmatrix} D_{ji}^e & D_{jj}^e & D_{jk}^e \\ D_{ki}^e & D_{kj}^e & D_{kk}^e \end{bmatrix} \Psi^e \equiv \mathbf{B}^e \Psi^e \tag{II.7.44}$$

where $D_{rs}^e$ refers to the elements of matrix $\mathbf{D}^e$.

Now we can substitute (II.7.43) and (II.7.44) into the functional (II.7.39), which leads us to

$$\Pi(\psi) \approx \sum_e \Pi^e(\psi^e) = \sum_e \iint_{S^e} \left( \frac{1}{2} \Psi^{e^T} \mathbf{B}^{e^T} \mathbf{B}^e \Psi^e - \mu_0 \mathbf{J}_t^{e^T} \mathbf{N}^{e^T} \mathbf{N}^e \Psi^e \right) dr \, dz \tag{II.7.45}$$

Here, the summation is applied over all elements and $\mathbf{J}_t^e$ is the array of nodal values of toroidal current density function $J_t$ over the nodes $i$, $j$, and $k$ of element $e$, and $S^e$ is the area of element $e$, which is obtained from

$$S^e = \frac{1}{2} \left| \det(\mathbf{D}^e) \right|^{-1} \tag{II.7.46}$$

The variational property of (II.7.39) requires that the functional (II.7.45) with respect to the array $\Psi$ of the nodal values of the unknown function be stationary. Therefore, we have

$$\frac{\partial}{\partial \Psi^e} \sum_e \Pi^e(\psi^e) = 0 \tag{II.7.47}$$

which turns into the set of linear algebraic equations

$$\sum_e \iint_{S^e} \frac{1}{r} dr dz \mathbf{B}^{e^T} \mathbf{B}^e \Psi^e = \mu_0 \sum_e \iint_{S^e} \mathbf{N}^{e^T} \mathbf{N}^e dr dz \mathbf{J}_t^e \tag{II.7.48}$$

Here, the partial stiffness matrix $\mathbf{K}^e$ and partial force vector $\mathbf{F}^e$ are defined as:

$$\mathbf{K}^e = \iint_{S^e} \frac{1}{r} dr dz \mathbf{B}^{e^T} \mathbf{B}^e \tag{II.7.49}$$

$$\mathbf{F}^e = \mu_0 \iint_{S^e} \mathbf{N}^{e^T} \mathbf{N}^e dr dz \mathbf{J}_t^e \equiv \mu_0 \mathbf{E}^e \mathbf{J}_t^e \tag{II.7.50}$$

It should be noted $\mathbf{K}^e$ and $\mathbf{E}^e$ are both symmetric, and fortunately there are simple closed form expressions for evaluation of $\mathbf{E}^e$. As well, the double integral in $\mathbf{K}^e$ can be directly evaluated through algebraic expansion of integral region. For instance, the basic triangular elements A- and B-type as illustrated in Fig. II.7.6, yields the following expression for A-type

$$\iint_{S^e} \frac{1}{r} dr dz = \left( z_k - z_i \right) \left[ \frac{r_j}{r_j - r_i} \ln \frac{r_j}{r_i} - 1 \right] \tag{II.7.51}$$

and similarly for B-type

$$\iint_{S^e} \frac{1}{r} dr dz = \left( z_k - z_i \right) \left[ \frac{r_i}{r_i - r_j} \ln \frac{r_j}{r_i} + 1 \right] \tag{II.7.52}$$

elements. For other triangular elements which are not in the form of A- or B-type elements, one can always present them in combinations of A- and B-type, as any arbitrary triangle can be set in rectangle, surrounded by A- and B-type triangles, as illustrated in Fig. II.7.7.

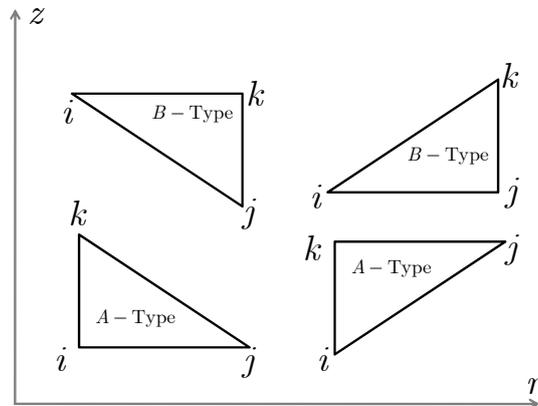

**Figure II.7.6:** Elementary triangular of A- and B-type.

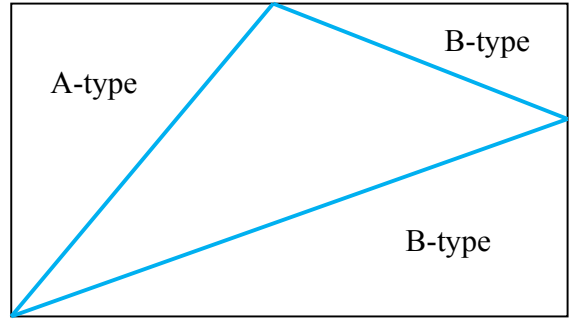

**Figure II.7.7:** Arbitrary triangle can be enclosed by three A- and B-type elements, forming a rectangle.

Therefore by subtraction of integrals belonging to the basic type elements from the surface integral on the rectangle, the unknown surface integral of the triangle is found. This technique helps us to get relieved from the excessive two-dimensional numerical integration needed over each elemental area, thus speeding up the calculations significantly.

The final system of equations by (II.7.48) can be hence written as

$$\mathbf{K}\psi = \mathbf{F} \qquad (II.7.53)$$

where the overall stiffness matrix $\mathbf{K}$ and force vector $\mathbf{F}$, have the dimensions $N \times N$ and $N \times 1$ (where $N$ is the number of nodes), respectively, and are generated by (II.7.49) and (II.7.50). The $N \times 1$ vector $\psi = \mathbf{K}^{-1}\mathbf{F}$ also denotes the array of unknown nodal values of the poloidal flux function.

### II.7.2.1 Problems with the Formulation

*a)  Singularity of (II.7.53)*
At first glance, the set of linear algebraic (II.7.53) due to the fact that the stiffness matrix $\mathbf{K}$ is singular cannot be solved. Because according to the GSE, the poloidal flux function $\psi$ is a potential and thus insensitive to the choice of an absolute reference. Therefore, at least one node must be subject to a boundary condition of Dirichlet type, so that $\mathbf{K}$ is not singular. It is now shown that $\psi$ must take on zero value on the $z$-axis.

As stated earlier, the GSE (II.6.1) allows Green's function solutions having the form

$$\psi(r,z) = \int_{-\infty}^{\infty}\int_{0}^{\infty} G(\mathbf{r},\mathbf{r}') J_t(r',z') dr' dz' \qquad (II.7.54)$$

in which the Green's function $G(\mathbf{r},\mathbf{r}')$ has the asymptotic expansion near the $z$-axis given by

$$G(\mathbf{r},\mathbf{r}') \approx \mu_0 \frac{r^2 r'^2}{4(r^2 + r'^2(z-z')^2)^{\frac{3}{2}}} \qquad (II.7.55)$$

from which we readily obtain the required boundary condition

$$\lim_{r \to 0} G(\mathbf{r}, \mathbf{r}') = 0 \tag{II.7.56}$$

Accordingly, the poloidal flux function $\psi(r,z)$ has to take on zero value at $r = 0$. This shows that the zero-boundary condition of Dirichlet type over the symmetry axis must be imposed to the system of equations (II.7.53), that is

$$\psi(0,z) = 0 \tag{II.7.57}$$

This elevates the singularity of $\mathbf{K}$.

*b)    Non-physical Neumann boundary condition*

Another problem with the system (II.7.53) is the occurrence of a non-physical boundary condition of homogeneous Neumann type over the boundary of the solution region. This difficulty happens in the form of normal magnetic surfaces or poloidal flux contours at the boundaries in the numerical solution. Mathematically it can be represented as

$$\frac{\partial}{\partial n}\psi = \hat{n} \cdot \nabla \psi = \hat{n} \cdot \hat{\psi} = 0 \tag{II.7.58}$$

where $\hat{n}$ stands for the normal vector to the boundaries. To show how this boundary condition implicitly appears in the variational formulation of the GSE (II.7.39), we directly take the variation of $\psi$ in (II.7.39), which yields

$$\delta \Pi(\psi) = \iint \left( \frac{1}{r} \nabla \psi \cdot \nabla \delta \psi - \mu_0 J_t \psi \right) dr\, dz \tag{II.7.59}$$

Using the identity

$$\frac{1}{r} \nabla \psi \cdot \nabla \delta \psi = \nabla \cdot \left( \frac{\delta \psi}{r} \nabla \psi \right) - \frac{\delta \psi}{r} \Delta^* \psi \tag{II.7.60}$$

equation (II.7.59) turns into

$$\delta \Pi(\psi) = -\iint \left( \frac{1}{r} \Delta^* \psi + \mu_0 J_t \right) \delta \psi\, dr\, dz + \iint \nabla \cdot \left( \frac{\delta \psi}{r} \nabla \psi \right) dr\, dz \tag{II.7.61}$$

The second integral in (II.7.61) can be written as

$$\iint \nabla \cdot \left( \frac{\delta \psi}{r} \nabla \psi \right) dr\, dz = \oint \frac{\delta \psi}{r} \frac{\partial \psi}{\partial n} ds \tag{II.7.62}$$

where the contour integration is done in a counter clockwise sense in the $(r,z)$ plane. Setting (II.7.61) to zero requires that the GSE hold. Therefore, in order to prevent the effect of (II.7.62) entering the solution, either $\psi$ should be fixed over the boundary, that is the case only for the left boundary at $r=0$ with (II.7.57), or its normal derivate should vanish, as stated in (II.7.58).

Physically, if the system is symmetric with respect to its equatorial plane at $z=0$, the solution region can be halved at the equatorial plane $z=0$. In this case, (II.7.58) must hold at the bottom of the solution region in order to maintain the mirror symmetry. However, the numerical solution over the right and upper borders would be meaningless, because of the fact that (II.7.58) is here non-physical. To stay away from this problem, the infinite elements provide excellent solution when used over the upper and right boundaries. The infinite elements virtually extend the solution region to infinity, where both $\psi$ and $\nabla\psi$ approach to zero and therefore (II.7.58) is automatically satisfied.

A typical infinite element is illustrated in Fig. II.7.8. The definition of an infinite element relies on taking three fixed reference points, which are not in a straight line. The first point can be chosen to be the origin of the system of coordinates at $(0,0)$. However, the second and third points vary with the position of the infinite element. In order to preserve the continuity of the solution, it is necessary to choose two consecutive boundary nodes to serve as these two points, e.g. at $(r_1, z_1)$ and $(r_2, z_2)$.

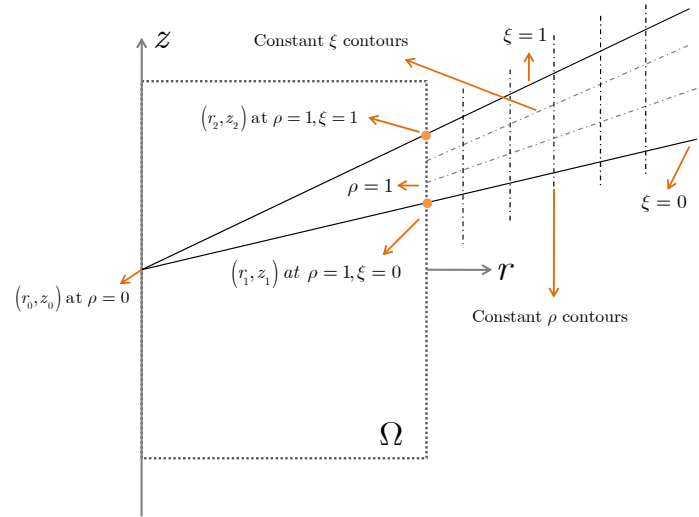

**Figure II.7.8:** Infinite element used in computation of magnetic poloidal flux.

The triangular system of coordinates $(\rho, \xi)$ for the infinite element $e$ are defined as

$$\begin{aligned} r &= \rho\left[r_1^e + \xi\left(r_2^e - r_1^e\right)\right] \\ z &= \rho\left[z_1^e + \xi\left(z_2^e - z_1^e\right)\right] \end{aligned} \quad \text{(II.7.63)}$$

This coordinate transformation will be utilized for mapping the infinite element into a rectangular region, so that the infinite element $e$ occupies the area extended from $\rho = 1$ to $\rho = \infty$, and from $\xi = 0$ to $\xi = 1$. This technique simplifies the evaluation of integrals. Moreover, the flux function is assumed to behave as

$$\psi^e(\rho,\xi) = \frac{1}{\rho}\left[\xi \psi_i^e + (1-\xi)\psi_j^e\right] \tag{II.7.64}$$

within the finite element. This special definition of variation of the unknown function on the infinite element guarantees continuity of the solution on all three borders of the element, as well as decaying the solution and its derivative at infinity. Now the contribution of the element integrals corresponding to infinite elements should be added to (II.7.48). Since $J_t = 0$ outside the solution region where the infinite elements are, therefore the infinite elements only affect the stiffness matrix $\mathbf{K}$. Hence, it would be necessary to compute only the corresponding partial stiffness matrices $\mathbf{K}^e$. One can show that

$$\mathbf{K}^e = \int_0^1 \int_0^\infty \frac{1}{\rho\left[r_1^e + \xi(r_2^e - r_1^e)\right]} \mathbf{B}^{e^T} \mathbf{B}^e \frac{\partial(r,z)}{\partial(\rho,\xi)} d\rho d\xi \tag{II.7.65}$$

where the Jacobian of the triangular system of coordinates is given by:

$$\frac{\partial(r,z)}{\partial(\rho,\xi)} = 2A^e \rho \tag{II.7.66}$$

in which $A^e$ is the area of the triangle formed by the three reference points. Note that the corresponding triangle should be formed in a counter-clock-wise sense ao that $A^e$ be positive. Also, the matrix $\mathbf{B}^e$ as a function of coordinates is given by

$$\mathbf{B}^e = \begin{bmatrix} \dfrac{\xi-1}{\rho}\dfrac{\partial \rho}{\partial r} - \dfrac{1}{\rho}\dfrac{\partial \xi}{\partial r} & -\dfrac{\xi}{\rho^2}\dfrac{\partial \rho}{\partial r} + \dfrac{1}{\rho}\dfrac{\partial \xi}{\partial r} \\ \dfrac{\xi-1}{\rho}\dfrac{\partial \rho}{\partial z} - \dfrac{1}{\rho}\dfrac{\partial \xi}{\partial z} & -\dfrac{\xi}{\rho^2}\dfrac{\partial \rho}{\partial z} + \dfrac{1}{\rho}\dfrac{\partial \xi}{\partial z} \end{bmatrix} \tag{II.7.67}$$

Thus, the evaluation of partial stiffness matrix needs numerical integration, but it is carried out only on the nodes over the boundary shared by infinite elements.

**II.7.2.2 Example**

In this section, the flux resulting from a magnetic quadrupole consisting of four poloidal coils located at $(r,z) = (1,2)$, $(2,1)$, $(1,-2)$, and $(2,-1)$ with toroidal currents $+1$, $-1$, $+1$ and $-1$, respectively, is considered. In Fig. II.7.9, the computation is done by the Variational Axisymmetric Finite Element

Method (VAFEM). It should be mentioned that since the system is symmetric with respect to the equatorial plane $z = 0$, only the upper half is shown. The resulting poloidal flux by the Green's function formalism through (II.6.3) is also illustrated in Fig. II.7.10 for comparison.

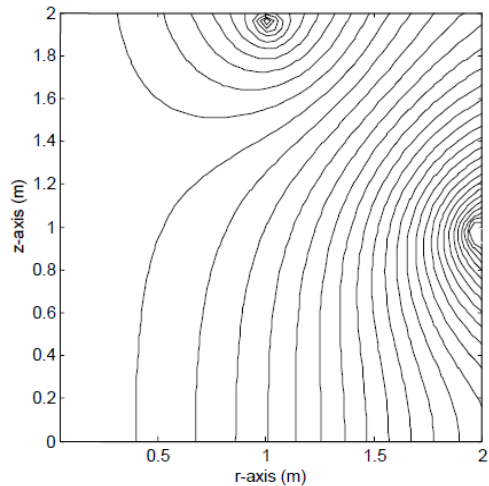

**Figure II.7.9:** Constant contours of the poloidal flux of the magnetic quadrupole computed by VAFEM.

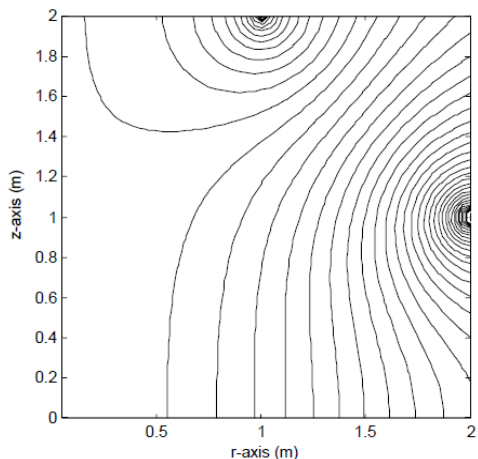

**Figure II.7.10:** Constant contours of the poloidal flux of the magnetic quadrupole computed by Green's function method.

# III. Plasma Stability

The most challenging problem in magnetic confinement of plasmas is instabilities. In order to achieve confinement, the plasma needs to be in equilibrium as well as in stable state. Otherwise, small perturbations would grow immoderately, causing catastrophic instabilities. Apart from the consideration of stability or instability, several classifications exist for plasma oscillation modes as follows:

a) *Ideal* and *resistive* MHD modes
b) *Internal* and *external* modes
c) *Pressure-driven* and *current-driven* modes
d) *Micro* and *macro* instabilities

The first classification deals with the finite resistivity of plasmas. Ideal MHD modes are described with the approximation of infinite conductivity for plasma, and therefore do not trigger tearing of magnetic surfaces. Most ideal MHD modes occur on short time scales, typically under 10μsec, and are normally controlled via passive mechanisms. In contrast, finite resistivity of plasma is usually responsible to cause major instabilities, which are accompanied with change of topology of magnetic surfaces and birth and growth of islands. As the growth rates of these instabilities are slow, they do not lead to a macroscopic loss of plasma, but instead they increase transport losses. Resistive MHD modes are associated with a typical time scale of 100μsec or larger, and need stabilization via active electronic control systems.

A second classification is based on the location of the instability where the instability starts to develop. If the corresponding mode grows without perturbing the plasma surface then it is referred to as internal modes; internal modes thus by definition affect the shape and location of closed magnetic surfaces inside the plasma, but do not cause change of topology. On the other hand, those modes that perturb the plasma boundary are called external modes. External kink modes cause large distortions in the shape of plasma column and need feedback control stabilization, otherwise they can easily lead to disruptions.

Another way to classify plasma instabilities is to notice the driving source of the plasma instability. In general, instabilities are driven by gradients in the pressure or the current density profiles. Pressure-driven modes have little role in equilibrium and stability of plasmas, while current-driven modes are usually responsible for nearly all ideal MHD instabilities.

Finally, one could categorize the instabilities with regard to the plasma volume affected. Instabilities that only affect a small portion of the plasma volume are called micro instabilities, while those associated with a large portion of the plasma volume are called macro instabilities.

Due to the fact that plasmas of thermonuclear fusion reactors can be seen as strongly non-linear, it is possible to make use of the infamous *Lyapunov Stability Theorem* to deal with such systems. In the next section we will assess this method.

## III. 1.  Lyapunov Stability in Nonlinear Systems

Any nonlinear system is subject to instability, even though it might be under equilibrium. In theory, there are several types of stability such as input-output stability, stability of periodic orbits, and the most important of all, stability of equilibrium points. Not every equilibrium configuration would result stable operation. The purpose of study of stability is to decide whether a given plasma equilibrium is stable or not, which modes are not stable and what methods should be employed to stabilize those.

In the context of nonlinear systems, *Lyapunov stability* occurs when all solutions of dynamical system which start near an equilibrium point $r_{eq}$ in the corresponding phase space, stay near it forever. Mathematically it can be written as

$$\forall \; \varepsilon > 0, \; \exists \; \delta = \delta(\varepsilon) > 0; \quad \left\| r_{eq}(0) \right\| < \delta \Rightarrow \left\| r_{eq}(t) \right\| < \varepsilon \;\; \forall t \geq 0 \qquad \text{(III.1.1)}$$

The nonlinear system at the equilibrium point $r_{eq}$, is said to be *asymptotically stable*, if all solutions that start out near $r_e$ converge to $r_{eq}$. Equivalently

$$\forall \left\| r_{eq}(0) \right\| < \delta; \quad \lim_{t \to \infty} \left\| r_{eq}(t) \right\| = 0 \tag{III.1.2}$$

For an asymptotic stable nonlinear system, the state may initially tend away from the equilibrium state but subsequently return to it. It should be noted that asymptotic stability does not imply anything about how long it takes to converge to a prescribed neighborhood of equilibrium point.

### III.1.1 Intuitive Interpretation (Ball and Wall Analogy)

Simple notions of stability often use the paradigm of the ball and curved surface as illustrated in Fig. III.1.1. This idea employs the concept of potential energy, which states that physical systems are stable when they are at their lowest energy.

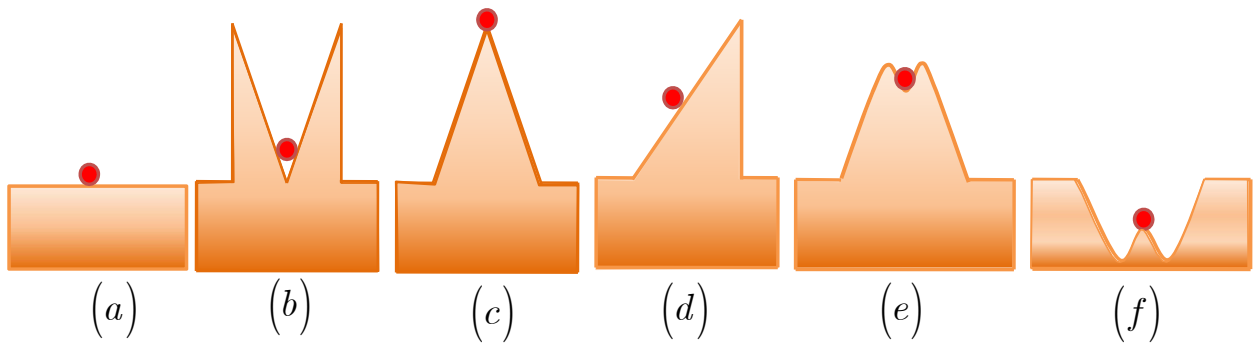

**Figure III.1.1:** Stability and equilibrium of different mechanical system consist of ball and curved surfaces.

As illustrated in Fig. III.1.1, various configurations of ball and curved surfaces lead to different states in stability and equilibrium of ball, which are listed in Table III.1.1.

**Table III.1.1:** Mechanical equilibrium and stability of a ball in a curve surface.

| Configuration | Equilibrium | Stability |
|---|---|---|
| (a) | ☺ | Marginally Stable |
| (b) | ☺ | Stable |
| (c) | ☺ | Unstable |
| (d) | No Equilibrium | Unstable |
| (e) | ☺ | Linearly Stable, Non-linearly Unstable |
| (f) | ☺ | Linearly Unstable, Non-linearly Stable |

This mechanical system is analogous to plasma in magnetic thermonuclear fusion, in which ball represents the plasma and form of the curve is a symbol of potential energy due to magnetic field configuration. When the ball is in *stable* position, any perturbation causes the ball to oscillate with reference to its equilibrium position. In contrast to ball in stable position, any small perturbation causes an unstable ball to incessantly move farther from the equilibrium point. When ball is marginally stable, it is on the border between stability and instability; any perturbation may cause switching between these two states. When the ball is linearly stable but non-linearly unstable, a small perturbation leaves the system at rest, but large perturbations kick the ball out of equilibrium. On the other hand, when the ball is linearly unstable but non-linearly stable, a large perturbation drives the system toward a stable state.

Difference of energy levels of ball between the initial and final states determines the stability of the ball, while its slope determines the equilibrium. Hence, the concept of energy principle has been evolved as a powerful mathematical tool to study the stability of equilibrium configurations.

## III. 2. Energy Principle

As stated earlier in discussion of MHD, the forces are in balance under equilibrium condition. Now, suppose that magnetic plasma is in its equilibrium state, where the potential energy of system is at a minimum. Let fluctuations cause the plasma to be physically displaced by an infinitesimal vector field $\xi$ out of its equilibrium point. Due to this fact, the net applied force **F**, on magnetic plasma is no longer equal to zero, the system is no more in equilibrium.

Assume the displacement $\xi$ and the force **F** are not in the same direction, so the force **F** tends to bring the plasma back to equilibrium. In this case, the net change in potential energy $\delta W$ is positive and the system is stable. Mathematically the extremum of the energy is a local minimum. Now if both the force **F** and displacement $\xi$ are in the same direction, then the force tends to move the system farther from its equilibrium position. One can conclude that the change in potential energy is negative and consequently the system is unstable. In this situation the extremum of the energy corresponds to a local maximum, or an inflection point.

Now, we exploit MHD theory to develop an expression for the change in potential energy $\delta W$ of plasma, when displaced from an equilibrium. We start with MHD equations

$$\rho \frac{\partial \mathbf{V}}{\partial t} + \mathbf{V} \cdot \nabla \mathbf{V} = -\nabla p + \frac{1}{\mu_0}(\nabla \times \mathbf{B}) \times \mathbf{B}$$

$$\frac{\partial \mathbf{B}}{\partial t} = \nabla \times (\mathbf{V} \times \mathbf{B})$$

$$\frac{\partial \rho}{\partial t} + \nabla \cdot (\rho \mathbf{V}) = 0$$

$$\left(\frac{\partial \rho}{\partial t} + \mathbf{V} \cdot \nabla\right)\frac{p}{\rho^\gamma} = 0$$

(III.2.1)

As MHD stability analysis is a complex nonlinear problem, linear perturbation method is the best mathematical tool that helps us to simplify the stability problem through linearization. The perturbation method leads us to an expression for the desired solution in terms of a power series in some small parameter, call perturbation. Due to the fact that the amplitude of the perturbation is infinitesimal, one can obtain the linear perturbation solution by truncating the series, usually by retaining the first two terms, referring to as the equilibrium solution and the first order perturbation correction. Hence we have

$$\rho(\mathbf{r},t) = \rho_0(\mathbf{r}) + \rho_1(\mathbf{r},t)$$
$$p(\mathbf{r},t) = p_0(\mathbf{r}) + p_1(\mathbf{r},t)$$
$$\mathbf{J}(\mathbf{r},t) = \mathbf{J}_0(\mathbf{r}) + \mathbf{J}_1(\mathbf{r},t)$$
$$\mathbf{B}(\mathbf{r},t) = \mathbf{B}_0(\mathbf{r}) + \mathbf{B}_1(\mathbf{r},t)$$

(III.2.2)

in which the terms marked with zero index $\rho_0(\mathbf{r})$, $p(\mathbf{r})$, $\mathbf{J}_0(\mathbf{r})$, and $\mathbf{B}_0(\mathbf{r})$ are respectively the mass density, pressure, current density and magnetic field, respectively; the zero subscript denotes the equilibrium values. Also, the terms marked with unity index, $\rho_1(\mathbf{r},t)$, $p_1(\mathbf{r},t)$, $\mathbf{J}_1(\mathbf{r},t)$, and $\mathbf{B}_1(\mathbf{r},t)$, represent the infinitesimal perturbation values.

Assume that the perturbed displacement from equilibrium position is represented by oscillatory time-dependent vector field $\mathbf{d}(\mathbf{r},t) = \boldsymbol{\xi}(\mathbf{r})\exp(-i\omega t)$, so that the velocity and all other perturbed quantities such as mass density, current density, pressure and magnetic field can be written as

$$\mathbf{V} = \frac{\partial \mathbf{d}(\mathbf{r},t)}{\partial t} = -i\omega\boldsymbol{\xi}(\mathbf{r})\exp(-i\omega t)$$
$$\delta\rho = \rho_1(\mathbf{r})\exp(-i\omega t)$$
$$\delta p = p_1(\mathbf{r})\exp(-i\omega t)$$
$$\delta\mathbf{B} = \mathbf{B}_1(\mathbf{r})\exp(-i\omega t)$$

(III.2.3)

where

$$\rho_1 = -\nabla\cdot(\rho_0\boldsymbol{\xi})$$
$$p_1 = -(\boldsymbol{\xi}\cdot\nabla)p_0 - \gamma p_0\nabla\cdot\boldsymbol{\xi}$$
$$\mathbf{J}_1 = \left(\frac{1}{\mu_0}\right)\nabla\times[\nabla\times(\boldsymbol{\xi}\times\mathbf{B})]$$
$$\mathbf{B}_1 = \nabla\times(\boldsymbol{\xi}\times\mathbf{B}_0)$$

(III.2.4)

The angular frequency $\omega$ in (III.2.3) may taken on complex values and appears as an eigenvalue in the formulation. It can be shown that the final eigenvalue problem appears as an eigenfunction problem belonging to the force field $\mathbf{F}(\boldsymbol{\xi})$, which is a self-adjoint operator and thus has real eigenvalues given by $\omega^2 \in \mathbb{R}$. Hence we have either non-negative $\omega^2$ corresponding to stable and oscillatory motion of the perturbation, or negative $\omega^2$ corresponding to a purely imaginary angular frequency $\omega$, thus exponentially growing perturbations and unstable equilibrium. A given equilibrium may be stable with regard to a some perturbation modes, while being unstable with regard to the rest. In practice for stable modes with real-valued $\omega$, some energy is lost along with the oscillations by various energy loss mechanisms of plasma, thereby damping the oscillation amplitudes gradually towards equilibrium.

We furthermore note that perturbation method requires smallness of perturbation amplitudes, that is

$$\begin{aligned} \rho_0(\mathbf{r}) &\gg \rho_1(\mathbf{r},t) \\ p_0(\mathbf{r}) &\gg p_1(\mathbf{r},t) \\ \mathbf{J}_0(\mathbf{r}) &\gg \mathbf{J}_1(\mathbf{r},t) \\ \mathbf{B}_0(\mathbf{r}) &\gg \mathbf{B}_1(\mathbf{r},t) \end{aligned} \qquad (\text{III.2.5})$$

Along with MHD equations (III.2.1), and perturbation expansions (III.2.2), one can easily obtain linear stability equations given by

$$\begin{aligned} \frac{\partial \mathbf{B}_1}{\partial t} &= \nabla \times (\mathbf{V} \times \mathbf{B}_0) \\ \nabla \times \mathbf{B}_1 &= \mu_0 \mathbf{J}_1 \\ \frac{\partial \rho_1}{\partial t} + \nabla \cdot (\rho_o \mathbf{V}_1) &= 0 \\ \rho_0 \frac{\partial V_1}{\partial t} + \nabla p &= \mathbf{J}_0 \times \mathbf{B}_1 + \mathbf{J}_1 \times \mathbf{B}_1 \\ \frac{\partial p_1}{\partial t} + \mathbf{V} \cdot \nabla p_0 + \frac{\gamma p_0}{\rho_0}\left(\frac{\partial \rho_1}{\partial t} + \mathbf{V} \cdot \nabla \rho_0\right) &= 0 \end{aligned} \qquad (\text{III.2.6})$$

One can decide on the stability of system with regard to a given perturbation or mode, by knowing the sign of $\delta W$ as

$$\begin{aligned} \delta W &> 0 \qquad \text{Stable} \\ \delta W &< 0 \qquad \text{Unstable} \end{aligned} \qquad (\text{III.2.7})$$

in which, the change in potential of system $\delta W$ caused by perturbation (here physical displacement) $\boldsymbol{\xi}$ is equal to

$$\delta W = \frac{-1}{2}\int \boldsymbol{\xi} \cdot \mathbf{F}(\boldsymbol{\xi})\, d\tau = \delta W_P + \delta W_V + \delta W_S \qquad (\text{III.2.8})$$

where $\delta W_P$, $\delta W_V$ and $\delta W_S$ are changes in the potential energy of the plasma, the vacuum magnetic field around the plasma and the plasma surface, given respectively by

$$\delta W_P = \frac{1}{2} \int_{\substack{\text{Plasma}\\\text{Volume}}} \left\{ \frac{B_1^2}{\mu_0} + (\nabla \cdot \boldsymbol{\xi})[\gamma p_0 \nabla \cdot \boldsymbol{\xi} + \boldsymbol{\xi} \cdot \nabla p_0] + \frac{1}{\mu_0}(\nabla \times \mathbf{B}_0) \cdot (\boldsymbol{\xi} \times \mathbf{B}_1) \right\} d\tau \quad \text{(III.2.9)}$$

$$\delta W_V = \int_{\substack{\text{Vacuum}\\\text{Region}}} \frac{B_1^2}{2\mu_0} d\tau \quad \text{(III.2.10)}$$

$$\delta W_S = \frac{1}{2} \int_{\substack{\text{Plasma}\\\text{Interface}}} \xi_n \left[ \nabla \left( p_0 + \frac{B_0^2}{2\mu_0} \right) \right] \cdot d\mathbf{S} = \frac{1}{2} \int_{\substack{\text{Plasma}\\\text{Interface}}} \xi_n^2 \left[ (\mathbf{B}_0 \cdot \nabla) \mathbf{B}_0 \right] \cdot d\mathbf{S} \quad \text{(III.2.11)}$$

In (III.2.9), the first term is the change in magnetic field energy caused by perturbation $\boldsymbol{\xi}$, and the forthcoming terms correspond to changes in energy due to perturbation in pressure and the work done against magnetic forces. As it can be seen, the two first terms in (III.2.9) are always positive, while the remaining two terms can take on negative values. Change in vacuum energy given by (III.2.10) is always positive and hence it contributes to stabilization of plasma. However, the interface energy between plasma and vacuum (III.2.11), which is due to surface current could have a destabilizing role.

### III.2.1. Application of Energy Principle
The simple configuration between plasma and vacuum is illustrated in Fig. III.2.1, where the magnetic field of plasma vanishes and pressure profile is uniform; on the other hand, pressure in vacuum is effectively zero.

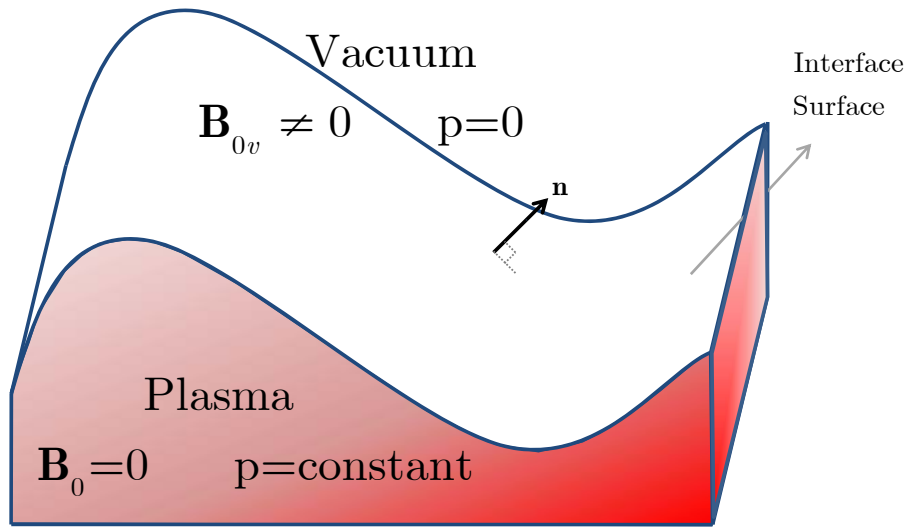

**Figure III.2.1:** Plasma–vacuum interface.

Potential energy inside the plasma is determined by (III.2.9), where in this situation the non-vanishing term is

$$\delta W_P = \frac{1}{2} \int_{\substack{\text{Plasma} \\ \text{Volume}}} \gamma p_0 \left( \nabla \cdot \boldsymbol{\xi} \right)^2 d\tau \tag{III.2.12}$$

It can be easily seen that

$$\delta W_P \geq 0 \tag{III.2.13}$$

For those modes satisfying $\nabla \cdot \boldsymbol{\xi} = 0$ then the total energy becomes

$$\delta W = \delta W_V + \delta W_S \tag{III.2.14}$$

in which according to (III.2.10) and (III.2.11) we obtain

$$\delta W = \frac{1}{2} \int_{\text{Interface}} \xi_n^2 \hat{n} \cdot \nabla \left( \frac{B_{0v}^2}{2\mu_0} \right) ds + \frac{1}{2\mu_0} \int_{\text{Vacuum}} \left| \nabla \times \mathbf{A}_v \right|^2 d\tau \tag{III.2.15}$$

As you can see, the stability is determined by the first term on the right-hand-sine of (III.2.15). Equivalently the sign of expression $\hat{n} \cdot \nabla B_{0v}^2 \big|_{\text{interface}} = \partial B_{0v}^2 / \partial n$ is the stability criterion. Therefore one can conclude that the system can be unstable when

$$\frac{\partial B_{0v}^2}{\partial n} < 0 \tag{III.2.16}$$

We notice that $\nabla B_{ov}^2$ plays an important role in the stability of system. Using the vector identity

$$\nabla \left( \mathbf{A} \cdot \mathbf{B} \right) = \left( \mathbf{A} \cdot \nabla \right) \mathbf{B} + \left( \mathbf{B} \cdot \nabla \right) \mathbf{A} + \mathbf{A} \times \left( \nabla \times \mathbf{B} \right) + \mathbf{B} \times \left( \nabla \times \mathbf{A} \right) \tag{III.2.17}$$

where by putting $\mathbf{A} = \mathbf{B} = \mathbf{B}_{0v}$ we get

$$\nabla B_{0v}^2 = 2 \left( \mathbf{B}_{0v} \cdot \nabla \right) \mathbf{B}_{0v} + 2 \mathbf{B}_{0v} \times \left( \nabla \times \mathbf{B}_{0v} \right) \tag{III.2.18}$$

In vacuum region, we have $\mu_0 \mathbf{J} = \nabla \times \mathbf{B}_{0v} = 0$, and hence

$$\nabla B_{0v}^2 = 2 \left( \mathbf{B}_{0v} \cdot \nabla \right) \mathbf{B}_{0v} \tag{III.2.19}$$

One can show that

$$\frac{\partial B_{0v}^2}{\partial n} = \hat{n} \cdot \nabla B_{0v}^2 = \hat{n} \cdot \left( \mathbf{B}_{0v} \cdot \nabla \right) B_{0v} = 2 \left( \hat{n} \cdot \mathbf{R}_c \right) \frac{B_{0v}^2}{R_c^2} \tag{III.2.20}$$

Here, the so-called curvature vector $\mathbf{R}_c$ points from the interface to the center of curvature, illustrated in Fig. III.2.2.

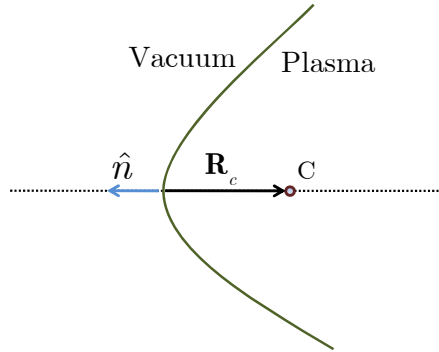

**Figure III.2.2:** Plasma-vacuum interface curvature and the curvature vector $\mathbf{R}_c$.

Substitution of (III.2.20) in (III.2.11) yields

$$\delta W_s = \frac{1}{4\mu_0} \int_{\substack{\text{Plasma}\\\text{Interface}}} (\hat{n} \cdot \boldsymbol{\xi})^2 (\hat{n} \cdot \mathbf{R}_c) \frac{B_{0v}^2}{R^2} ds \qquad (\text{III.2.21})$$

According to (III.2.21), the dot product of normal and curvature vectors $\hat{n} \cdot \mathbf{R}_c$ determines the stability as follows:

*Case 1*: $\hat{n} \cdot \mathbf{R}_c > 0$ and surface energy is stabilizing. The plasma and vacuum configuration at the interface is shown in Fig. III.2.3, in which is known as good curvature.

*Case 2*: $\hat{n} \cdot \mathbf{R}_c < 0$ and surface energy is destabilizing. The plasma and vacuum configuration at the interface is shown in Fig. III.2.4, in which is known as bad curvature.

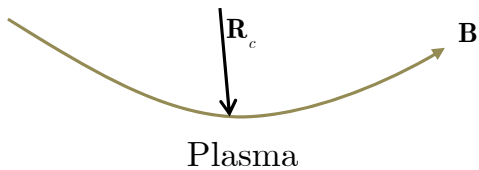

**Figure III.2.3:** Good curvature.

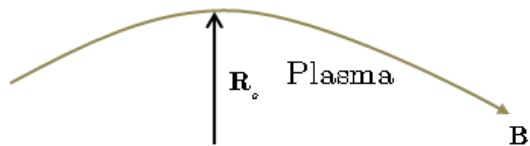

**Figure III.2.4:** Bad curvature.

In the next section we exploit energy principle to analyze the stability properties of the $\theta$-pinch, the $z$-pinch, and the general screw pinch.

## III.3. Modal Analysis

In this section we present the application of energy principle to analyze the stability characteristics of $\theta$-pinch and $z$-pinch. With the same method The kink instability is being studied.

### III.3.1 $\theta$-pinch

Since the equilibrium is symmetric with respect to both $\theta$- and $z$-coordinates, the perturbation can have the following form

$$\boldsymbol{\xi}(\mathbf{r}) = \boldsymbol{\xi}(r)\exp\left[i(m\theta + kz)\right] \tag{III.3.1}$$

where $m$ and $k$ are called poloidal and toroidal (or axial) mode numbers, respectively. While $m$ must be an integer, $k$ is a continuous variable if the system be infinitively long. For a cylinder with finite length $k$ can take on discrete values. Different values of mode numbers $m$ and $k$ lead to various perturbations, as illustrated in Fig. III.3.1.

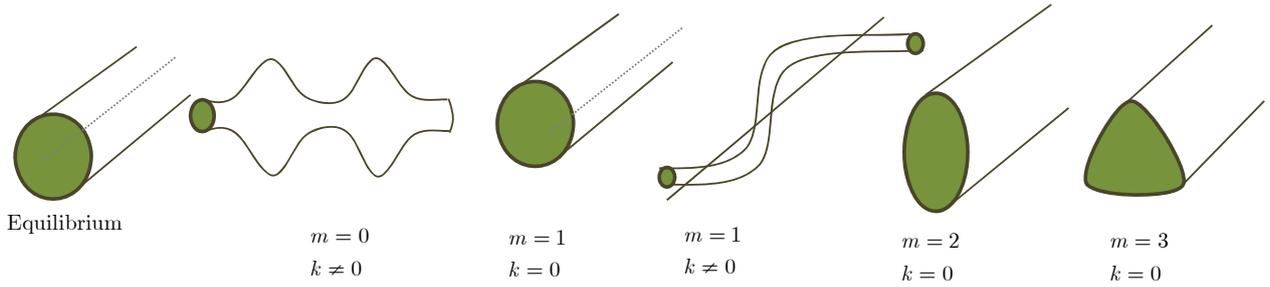

**Figure III.3.1:** Different perturbation correspond with various values of $m$ and $n$.

According to Fig. III.3.1, the mode with $m=0$ and $k\neq 0$, called *sausage mode*, usually arise from thermal disturbances, which can cause the incompressible plasma to develop axially periodic constrictions and bulges. The $m=1$ and $k=0$ mode, only shifts the plasma column with respect to its axis. Helical kink instabilities occurs in mode with $m=1$ and $k\neq 0$. In this instability, the concave surfaces of the plasma experience concentration of the azimuthal field resulting in a magnetic pressure that increases the concavity. Likewise at the convex surfaces, the azimuthal field is weaker so that the convex bulge will tend to increase. The plasma cross section at $m=2$ mode becomes elliptical , while for $m=3$ mode, the cross section becomes triangular, and so on.

One can obtain the expression for potential energy when $k\neq 0$ as

$$\frac{\delta W}{L} = \frac{\pi}{\mu_0}\int_0^a \frac{k^2 B_z^2}{m^2 + k^2 r^2}\left[\left|r\frac{\partial \xi_r}{\partial r}\right|^2 + \left(m^2 + k^2 r^2\right)\left|\xi_r\right|^2\right] r\,dr \tag{III.3.2}$$

It can be understood from (III.3.2) that for every choice of mode numbers, we have $\delta W > 0$; therefore $\theta$-pinch is stable with regard to all MHD modes having finite $k$. One reason that $\theta$-pinch is stable for all MHD modes, is that $\theta$-pinch has no curvature field lines. Another important factor that makes $\theta$-pinch so much resistant to MHD modes is that there is no axial current, i. e. $\mathbf{J}_z = 0$, and hence no current driven instabilities. The magnetic field lines of a typical $\theta$-pinch is depicted in Fig. III.3.2.

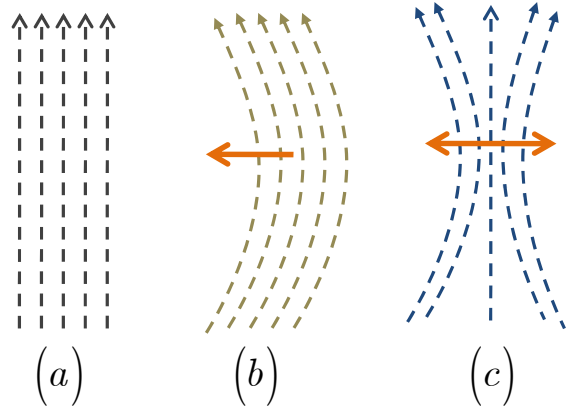

**Figure III.3.2:** magnetic field lines of a typical $\theta$-pinch.

According to Fig. III.3.2a, magnetic field lines in $\theta$-pinch are straight, bending them, Fig. III.3.2b, will lead to a magnetic field tension, and consequently to a force that makes the field straight again. Meanwhile, squeezing field lines as in Fig. III.3.2c, will lead to an increase in the magnetic field pressure and consequently to a force that prevents further squeezing.

### III.3.2 *z*-Pinch

#### III.3.2.1 *z*-Pinch, $m \neq 0$ Modes

The equilibrium condition for *z*-pinch was mentioned in (II.4.10), where we display it here again

$$\frac{\partial p_0}{\partial r} = -\frac{B_\theta}{\mu_0} \frac{\partial}{\partial r}\left(r B_\theta\right) \qquad \text{(III.3.3)}$$

The potential energy of a *z*-pinch with $m \neq 0$ condition may be shown to be

$$\frac{\delta W}{L} = \frac{\pi}{\mu_0} \int_0^a \left[\left(2\mu_0 r \frac{\partial p}{\partial r} + m^2 B_\theta^2\right)\frac{|\xi_r|^2}{r} + \frac{m^2 r^2 B_\theta^2}{m^2 + r^2 k^2}\left|\frac{\partial}{\partial r}\left(\frac{1}{r}\xi_r\right)\right|\right] r\, dr \qquad \text{(III.3.4)}$$

The worst situation is achieved by letting $k \to \infty$. Therefore the stability is determined by

$$\frac{\delta W}{L} = \frac{\pi}{\mu_0} \int_0^a \left[ 2\mu_0 r \frac{\partial p}{\partial r} + m^2 B_\theta^2 \right] |\xi_r|^2 dr \qquad \text{(III.3.5)}$$

In order for the system to be stable for all point inside the plasma the integrand should be positive, hence

$$m^2 B_\theta^2 > -2\mu_0 r \frac{\partial p_0}{\partial r} \qquad \text{(III.3.6)}$$

Substituting (III.3.3) in (III.3.6) gives

$$m^2 B_\theta^2 > 2 B_\theta \frac{\partial}{\partial r}(rB_\theta) \qquad \text{(III.3.7)}$$

The right-hand-side of (III.3.7) can be written as

$$B_\theta \frac{\partial}{\partial r}(rB_\theta) = B_\theta \frac{\partial}{\partial r}\left[r^2 \frac{B_\theta}{r}\right] = r^2 B_\theta \frac{\partial}{\partial r}\left(\frac{B_\theta}{r}\right) + 2B_\theta^2 \qquad \text{(III.3.8)}$$

or equivalently

$$B_\theta \frac{\partial}{\partial r}(rB_\theta) = r\frac{\partial}{\partial r}\left(\frac{B_\theta^2}{2}\right) + B_\theta^2 = \frac{\partial}{\partial r}\left(\frac{rB_\theta^2}{2}\right) + \frac{B_\theta^2}{2} \qquad \text{(III.3.9)}$$

Therefore by using (III.3.8) or (III.3.9) and substitution in the stability criterion (III.3.7), we arrive at

$$\frac{1}{2}(m^2 - 4) > \frac{r^2}{B_\theta}\frac{\partial}{\partial r}\left(\frac{B_\theta}{r}\right) \qquad \text{(III.3.10)}$$

or

$$\frac{1}{2}(m^2 - 1) > B_\theta^{-2}\frac{\partial}{\partial r}(rB_\theta^2) \qquad \text{(III.3.11)}$$

Typical magnetic field of a $z$-pinch is illustrated in Fig. III.3.3. According to Fig. III.3.3, for $r \to 0$ the magnetic field in $z$-pinch is proportional to $r$. Therefore the stability condition (III.3.10) simply becomes

$$m^2 > 4 \qquad \text{(III.3.12)}$$

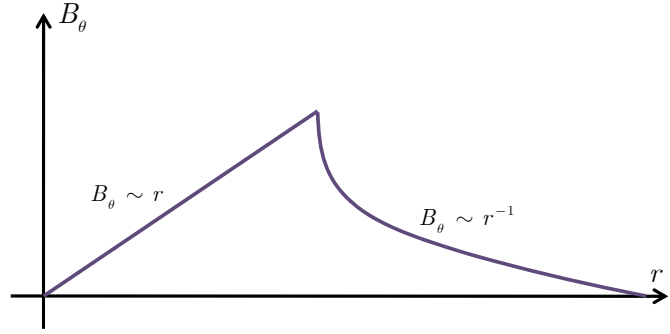

**Figure III.3.3:** *z*-pinch profile.

This is while for $r \to \infty$ we have $B_\theta \sim r^{-1}$, and the stability condition (III.3.10) becomes

$$\frac{1}{2}(m^2 - 4) > \frac{r^2}{B_\theta} \frac{\partial}{\partial r}\left(\frac{B_\theta}{r}\right) > r^3 \frac{\partial}{\partial r}\left(\frac{1}{r^2}\right) = -2 \qquad (\text{III.3.13})$$

or $m^2 > 0$. Hence the stability condition $m > 2$ is dominant. Similarly, the stability condition (III.3.11) for $r \to 0$ and $m = 1$ gives

$$r > 0 \qquad (\text{III.3.14})$$

In which, it means that for core plasma with small $r$, *z*-pinch is unstable. For plasma boundary of a thick *z*-pinch with $r \to \infty$, the stability condition is simply

$$m^2 > 1 \qquad (\text{II.3.1})$$

As in *z*-pinch the azimuthal current is zero $\mathbf{J}_\theta = 0$, the instability for $m = 1$ is caused by bad curvature of magnetic field lines.

### III.3.2.1 *z*-pinch, $m = 0$ Mode

Potential energy of the *z*-pinch for $m = 0$ mode equals to

$$\frac{\delta W}{L} = \frac{\pi}{\mu_0} \int_0^a \frac{\xi_r^2}{r}\left[\frac{r\gamma p_0 B_\theta^2}{\gamma \mu_0 p_0 + B_\theta^2} + 2r\frac{\partial p_0}{\partial r}\right]dr \qquad (\text{III.3.16})$$

where

$$\xi_z = \frac{i}{\gamma p_0 + B_\theta^2/\mu_0}\left[\frac{rB_\theta^2}{\mu_0}\frac{\partial}{\partial r}\left(\frac{\xi_r}{r}\right) + \frac{\gamma p_0}{r}\frac{\partial}{\partial r}(r\xi_r)\right] \qquad (\text{III.3.17})$$

In order for the z-pinch to be stable for $m = 0$ mode, the integrand of (III.3.16) should be positive, that is

$$-\frac{r}{p_0}\frac{\partial p_0}{\partial r} < \frac{4\gamma}{2 + \gamma\left(2\mu_0 p_0/B_\theta^2\right)} \tag{III.3.18}$$

At the plasma edge the pressure rapidly goes to zero that makes the radial pressure gradient $\partial p_0/\partial r$ to increase dramatically. This situation does not satisfy the stability condition (III.3.18). If the plasma is to be confined well by magnetic field, the upper limit on which the pressure can be decreased becomes

$$-\frac{r}{p}\frac{dp}{dr} < 2\gamma \tag{III.3.19}$$

or equivalently

$$-\frac{dp}{p} < 2\gamma\frac{dr}{r} \tag{III.3.20}$$

Integration of both sides of (III.3.20) and noting that $\gamma \approx 5/3$ gives

$$p(r) > r^{-\frac{10}{3}} \tag{III.3.21}$$

The above result states that, pressure must vary no faster than $r^{-\frac{10}{3}}$.

**III.3.3 Kink Instability**
The kink instability is an ideal MHD instability which at low $\beta$ is driven by the current gradient and at high $\beta$, by pressure gradients. It usually happens when between the plasma and the conducting wall there is a vacuum region. As stated earlier, in order to examine stability of plasma we perturb plasma from its equilibrium position, and determine whether a small perturbation will grow to disrupt the plasma or tends back to equilibrium. The perturbation in primitive toroidal coordinates may be written as

$$\boldsymbol{\xi}(\mathbf{r},t) = \xi(r)\exp\left[i(m\theta - n\varphi - \omega t)\right] \tag{III.3.22}$$

in which $\varphi$ and $\theta$ are the toroidal and poloidal angles, respectively. Under equilibrium, the plasma region is located at $r < a$, and the vacuum region is $a < r < b$, where $b$ is the radius of perfectly conducting wall. Plasma potential energy $\delta W_p$ for this configuration becomes

$$\delta W = \frac{B_\varphi^2}{\mu_0 R_0} \int_0^a \left[ \left(m^2 - 1\right)\xi^2 + r^2 \left(\frac{\partial \xi}{\partial r}\right)^2 \right] \left(\frac{n}{m} - \frac{1}{q}\right)^2 r dr +$$
$$+ \frac{B_\varphi^2 a^2 \xi_a^2}{\mu_0 R_0} \left[ \frac{2}{q(a)} \left(\frac{n}{m} - \frac{1}{q(a)}\right) + \left(\frac{n}{m} - \frac{1}{q(a)}\right)^2 \right]$$
(III.3.23)

where $q$ is the safety factor. On the other hand, the potential energy of vacuum is obtained as

$$\delta W_v = \frac{\pi^2 R}{\mu_0} \left[\frac{n}{m} - \frac{1}{q(a)}\right]^2 m \lambda a^2 \xi_a^2$$
(III.3.24)

where

$$\lambda = \frac{1 + (a/b)^{2m}}{1 - (a/b)^{2m}}$$
(III.3.25)

Using (III.3.24) and (III.3.23), one can obtain the total change in potential energy as

$$\delta W = \frac{\pi^2 B_\varphi^2}{\mu_0 R} \int_0^a \left[ \left(r \frac{\partial \xi}{\partial r}\right)^2 + \left(m^2 - 1\right)\xi^2 \right] \left(\frac{n}{m} - \frac{1}{q}\right)^2 r dr$$
$$+ \frac{2\pi^2 B_\varphi^2 a^2 \xi_a^2}{\mu_0 q(a) R} \left[ \left(\frac{n}{m} - \frac{1}{q(a)}\right) + (1 + m\lambda)\left(\frac{n}{m} - \frac{1}{q(a)}\right) \right]$$
(II.3.2)

From (III.3.26) one can conclude that if the vacuum region could be removed and the conducting wall would touch the plasma boundary, then $\xi_a$ would vanish, the potential energy difference would become positive, and in this case the plasma column would be stable.; clearly, this condition is not practical. Otherwise the stability condition for $(m, n)$ mode is satisfied by $q_a > m/n$.

### III.3.3 Interchange Instability
When two types of fluids in contact are situated with an external force such that the potential energy is not a minimum, *interchange instability* occurs and the two fluids will then interchange locations to bring the potential energy to a minimum. In plasmas with magnetic fields, the plasma may interchange position with the magnetic field. A prime example is the flute instability in mirror machines, in which the perturbation is uniform parallel to the magnetic field.

Two neighboring magnetic flux tubes with $p_1$ and $p_2$ as initial pressures, and $V_1$ and $V_2$ as volumes of tubes are shown in Fig. III.3.4.

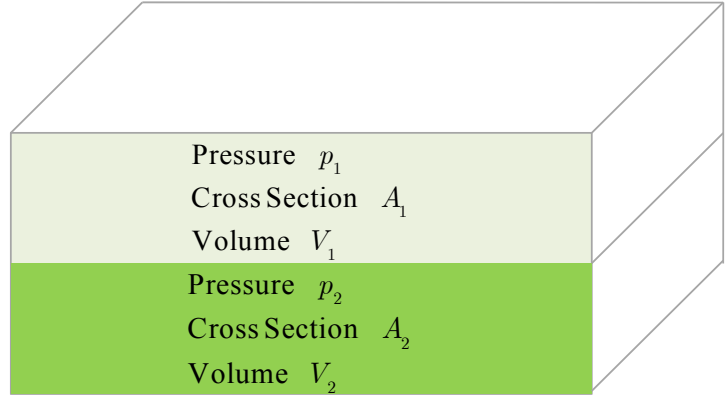

**Figure III.3.4**: Two adjacent magnetic flux tubes.

As magnetic fluxes are assumed to be equal, we have:

$$\phi = B_1 A_1 = B_2 A_2 \quad \text{(III.3.27)}$$

where $B_1$ and $B_2$ are the magnetic fields, and $A_1$ and $A_2$ are the cross sections of two flutes. Plasma of volume $V$ is adiabatic when

$$pV^\gamma = \text{cte} \quad \text{(III.3.28)}$$

After interchanging the new pressures will be

$$p_1' = p_1 \left(\frac{V_1}{V_2}\right)^\gamma \quad \text{(III.3.29)}$$

$$p_2' = p_2 \left(\frac{V_2}{V_1}\right)^\gamma \quad \text{(III.3.30)}$$

The difference in final and initial potential energy and of two tubes is therefore

$$\delta W = \frac{1}{\gamma - 1}\left[p_1\left(\frac{V_1}{V_2}\right)^\gamma V_2 + p_2\left(\frac{V_2}{V_1}\right)^\gamma V_1 - p_1 V_1 - p_2 V_2\right] \quad \text{(III.3.31)}$$

Now let

$$\begin{aligned}\delta p &= p_2 - p_1 \\ \delta V &= V_2 - V_1\end{aligned} \quad \text{(III.3.32)}$$

Using (III.2.32), the change in potential energy becomes

$$\delta W = \delta p \delta V + \gamma p \frac{1}{V} \delta V^2 \qquad \text{(III.3.33)}$$

The second term in right-hand-side of (III.3.33) is always positive, and it can be ignored at plasma edge where the pressure is too small. Therefore the stability condition simply becomes

$$\delta p \delta V > 0 \qquad \text{(III.3.34)}$$

$\delta p$ for a confined plasma is negative because of outward decay pressure profile. Therefore in order to make the plasma stable, it is required to have negative $\delta V$ as well. But $\delta V$ can be written as

$$\delta V = \delta \left( \int A dl \right) \qquad \text{(III.3.35)}$$

Using $\phi = AB$, one can rewrite (III.3.35) as

$$\delta V = \delta \left( \int A dl \right) = \phi \, \delta \left( \int \frac{dl}{B} \right) \qquad \text{(III.3.36)}$$

Hence for stability we need to have

$$\delta \left( \int \frac{dl}{B} \right) < 0 \qquad \text{(III.3.37)}$$

## III.4. Simplifications for Axisymmetric Toroidal Machines

The Change in potential energy, which determines the stability of system (III.2.8), can also be evaluated in axisymmetric toroidal system. To derive $\delta W$ in axisymmetric system, it is convenient to employ flux coordinate system $(\psi, \zeta, \varphi)$, which is shown in Fig. (III.4.1).

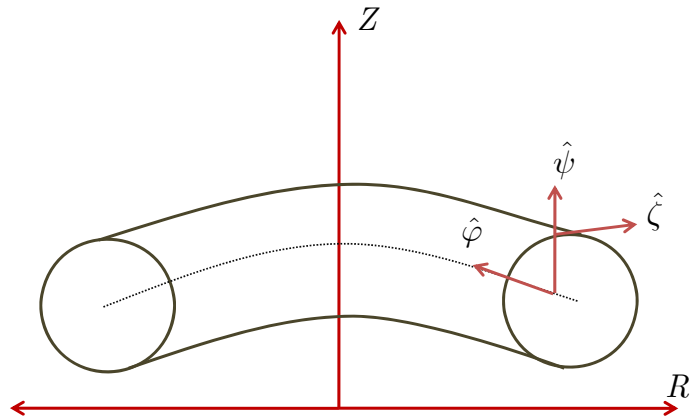

**Figure III.4.1:** Flux orthogonal coordinate system.

$\psi$ is the flux function which is defined by

$$\psi = -RA_\varphi \tag{III.4.1}$$

Also $\zeta$ and $\varphi$ are poloidal and toroidal angels, respectively. Magnetic field and a field line in flux coordinates can be written as

$$\mathbf{B} = \hat{\varphi} \times \hat{\psi} + \overline{I}(\psi)\hat{\varphi} \tag{III.4.2}$$

$$\frac{Rd\varphi}{JB_\zeta d\zeta} = \frac{B_\varphi}{B_\zeta} = \frac{\overline{I}(\psi)}{RB_\zeta} \tag{III.4.3}$$

where $J(\psi)$ is the Jacobian determinant, which is obtained by using (II.2.5) and the flux function $\overline{I}(\psi)$ is defined as

$$\overline{I}(\psi) = \frac{\mu_0 I(\psi)}{2\pi} \tag{III.4.4}$$

Safety factor in flux coordinates can also be defined using the path integral as

$$q(\psi) = \oint \frac{B_\varphi}{RB_\zeta} ds \tag{III.4.5}$$

in which the integral is taken along a closed path encircling the minor axis and lying on a specific magnetic surface. Substituting (III.4.3) in (III.4.5) yields

$$q(\psi) = \frac{1}{2\pi} \oint \frac{J(\psi)\overline{I}(\psi)}{R^2} d\zeta \tag{III.4.6}$$

The change in the potential energy of system can be written as

$$W = \frac{1}{2} \int_V \left[ \frac{B_1^2}{\mu_0} + \gamma p |\nabla \cdot \boldsymbol{\xi}|^2 + (\boldsymbol{\xi} \cdot \nabla p)(\nabla \cdot \boldsymbol{\xi}^*) - \boldsymbol{\xi}^* \cdot (\mathbf{J} \times \mathbf{B}_1) \right] d\tau \tag{III.4.7}$$

But the perturbation vector in flux coordinates can be represented in the covariant form of

$$\boldsymbol{\xi} = \xi_\psi \hat{\psi} + \xi_\zeta \hat{\zeta} + \xi_\varphi \hat{\varphi} \tag{III.4.8}$$

with the components

$$\xi_\psi = \frac{K}{RB_\zeta}$$

$$\xi_\zeta = B_\zeta L \quad \text{(III.4.9)}$$

$$\xi_\varphi = RM + \frac{\overline{I}}{R}L$$

Here, $M$ defined as

$$M \equiv \frac{1}{RB_\zeta}\left(B_\zeta \xi_\varphi - B_\varphi \xi_\zeta\right) \quad \text{(III.4.10)}$$

The first term in (III.4.7) can be written as

$$\frac{B_1^2}{2\mu_0} = \frac{\mathbf{B}_1 \cdot \mathbf{B}_1}{2\mu_0} = \frac{1}{2\mu_0}\left[\left|B_{1\zeta}\right|^2 + \left|B_{1\varphi}\right|^2\right] \quad \text{(III.4.11)}$$

One should thus obtain expressions for $B_{1\zeta}^2$ and $B_{1\varphi}^2$ in flux coordinates. We first note by (III.2.4) that

$$\begin{aligned}B_1 &= \nabla \times (\boldsymbol{\xi} \times \mathbf{B}) = \nabla \times \left[\left(\xi_\zeta B_\varphi - \xi_\varphi B_\zeta\right)\hat{\psi} - \xi_\psi B_\varphi \hat{\zeta} + \xi_\psi B_\zeta \hat{\varphi}\right] \\ &= B_{1\psi}\hat{\psi} + B_{1\zeta}\hat{\zeta} + B_{1\varphi}\hat{\varphi}\end{aligned} \quad \text{(III.4.12)}$$

Where

$$\begin{aligned}B_{1\psi} &= \frac{i}{B_\zeta R} B k_\parallel K \\ B_{1\zeta} &= -B_\zeta \left(inM + \frac{\partial K}{\partial \psi}\right) \\ B_{1\varphi} &= \frac{R}{J}\left[-\frac{\partial}{\partial \psi}\left(\frac{J\overline{I}}{R^2}K\right) + \frac{\partial M}{\partial \zeta}\right]\end{aligned} \quad \text{(III.4.13)}$$

Consequently, we have the followings

$$\begin{aligned}\frac{\left|B_{1\zeta}\right|^2}{2\mu_0} &= \frac{B_\zeta^2}{2\mu_0}\left|inM + K'\right|^2 = \frac{B_\zeta^2}{2\mu_0}\left|inM + K' - \frac{\mu_0 J_\varphi}{RB_\zeta^2}K\right| \\ &+ \left(inMK^* - inM^*K\right)\frac{J_\varphi}{2R} + \left(K'K^* + K^{*\prime}\right)\frac{J_\varphi}{R} - \frac{\mu_0 J_\varphi^2}{2R^2 B_\zeta^2}KK^*\end{aligned} \quad \text{(III.4.14)}$$

$$\frac{|B_{1\varphi}|^2}{2\mu_0} = \frac{R^2}{2\mu_0 J^2}\left|\frac{\partial M}{\partial \zeta} - \overline{I}\left(\frac{JK}{R^2}\right)' - \frac{JK}{R^2}\overline{I}'\right|^2 = \frac{R^2}{2\mu_0 J^2}\left|\frac{\partial M}{\partial \zeta} - I\left(\frac{JK}{R^2}\right)'\right|^2 + \frac{\overline{I}'^2}{\mu_0 R^2}KK^*$$
$$- \frac{\overline{I}'}{2\mu_0 J}\left(\frac{\partial M}{\partial \zeta}K^* - \frac{\partial M^*}{\partial \zeta}K\right) + \frac{\overline{II}'}{2\mu_0 R^2}\left(K'K^* + K^{*\prime}K\right) + \frac{\overline{II}'}{\mu_0 J}\left(\frac{J'}{R^2} - \frac{2R'}{R^3}J\right)KK^*$$
(III.4.15)

Using (III.4.8), the term $\nabla \cdot \boldsymbol{\xi}$ in (III.4.7) in flux coordinate system can be expressed as

$$\nabla \cdot \boldsymbol{\xi} = \frac{1}{J}\left[\frac{\partial}{\partial \psi}\left(JB_\zeta R\xi_\psi\right) + \frac{\partial}{\partial \zeta}\left(\frac{\xi_\zeta}{B_\zeta}\right) + \frac{\partial}{\partial \varphi}\left(\frac{J\zeta_\varphi}{R}\right)\right]$$
$$= \frac{1}{J}\left[\frac{\partial}{\partial \psi}(JK) + \frac{\partial}{\partial \zeta}(L) + \frac{\partial}{\partial \varphi}J\left(M + \frac{\overline{I}}{R^2}L\right)\right]$$
(III.4.16)

Using (III.4.9) and letting $\boldsymbol{\xi}(\mathbf{r}) = \boldsymbol{\xi}(\psi,\zeta)\exp(in\varphi)$, (III.4.16) turns into

$$\nabla \cdot \boldsymbol{\xi} = \frac{1}{J}(JK)' + iBk_\parallel L + inM$$
(III.4.17)

where

$$k_\parallel = -\left(\frac{\overline{I}}{BR^2}n + i\frac{1}{JB}\frac{\partial}{\partial \zeta}\right)$$
(III.4.18)

The term $\boldsymbol{\xi} \cdot \nabla p$ in (III.4.7) in flux coordinates also takes the form

$$\boldsymbol{\xi} \cdot \nabla p = \xi_\psi RB_\xi p' = Kp'$$
(III.4.19)

From GSE (II.5.20), one can obtain

$$\boldsymbol{\xi} \cdot \nabla p = Kp' = -K\left(\frac{J_\varphi}{R} + \frac{\overline{II}'}{\mu_0 R^2}\right)$$
(III.4.20)

Multiplying (III.4.17) by (III.4.20) yields

$$(\boldsymbol{\xi} \cdot \nabla p)(\nabla \cdot \boldsymbol{\xi}) = -K\left(\frac{J_\varphi}{R} + \frac{\overline{II}'}{\mu_0 R^2}\right)\left[\frac{1}{J}(JK)' + iBk_\parallel L + inM\right]$$
(III.4.21)

Therefore, we similarly obtain

$$(\boldsymbol{\xi} \cdot \nabla p)(\nabla \cdot \boldsymbol{\xi}^*) = -K\left(\frac{J_\varphi}{R} + \frac{\overline{II}'}{\mu_0 R^2}\right)\left[\frac{1}{J}(JK^*)' - iBk_\parallel L^* - inM^*\right] \quad \text{(III.4.22)}$$

The next term in (III.4.7) that should be manipulated in order to be expressible in flux coordinates is $\boldsymbol{\xi}^* \cdot (\mathbf{J} \times \mathbf{B}_1)$. Using the vector identity $(\mathbf{A} \times \mathbf{B}) \cdot \mathbf{C} \equiv (\mathbf{B} \times \mathbf{C}) \cdot \mathbf{A} \equiv (\mathbf{C} \times \mathbf{A}) \cdot \mathbf{B}$ we have

$$\boldsymbol{\xi}^* \cdot (\mathbf{J} \times \mathbf{B}_1) = \mathbf{J} \cdot (\mathbf{B}_1 \times \boldsymbol{\xi}^*) \quad \text{(III.4.23)}$$

Substituting (III.4.12) and (III.4.8) in (II.4.23) yields

$$\begin{aligned}\mathbf{J} \times (\mathbf{B}_1 \times \boldsymbol{\xi}^*) &= \frac{\overline{I}'}{\mu_0 J}\left(\frac{\overline{I}J}{R^2}K\right)' K^* - \frac{\overline{I}'}{\mu_0 J}\frac{\partial M}{\partial \zeta}K^* + \frac{K^* K'}{R}J_\varphi \\ &+ i\frac{Bk_\parallel}{R}K\left[L^* j_\varphi + \frac{\overline{I}'}{\mu_0}\left(RM^* + \frac{\overline{I}}{R}L^*\right)\right] + i\frac{nMK^*}{R}J_\varphi\end{aligned} \quad \text{(III.4.24)}$$

where $J$ is the Jacobian determinant and $J_\varphi$ is toroidal current. Now, by substituting (II.4.14), (II.4.15), (II.4.22), and (II.4.24) in (II.4.7) we get:

$$\begin{aligned}W = \int_V &\left[\frac{1}{2\mu_0}\frac{B^2 k_\parallel^2}{B_\zeta^2 R^2}|K|^2 + \frac{1}{2\mu_0}\frac{R^2}{J^2}\left|\frac{\partial M}{\partial \zeta} - I\left(\frac{JK}{R^2}\right)'\right|^2 - UKK^* \right. \\ &\left. + \frac{B_\zeta^2}{2\mu_0}\left|inM + K' - \frac{\mu_0 J_\varphi}{RB_\zeta^2}K\right|^2 + \frac{1}{2}\gamma p\left|\frac{1}{J}(JK)' + iBk_\parallel L + inM\right|^2\right]d\tau\end{aligned} \quad \text{(III.4.25)}$$

where $\gamma$ is the plasma density, and $d\tau$ in flux coordinates is

$$d\tau = J d\psi d\zeta d\varphi \quad \text{(III.4.26)}$$

Also, $U$ in (III.4.25) is defined as

$$U \equiv \frac{\overline{II}'}{\mu_0 R^2}\frac{R'}{R} + \frac{J_\varphi}{2R}\left(\frac{J'}{J} + \frac{\mu_0 J_\varphi}{RB_\zeta^2}\right) = \frac{\overline{II}'}{\mu_0 R^2}\frac{R'}{R} + \frac{J_\varphi}{R}\left(\frac{J'}{J} + \frac{B_\zeta'}{B_\zeta}\right) \quad \text{(III.4.27)}$$

# IV. Plasma Transport

In sections II and III, we studied equilibrium and stability of a plasma, respectively. The quality of plasma confinement with regard to the maximum plasma temperature, density, and confinement time is limited by the transport of heat and particles across the magnetic surfaces. In most equilibrium configurations transport contributes to significant loss of energy from the plasma core. Gradients in particle density, as well as electron and ion temperatures known as potentials, drive fluxes known as transport, in such a way to counter the gradients, thus lowering maximum achievable performance of plasma confinement. Moreover, the toroidal shape of magnetic surfaces result in excessive transport than what is predicted by the so-called classical transport for cylindrical plasma of the hypothetical straight tokamak with zero curvature, which is normally referred to neo-classical transport. It is known that even the theory of neo-classical transport fails to describe the confinement behavior of thermonuclear plasmas where other mechanisms, such as turbulence, play a dominant role. In inertially confined plasmas, radiation transport adds up to the major transport mechanisms, which needs a very detailed and elaborate consideration. In this section we limit the discussion to classical and neo-classical transport and leave the discussion of turbulence and radiation transport to references.

The Boltzmann transport equation in phase space can be derived by considering how a distribution function changes in time. The classical and neo-classical theories of transport are best understood when their respective formulations are based on Boltzmann equation.

## IV.1. Boltzmann Equation

Plasma consists of numerous charged and uncharged particles. At any given moment, every particle has a precise position $\mathbf{r}$ and velocity $\mathbf{v}$ in the phase space $(\mathbf{r},\mathbf{v})$, and hence follows a trajectory expressible via a parametric curve as $\vec{C}(t) = [\mathbf{r}(t), \mathbf{v}(t)]$. Knowing the exact trajectory $\vec{C}_{i,s}(t)$ for all particles indexed by $i$ belonging to the species $s$ enable us to characterize the plasma accurately at all times. This can be only done through extensive particle simulations; even though powerful supercomputers are utilized for this purpose, it is impossible to simulate a real thermonuclear plasma with its full number of particles.

The alternative solution is to make a local average over all particles belonging to the species $s$ at a given time and within the neighborhood of a given phase space point $(\mathbf{r},\mathbf{v})$. This averaged quantity known as the *distribution function* $f_s(\mathbf{r},\mathbf{v},t)$ thus gives information about the phase-space density of species $s$ at the time $t$; hence, $dn = f_s(\mathbf{r},\mathbf{v},t)\,dr^3 dv^3$ represents the time-dependent number of particles which at the neighborhood of $\mathbf{r}$ have velocities close to $\mathbf{v}$. Since plasma can be considered almost free of neutral particles, the governing equation for the evolution of *distribution function* $f_s(\mathbf{r},\mathbf{v},t)$, or the so-called Boltzmann's equation, is only written for ions and electrons.

Boltzmann's equation is

$$\frac{Df}{Dt} = \frac{\partial f}{\partial t} + \mathbf{v}\cdot\frac{\partial f}{\partial \mathbf{r}} + \mathbf{a}\cdot\frac{\partial f}{\partial \mathbf{v}} = \left(\frac{df}{dt}\right)_{\text{collision}} = \hat{C}[f] \qquad (\text{IV}.1.1)$$

where $D/Dt$ is total time derivative, and $\mathbf{a}$ is the particle acceleration which is given by Lorentz force as

$$\mathbf{a} = \frac{q}{m}\left(\mathbf{E} + \mathbf{v} \times \mathbf{B}\right) \qquad (\text{IV}.1.2)$$

Also $\left(df/dt\right)_{\text{collision}}$ and $\hat{C}[f]$ in (IV.1.1) represent the collision term and *collision operator*, respectively. Inserting Coulomb collision in a plasma leads to the *Fokker-Plank's equation*. On the other hand, in a collisionless plasma the collision term $\left(df/dt\right)_{\text{collision}}$ becomes zero and Boltzmann's equation turns into the *Vlasov's equation*

$$\frac{\partial f}{\partial t} + \mathbf{v} \cdot \frac{\partial f}{\partial \mathbf{r}} + \mathbf{a} \cdot \frac{\partial f}{\partial \mathbf{v}} = 0 \qquad (\text{IV}.1.3)$$

which is valid for high temperatures and low densities.

In a fluid description of a plasma motion, the distribution function $f_s(\mathbf{r}, \mathbf{v}, t)$ can be used to define a number of macroscopic quantities as follows

1. Density of species $s$

$$n_s(\mathbf{r}, t) \equiv \int f_s(\mathbf{r}, \mathbf{v}, t) d^3 v \qquad (\text{IV}.1.4)$$

2. Average velocity of species $s$

$$\mathbf{V}_s \equiv n_s^{-1} \int \mathbf{v} f_s(\mathbf{r}, \mathbf{v}, t) d^3 v \qquad (\text{IV}.1.5)$$

3. Pressure tensor

$$\overleftrightarrow{\mathbf{p}}_s \equiv \int m_s f_s(\mathbf{r}, \mathbf{v}, t)(\mathbf{v} - \mathbf{V}_s)(\mathbf{v} - \mathbf{V}_s) d^3 v \qquad (\text{IV}.1.6)$$

4. Trace of pressure tensor, or simply the isotropic pressure

$$p_s \equiv \frac{1}{3} \int m_s \left|\mathbf{v} - \mathbf{V}_s\right|^2 f_s(\mathbf{r}, \mathbf{v}, t) d^3 v \qquad (\text{IV}.1.7)$$

5. Kinetic temperature of species $s$

$$T_s \equiv \frac{p_s}{n_s} \qquad (\text{IV}.1.8)$$

6- Stress Tensor

$$\overset{\leftrightarrow}{\mathbf{P}}_s \equiv \int m_s \mathbf{vv} f_s(\mathbf{r},\mathbf{v},t) d^3v \qquad \text{(IV.1.9)}$$

where the relation between $\overset{\leftrightarrow}{\mathbf{P}}_s$ and $\overset{\leftrightarrow}{\mathbf{p}}_s$ is

$$\overset{\leftrightarrow}{\mathbf{P}}_s = \overset{\leftrightarrow}{\mathbf{p}}_s + m_s n_s \mathbf{V}_s \mathbf{V}_s \qquad \text{(IV.1.10)}$$

7- Energy flux of species $s$

$$\mathbf{Q}_s = \int \frac{1}{2} m_s v^2 \mathbf{v} f_s(\mathbf{r},\mathbf{v},t) d^3v \qquad \text{(IV.1.11)}$$

8- Heat flux of species $s$

$$\mathbf{q}_s \equiv \int \frac{1}{2} m_s |\mathbf{v}-\mathbf{V}_s|^2 (\mathbf{v}-\mathbf{V}_s) f_s(\mathbf{r},\mathbf{v},t) d^3v \qquad \text{(III.1.1)}$$

where the relation between $\mathbf{Q}_s$ and $\mathbf{q}_s$ is

$$\mathbf{Q}_s = \mathbf{q}_s + \mathbf{V}_s \cdot \overset{\leftrightarrow}{\mathbf{p}}_s + \frac{3}{2} p_s \mathbf{V}_s + \frac{1}{2} m_s n_s V_s^2 \mathbf{V}_s \qquad \text{(IV.1.13)}$$

9- Energy-weighted stress

$$\mathbf{R}_s \equiv \frac{1}{2} \int m_s v^2 \mathbf{vv} f_s(\mathbf{r},\mathbf{v},t) d^3v \qquad \text{(IV.1.14)}$$

10- Energy-weighted friction

$$\mathbf{G}_s = \frac{1}{2} \int m_s v^2 \mathbf{v} \hat{C}[f_s] d^3v \qquad \text{(IV.1.15)}$$

where $\hat{C}[f]$ is the collision operator.

11- Energy exchange

$$W = \frac{1}{2} \int m_s |\mathbf{v}-\mathbf{V}_c|^2 \hat{C}[f] \qquad \text{(IV.1.16)}$$

12- Friction force

$$\mathbf{F}_s = \int m_s \mathbf{v} \hat{C}[f] d^3v \qquad \text{(IV.1.17)}$$

13- Collisional friction

$$R_s^n = \int C_s z_n(\mathbf{v}) d^3v \qquad \text{(IV.1.18)}$$

where $z_n$ is defined as

$$\begin{aligned} z_0 &= 1 & z_1 &= m\mathbf{v} \\ z_2 &= \frac{1}{2}m(\mathbf{v}\cdot\mathbf{v}) & z_3 &= \frac{1}{2}m(\mathbf{v}\cdot\mathbf{v})\mathbf{v} \end{aligned} \qquad \text{(IV.1.19)}$$

### IV.1.1 Moments Equations

While the microscopic distribution depends on $\mathbf{r}$, $\mathbf{v}$, and $t$, macroscopic physical parameters such as density or temperature, depend only on $\mathbf{r}$ and $t$, and consequently are obtained by integration over the entire velocity space, which are called as *moments*. The $i$-th moment is defined as

$$\mathrm{M}_i(\mathbf{r},t) = \int f(\mathbf{r},\mathbf{v},t)\mathbf{v}^i d^3v, \; i \in \mathbb{Z}^+ \qquad \text{(IV.1.20)}$$

in which $\mathbf{v}^i = \mathbf{v}.\mathbf{v}....\mathbf{v}$ denotes the $i$-fold dyadic product. The zeroth-order moment of (IV.1.1) yields the equation of continuity

$$\frac{\partial n}{\partial t} + \nabla \cdot (n\mathbf{v}) = 0 \qquad \text{(IV.1.21)}$$

First- and second-order moments of the Boltzmann equation yield

$$\mathbf{F} = \frac{\partial}{\partial t}mn\mathbf{v} + \nabla \cdot \overleftrightarrow{\mathbf{P}} - en(\mathbf{E} + \mathbf{v} \times \mathbf{B}) \qquad \text{(IV.1.22)}$$

$$\frac{\partial}{\partial t}\left(\frac{3}{2}p + \frac{1}{2}mnv^2\right) + \nabla \cdot \mathbf{Q} = W + \mathbf{v} \cdot (\mathbf{F} + ne\mathbf{E}) \qquad \text{(IV.1.23)}$$

The fourth moment equation is obtain by multiplying Boltzmann equation by $\mathbf{v}^3$ and integrating

$$\frac{\partial \mathbf{Q}}{\partial t} + \nabla \cdot \overleftrightarrow{\mathbf{R}} - \frac{3}{2}\frac{e}{m}p\mathbf{E} - \frac{1}{2}env^2\mathbf{E} - \frac{e}{m}\mathbf{E}\cdot\overleftrightarrow{\mathbf{P}} - \frac{e}{mc}\mathbf{Q}\times\mathbf{B} = \mathbf{G} \qquad \text{(IV.1.24)}$$

### IV.1.2 Application of Boltzmann Equation

Consider a distribution function with $x$-direction dependence in position and velocity $f(x,v_x,t)$. The Boltzmann equation then becomes

$$\left[\frac{\partial f(x,v_x,t)}{\partial t}\right]_{\text{collision}} = \frac{\partial x}{\partial t}\frac{\partial f(x,v_x,t)}{\partial x} = v_x \frac{\partial f(x,v_x,t)}{\partial x} \qquad (\text{IV}.1.25)$$

in which $v_x = \partial x/\partial t$. But the left-hand-side of (IV.1.25) equals to

$$\left[\frac{\partial f(x,v_x,t)}{\partial t}\right]_{\text{collision}} = -\frac{f(x,v_x,t) - f_{eq}(x,v_x)}{\tau} \qquad (\text{IV}.1.26)$$

where $f_{eq}(x,v_x)$ is the time-independent distribution function in equilibrium and $\tau$ is the relaxation time. Thus

$$v_x \frac{\partial f(x,v_x,t)}{\partial x} = -\left[\frac{f(x,v_x,t) - f_{eq}(x,v_x)}{\tau}\right] \qquad (\text{IV}.1.27)$$

The first order solution to (IV.1.27) is hence

$$f_1(x,v_x) = f_{eq}(x,v_x) - v_x \tau \frac{\partial f_{eq}}{\partial x} \qquad (\text{IV}.1.28)$$

Higher order solutions can be obtained by iterating. Hence the second order solution is

$$f_2(x,v_x) = f_{eq}(x,v_x) - v_x \tau \frac{\partial f_1}{\partial x} = f_{eq} - v_x \tau \frac{\partial f_{eq}}{\partial x} + v_x^2 \tau^2 \frac{\partial^2 f_{eq}}{\partial x^2} \qquad (\text{IV}.1.29)$$

The iteration is useful in considering nonlinear effects.

## IV.2. Flux-Surface-Average Operator

A flux-surface averaged of some quantity such as particle flux and heat flux, is a very useful concept for transport analysis of a toroidal plasma. The flux-surface average of a function is defined by the volume average over an infinitesimally small shell with volume $\Delta V$ as,

$$\langle A \rangle = \lim_{\Delta V \to 0} \frac{1}{\Delta V} \int_{\Delta V} A \, d^3 r \qquad (\text{IV}.2.1)$$

where $\Delta V$ lies between two neighboring flux surfaces with volume $V$ and $V + \Delta V$. To be strict, $V$ denotes the volume, while $\mathbf{v}$ represents the velocity coordinate in phase space.

It is physically more appealing to take average over a flux layer instead of taking average over geometric surface. Labeling flux surface by $\psi$, leads to:

$$\langle A \rangle = \frac{d\psi}{dV} \int \frac{f dS}{|\nabla \psi|} = \frac{d\psi}{dV} \int \frac{f dS}{|\hat{\psi}|} \tag{IV.2.2}$$

Here, $V$ is the volume enclosed by the flux surface. One can rewrite (IV.2.1) in flux coordinate as

$$\langle f \rangle = \frac{1}{V'} \oint \sqrt{g} d\theta d\zeta \tag{IV.2.3}$$

where $g$ is by (II.2.23) equal to the inverse of square of Jacobian.

There a number of important properties associated with the flux-surface average operator as

1- The flux-surface average of the divergence of a vector **A**

$$\langle \nabla \cdot \mathbf{A} \rangle = \lim_{\Delta V \to 0} \frac{1}{\Delta V} \oint_S \mathbf{A} \cdot d\mathbf{S} = \frac{1}{V'} \frac{d}{d\psi} \left\{ V' \langle \mathbf{A} \cdot \hat{\psi} \rangle \right\} \tag{IV.2.4}$$

where $V' = dV/d\psi$.

2- The flux-surface average annihilates the operator $\langle \mathbf{B} \cdot \nabla \rangle$

$$\langle \mathbf{B} \cdot \nabla A \rangle \equiv 0 \tag{IV.2.5}$$

3- The identity of flux-surface average

$$\langle \nabla \psi \cdot \nabla \times \mathbf{G} \rangle \equiv 0 \tag{IV.2.6}$$

which holds for any vector field **G**.

In order to achieve the flux-surface averaged form of the equation of Continuity (IV.1.19), we apply the flux-surface average operator to obtain

$$\left\langle \frac{\partial n}{\partial t} \right\rangle + \langle \nabla \cdot (n\mathbf{v}) \rangle = 0 \tag{IV.2.7}$$

or equivalently

$$\frac{\partial}{\partial t} \langle n \rangle = -\frac{d}{dV} \langle n\mathbf{v} \cdot \nabla V \rangle = -\frac{1}{V'} \frac{d}{d\psi} \langle n\mathbf{v} \cdot \nabla V \rangle \tag{IV.2.8}$$

One also can rewrite (IV.2.8) as

$$\frac{\partial n}{\partial t} + \frac{1}{V'}\left(V'\langle nv^\psi \rangle\right)' = 0 \qquad (IV.2.9)$$

In which the $nv^\psi$ is the contravariant component of particle flux in direction of $\psi$ and its flux-surface average is radial particle flux, usually denoted by $\Gamma$. Moreover, prime denotes differentiation with respect to the magnetic poloidal flux.

In the next section we will study classical and non-classical transport in axisymmetric toroidal system.

## IV.3. Classical and Non-classical Transport

*Classical transport* refers to those transport fluxes that happen in straight and uniform magnetic field lines. Classical transport of particles is due to Coulomb collisions and one should take into the account the gyrations of particles in the magnetic field. But when the geometry change into torus the dominant diffusive transport is most due to drifts across particle guiding center orbits. In particular, the collision and particle displacements are enhanced because the gyrocenter displacement from the magnetic surface gets larger than the gyroradius itself. This type of transport is faster than classical transport and is called *Neoclassical* (non-classical) *Transport*. Therefore, geometrical effects cause to complicate particle orbits and drifts in neoclassical model, where they are routinely ignored in the classical model. *Banana orbits, potato orbits,* and *bootstrap current* arise from the neoclassical transport model.

We first study the classical theory of collisions in cylindrical plasma and next we consider the neoclassical transport.

### IV.3.1 Classical Collisional Transport
Equations of transport are

$$\begin{aligned}
\frac{\partial n_s}{\partial t} + \nabla \cdot \mathbf{\Gamma}_s &= \text{Source- Sink} \\
\frac{\partial}{\partial t}\left(\tfrac{3}{2} n_s kT_s\right) + \nabla \cdot \mathbf{q}_s &= \text{Source- Sink} \\
p\frac{\partial \mathbf{v}}{\partial t} + \nabla \cdot \overset{\leftrightarrow}{\Pi} &= \text{Source- Sink} \\
\mathbf{E} + \mathbf{v} \times \mathbf{B} &= \eta \mathbf{J} = \eta_\perp \mathbf{J}_\perp + \eta_\parallel \frac{J_\parallel}{B}\mathbf{B}
\end{aligned} \qquad (IV.3.1)$$

which become complete along with Maxwell's equations

$$\begin{aligned}
\frac{\partial B_z}{\partial t} &= -\frac{1}{r}\frac{\partial}{\partial r}(rE_\varphi) & \frac{\partial B_\varphi}{\partial t} &= \frac{\partial E_z}{\partial r} \\
\mu_0 J_z &= \frac{1}{r}\frac{\partial}{\partial r}(rB_\varphi) & \mu_0 J_\varphi &= -\frac{\partial B_z}{\partial r}
\end{aligned} \qquad (IV.3.2)$$

Here, the subscript $s$ refers to ion or electron species, and $\mathbf{\Gamma}_s$, $\mathbf{q}_s$ and $\overleftrightarrow{\Pi}$ are *particle flux*, *heat flux* and *viscous tensor* respectively, defined as

$$\mathbf{\Gamma}_s = -D\nabla n_s + n_s \mathbf{V}_c$$
$$\mathbf{q}_s = -n_s \chi_s \nabla T_s + \frac{5}{2}\mathbf{\Gamma}_s k T_j + \mathbf{q}_{\text{conv}} \tag{IV.3.3}$$

Also $\mathbf{J}_\perp$ and $J_\parallel$ in (III.3.1) are given by:

$$\mathbf{J}_\perp = \frac{J_\varphi B_z - J_z B_\varphi}{B^2}\left(B_z \hat{\varphi} - B_\varphi \hat{z}\right) = \frac{1}{B^2}\frac{\partial p}{\partial r}\left(B_z \hat{\varphi} - B_\varphi \hat{z}\right) \tag{IV.3.4}$$

$$J_\parallel = \frac{J_\varphi B_\varphi - J_z B_z}{B} \tag{IV.3.5}$$

### IV.3.1.1 Random Walk Model
*Random Walk Model* is the simplest model that can be used to determine transport coefficients, and it is dependent on the mean collision time and the mean free path associated with the random motion of particles. The random motion of a particle is shown in Fig. IV.3.1.

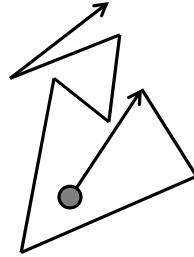

**Figure IV.3.1:** Random walk.

In this model the diffusion coefficient is simply given by

$$D = l^2/\tau \tag{IV.3.6}$$

where $l$ and $\tau$ are the average step size and average time between collisions, respectively.

### IV.3.1.2 Particle Diffusion in Fluid Picture
We may take the cross product of Ohm's law with magnetic field $\mathbf{B}$ to yield

$$\mathbf{E}\times\mathbf{B} + \left(\mathbf{v}\times\mathbf{B}\right)\times\mathbf{B} = \eta\mathbf{J}\times\mathbf{B} = \eta_\perp \nabla p \tag{IV.3.7}$$

which upon simplification takes the form

$$\mathbf{E} \times \mathbf{B} - \mathbf{v}_\perp B^2 = \eta_\perp \nabla p \qquad (IV.3.8)$$

with the perpendicular velocity given by

$$\mathbf{v}_\perp = \frac{(\mathbf{E} \times \mathbf{B})}{B^2} - \left(\frac{\eta_\perp}{B^2}\right)\nabla p \qquad (IV.3.9)$$

The first term on the right-hand-side of (IV.3.9) is $\mathbf{E} \times \mathbf{B}$ drift of particles and the second term is diffusion velocity in direction of $\nabla p$. Now, letting $T$ to be constant, we get

$$\nabla p = T \nabla n \qquad (IV.3.10)$$

Hence, the radial particle flux is derived as

$$\Gamma_\perp = n\mathbf{v}_\perp = \left(\eta n T/B^2\right)\nabla n = D_\perp \nabla n \qquad (IV.3.11)$$

where $D_\perp = \eta n T/B^2$ is the particle diffusion coefficient. When the electric field is applied to the plasma, electrons accelerate to the drift velocity $v_d$. In this situation the force of electric field is balanced by collision force, in this manner we have:

$$eE = m_e v_d / \tau_c \qquad (IV.3.12)$$

Here, $\tau_c$ is momentum loss time. Hence the scalar resistivity is obtained as

$$\eta = |\mathbf{E}|/|\mathbf{J}| = m_e / n_e e^2 \tau_c \approx m_e / n_e e^2 \tau_e \qquad (IV.3.13)$$

with $\tau_e$ being the electron collision time. Substituting (IV.3.13) in (IV.3.6) yields the expression for electron diffusion coefficient (the perpendicular subscript denotes transport across magnetic surfaces) as

$$D_\perp = \frac{p}{B^2} \frac{m_e}{n_e e^2 \tau_e} \qquad (IV.3.14)$$

### IV.3.2 Neoclassical Collisional Transport
### IV.3.2.1 Trapped Particles and Banana Orbit

Since the toroidal field cannot individually confine the plasma of tokamak at equilibrium, a combination of toroidal and poloidal magnetic fields, together with a toroidal current, is necessary to form closed magnetic surfaces. Therefore, the magnetic field lines are helically wound on toroidally nested surfaces and charged particles follow helical field lines. Now, let $R$ be the distance from the major axis in toroidal geometry; then the magnitude of toroidal magnetic field falls off with distance from the major axis of torus $R$, according to the Solov'ev equilibrium (II.7.25). Therefore, the guiding centers of particles as they follow along the magnetic field feel a change in the strength of the magnetic field. This means that particles moving slowly along the magnetic field are reflected and subsequently, when they attempt to travel across the torus in the reverse direction, they are reflected back again. These are the trapped particles in the so-called Banana orbits. The name of Banana comes from the fact that poloidal projections of trapped particle onto constant $\zeta$-surface are similar to Banana as shown in Fig. IV.3.2.

On the other hand, we have passing particles in contrast to trapped particles. Passing particles are not trapped and thus not reflected, and follow spiral paths around the torus following the helical path of the field lines. Hence the particles whose velocity components along the field are low contribute to the population of trapped prticles, while particles with higher velocities parallel to the field cycle around the torus and increse the population of passing particles.

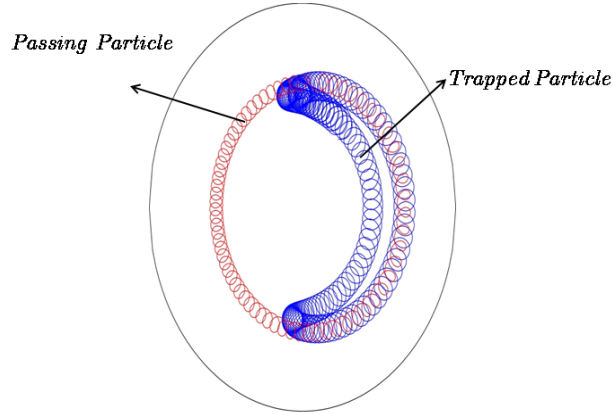

**Figure IV.3.2:** Trapped particles in Banana orbits and passing particles [4].

The condition for particles to be trapped in a large aspect ratio tokamak is obtained by using the conservation of energy and magnetic moment as

$$\frac{v_\parallel}{v^2} < 1 - \frac{B_{\min}}{B_{\max}} \qquad \text{(IV.3.15)}$$

Now, since according to (II.7.25) we roughly have $B \sim 1/R$ we have:

$$\frac{B_{\min}}{B_{\max}} = \frac{R_0 - r}{R_0 + r} \approx 1 + \frac{2r}{R_0} \qquad \text{(IV.3.16)}$$

Thus, requirement for trapping simply becomes

$$\frac{v_\parallel}{v^2} < 2\varepsilon \qquad \text{(IV.3.17)}$$

where $\varepsilon = r/R_0$ is the inverse aspect ratio. Integration of the equation of motion leads us to the Banana width orbit $\Delta b$

$$\Delta b = \frac{mv}{qB_\theta}\varepsilon^{1/2} = \frac{\rho\varepsilon^{-1/2}}{\iota} \qquad (IV.3.18)$$

in which $\iota$ is the rotational transform and is given by

$$\iota = \frac{RB_\theta}{rB_\varphi} \qquad (IV.3.19)$$

Similarly the displacement of the guiding centre from the flux surface for passing particles is

$$\Delta p = \frac{mv}{\iota|q|B_\varphi} \qquad (IV.3.20)$$

One can illustrate the boundary between trapped and untrapped particles in velocity phase space as shown in Fig IV.3.3.

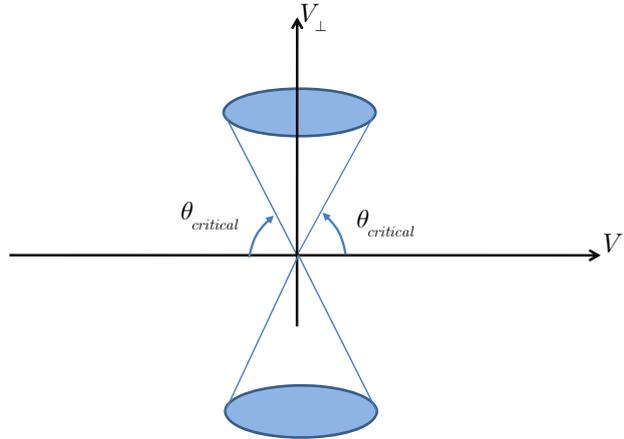

**Figure IV.3.3:** Boundary between trapped and untrapped particles.

Critical angle $\theta_c$ in Fig. IV.3.3 is determined by

$$\theta_c = \cos^{-1}\frac{v_\parallel}{v} \approx \cos^{-1}\sqrt{2r/R_0} \qquad (IV.3.21)$$

For a Maxwellian distribution function, one can then easily obtain the fraction of trapped particles, as

$$f = \frac{2\pi}{n}\int_{\theta_c}^{\pi-\theta_c}\int_0^\infty F_M(v)v^2\sin\theta\, dv\, d\theta = \cos\theta_c = \sqrt{\frac{2r}{R_o}} = \sqrt{2\varepsilon} \qquad (IV.3.22)$$

## IV.3.2.2 Different regimes

Diffusion coefficients in neo-classical transport significantly vary in Banana, Plateau and Pfirsch-Schlüter regimes, depending on the strength of collisionality as illustrated in Fig. IV.3.4.

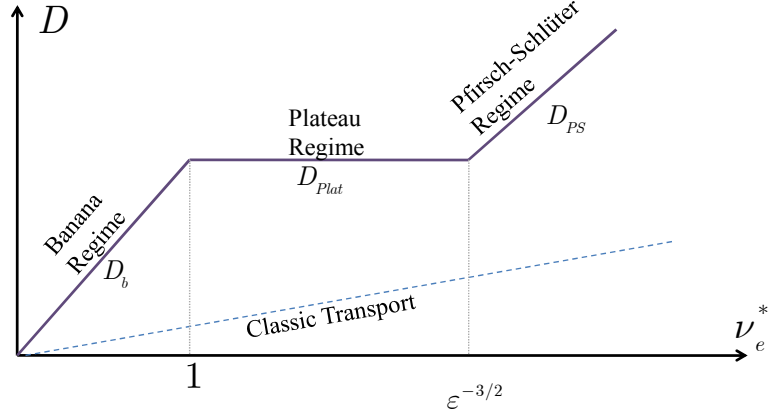

**Figure IV.3.4:** Different transport regimes.

The dimensionless *collisionality* $\nu^*$ in Banana regime is defined as

$$\nu_e^* = \frac{\tau_b}{\tau_e} = \frac{R_o q}{\tau_e v_{th} \varepsilon^{3/2}} \tag{IV.3.23}$$

In Banana regime where $\nu_e^* < 1$ electrons can complete their Banana orbits many times before colliding; hence, only trapped particles contribute to the transport. Therefore one can use Banana-orbit width $\Delta b$ as the step size in random walk model and obtain

$$D_b = \frac{f_{trap} \Delta b^2}{\tau_e} \sim \frac{1}{\varepsilon^{3/2}} \frac{q^2 \lambda_e^2}{\tau_e} \sim \frac{1}{\varepsilon^{3/2}} q^2 D_{classic} \tag{IV.3.24}$$

Pfirsch-Schlüter transport arises from $\mathbf{E} \times \mathbf{B}$ term for $v_\perp$ in (IV.3.9). When $\nu_e^* > \varepsilon^{-3/2}$ collisions prevent the particles completing Banana orbits and Pfirsch–Schlüter diffusion reads

$$D_{PS} = \left(1 + \alpha q^2\right) D_{classic} \tag{IV.3.25}$$

where $\alpha$ is a numerical factor having the order of unity.

The intermediate regime bounded by Banana and Pfirsch–Schlüter regimes is Plateau regime, for which we have $1 < \nu_e^* < \varepsilon^{-3/2}$. In this regime, particles make about one collision after completing one Banana orbit. One determines the plateau diffusion as

$$D_{Plat} \sim q T_e \lambda_e \tag{IV.3.26}$$

## IV.3.2.3 Transport Matrix

The current density, particle, electron and ion heat transport fluxes are functions of driving gradients $(\nabla n, \nabla T_i, \nabla T_e, \nabla V_l)$ in which the parallel electric field is $\mathbf{E}_\parallel = -\nabla V_l$, and $V_l$ is the plasma's electric potential around the torus. The neoclassical transport is described by a *transport matrix* as below:

$$\begin{pmatrix} \mathbf{\Gamma} \\ \mathbf{q}_e \\ \mathbf{q}_i \\ \mathbf{J} \end{pmatrix} = -\begin{pmatrix} D & M_{12} & M_{13} & \omega \\ M_{21} & n\chi_e & M_{23} & M_{24} \\ M_{31} & M_{32} & n\chi_i & M_{34} \\ b_n & b\tau_e & b\tau_i & \sigma \end{pmatrix} \begin{pmatrix} \nabla n \\ \nabla T_e \\ \nabla T_i \\ \nabla V_l \end{pmatrix} \qquad \text{(IV.3.27)}$$

where $\omega \sim \varepsilon^{1/2} n/B_\theta$. The above equation reveals that every type of transport can be driven by any of the potential gradients. This fact complicates the study of neo-classical transport phenomena in plasmas. This minus sign stresses on the fact that transport opposes gradients. The above can also be written as

$$\{\mathbf{F}_j\} = -[O_{ij}]\nabla\{P_i\} \qquad \text{(IV.3.28)}$$

in which $\mathbf{F}_j$, $O_{ij}$, and $P_i$ are respectively transport fluxes, Onsager coefficients, and potential functions. Onsager coefficients are functions of magnetic field $\mathbf{B}$ and may be shown to satisfy the symmetry

$$O_{ij}(\mathbf{B}) = \pm O_{ji}(-\mathbf{B}) \qquad \text{(IV.3.29)}$$

Hinton and Hazeltine gave mathematical derivation of neo-classical flux parameters as

$$\Gamma^r = -D_b n \left[ 1.12\left(\frac{T_e+T_i}{T_e}\right)\nabla\mathcal{N} - 0.43\nabla\mathcal{T}_e - 0.19\nabla\mathcal{T}_i + 2.44\sqrt{\frac{r}{R}}\mathbf{v}_f \right] \cdot \hat{r}$$

$$q_e^r = -D_b p_e \left[ -1.53\left(\frac{T_e+T_i}{T_e}\right)\nabla\mathcal{N} + 1.81\nabla\mathcal{T}_e + 0.27\frac{T_i}{T_e}\nabla\mathcal{T}_i + 1.75\sqrt{\frac{r}{R}}\mathbf{v}_f \right] \cdot \hat{r}$$

$$q_i^r = -0.48 D_b n \sqrt{\frac{m_i T_e T_i}{m_e}}\nabla\mathcal{T}_i \cdot \hat{r} \qquad \text{(IV.3.28)}$$

$$J_\parallel^r = \frac{p_e}{B_p}\sqrt{\frac{r}{R}}\left[ -2.44\left(\frac{T_e+T_i}{T_e}\right)\nabla\mathcal{N} - 0.69\nabla\mathcal{T}_e + 0.42\frac{T_i}{T_e}\nabla\mathcal{T}_i - \left(1 - 1.95\sqrt{\frac{r}{R}}\right)\sigma_\parallel \nabla V_l \right] \cdot \hat{r}$$

Here, $r$ superscript denotes the radial contravariant component obtained by inner product with $\hat{r}$. Also, $\mathbf{v}_f$ is the radial flux surface velocity, and $\mathcal{N} = \ln n$ and $\mathcal{T}_s = \ln T_s, s = e, i$ are dimensionless density and species temperature. Furthremore, $\sigma_\parallel$ is *Spitzer conductivity* given by

$$\sigma_\parallel = 1.98\,\tau_e e^2 n/m_e \qquad \text{(IV.3.29)}$$

Hence, (IV.3.28) can be written in a similar form to (IV.3.27) as

$$\begin{bmatrix} \Gamma^r \\ q_e^r \\ q_i^r \\ J_\parallel^r \end{bmatrix} = -\begin{bmatrix} 1.12 D_b n (1+\gamma_{ie}) & 0.43 D_b n & 0.19 D_b n & 0 \\ 1.53 D_b p_e (1+\gamma_{ie}) & 1.81 D_b p_e & 0.27 D_b p_e \gamma_{ie} & 0 \\ 0 & 0.48 D_b p_e \sqrt{\dfrac{m_i \gamma_{ie}}{m_e}} & 0 & 0 \\ 2.44(1+\gamma_{ie})\dfrac{p_e}{B_p}\sqrt{\varepsilon} & 0.69 \dfrac{p_e}{B_p}\sqrt{\varepsilon} & -0.42\gamma_{ie}\dfrac{p_e}{B_p}\sqrt{\varepsilon} & (1-1.95\sqrt{\varepsilon})\sigma_\parallel 0.42\gamma_{ie} \end{bmatrix} \nabla \begin{bmatrix} \mathcal{N} \\ \mathcal{T}_e \\ \mathcal{T}_i \\ V_l \end{bmatrix}$$

$$-\begin{bmatrix} 2.44 n \\ 1.75 p_e \\ 0 \\ 0 \end{bmatrix} D_b \sqrt{\varepsilon} v_f^r \qquad\qquad (IV.3.30)$$

in which $\gamma_{ie} = T_i/T_e$. The Onsager symmetry in (IV.3.30) is not apparent since the radial velocity $v_f^r$ should also first be expressed in terms of other potential gradients. However, the above form is more preferred in computations where fluxes across magnetic surfaces are required. Typical solution of plasma equilibrium and transport for Damavand tokamak is depicted in Fig. IV.3.5.

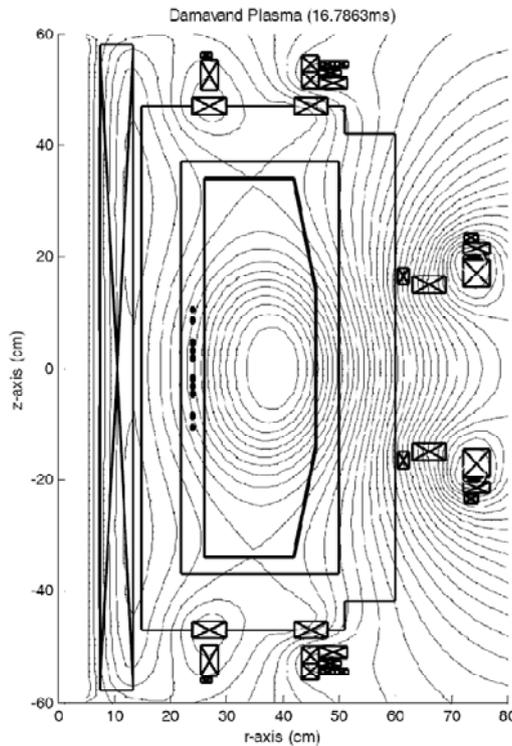

**Fig. IV.3.5:** Separatrix plasma configuration in Damavand tokamak.

### IV.3.3 Bootstrap Current

In the Banana region, radial diffusion induces a current in the toroidal direction known as the *Bootstrap current* $J_{BS}$, which is in parallel to the magnetic field. Unlike the Ohmic current $J_{Ohmic}$, this current does not require any external electric field and occurs naturally due to gradients in plasma profiles of temperature and density. From the fourth equation of (IV.3.30) we have

$$J_\parallel^r = -\frac{p_e}{B_p}\sqrt{\varepsilon}\left[2.44(1+\gamma_{ie})\nabla\mathcal{N} - 0.69\nabla\mathcal{T}_e + 0.42\gamma_{ie}\nabla\mathcal{T}\right] + (1-1.95\sqrt{\varepsilon})\sigma_\parallel 0.42\gamma_{ie}E_\parallel$$
$$= J_{BS} + J_{Ohmic} \tag{IV.3.29}$$

In fact, there is a fraction $\varepsilon^{1/2}$ of trapped particles having a parallel velocity as $\varepsilon^{1/2}v_{th} = \varepsilon^{1/2}\sqrt{k_B T_e/3m_e}$ where execute a Banana orbit of width $w_b = q\rho_L\varepsilon^{-1/2}$. Therefore, when a radial density gradient exists, these particles produce a current analogous to the diamagnetic current of untrapped, which reads as

$$J_{trapped} \sim -ew_b\frac{dn}{dr}\varepsilon^{1/2}\left(\varepsilon^{1/2}v_{th}\right) \sim -q\frac{\varepsilon^{1/2}}{B}T\frac{dn}{dr} \tag{IV.3.31}$$

There is a momentum transfer from the trapped to passing particles of both ions and electrons, due to this fact that both species produce such a current, which modifies the velocity of the passing particles. The difference in modified velocities of passing particles produces the toroidal bootstrap current $J_{BS}$. Now, the momentum exchange between passing ions and electrons is $m_e J_{BS}/e\tau_{ei}$. The passing electrons are affected by a momentum exchange with the trapped electrons. The trapped electrons are localized to a part $\sim \varepsilon^{1/2}$ of velocity space and the effective collision frequency is ascertained by the time needed to scatter out of this region as $\tau_{eff} \sim \varepsilon\tau_{ee}$. Thus, the momentum exchange rate between trapped and passing electrons is $m_e J_{trapped}/e\varepsilon\tau_{ee}$. The bootstrap current originates form balancing the momentum exchange of passing electrons with passing ions and with trapped electrons, approximately given by

$$J_{BS} \simeq \frac{\tau_{ei}}{\tau_{ee}}\frac{J_{trapped}}{\varepsilon} \simeq -\frac{\varepsilon^{1/2}}{B_\theta}T\frac{dn}{dr} \tag{IV.3.32}$$

This is while the precise expression to $O(\varepsilon^{1/2})$ according to (IV.3.29) is

$$J_{BS} = -\frac{\varepsilon^{1/2}n}{B_\theta}\left[2.44(T_e+T_i)\frac{1}{n}\frac{dn}{dr} + 0.69\frac{dT_e}{dr} - 0.42\frac{dT_i}{dr}\right] \tag{IV.3.33}$$

which indicates that the bootstrap current fraction of the total current scales as

$$\frac{I_{BS}}{I} = c\varepsilon^{1/2}\beta_p \tag{IV.3.34}$$

with $c$ being a dimensionless constant about $\frac{1}{3}$. In the low-aspect-ratio limit $\varepsilon \to 1$, when most particles are trapped, the bootstrap current is however determined by

$$J_{BS} \approx -\frac{1}{B_\theta}\frac{dp}{dr} \qquad (IV.3.35)$$

Here, the bootstrap current is driven entirely by the pressure rather than the density gradient.

## IV.3.4 Confinement Times

The particle confinement time for ions can be defined as

$$\tau_p \equiv \frac{\text{Number of Ions in Plasma}}{\text{Ion Loss Rate}} \simeq \frac{\text{Number of Ions in Plasma}}{\text{Ion Production Rate at Equilibrium}} \qquad (IV.3.36)$$

where

$$\begin{aligned}\text{Number of Ions in Plasma} &= \int_{\text{Plasma}} n\, dr^3 \\ \text{Ion Loss Rate} &= \int_{\text{Surface}} \mathbf{\Gamma}\cdot d\mathbf{S}\end{aligned} \qquad (IV.3.37)$$

If the plasma is at the steady state equilibrium then the production rate equals the loss rate. Then the electron particle confinement time is the same due to quasi-neutrality condition. The energy confinement time for electrons $\tau_{Ee}$ is obtained by

$$\tau_{Ee} \equiv \frac{\text{Electron Energy in Plasma}}{\text{Electron Energy Loss Rate}} \simeq \frac{\text{Electron Energy in Plasma}}{\text{Electron Heating Rate at Equilibrium}} \qquad (IV.3.38)$$

where

$$\begin{aligned}\text{Electron Energy in Plasma} &= \frac{3}{2}\int_{\text{Plasma}} nT_e\, dr^3 \\ \text{Electron Energy Loss Rate} &= \int_{\text{Surface}}\left(\mathbf{q}_e + 2.5 T_e \mathbf{\Gamma}_e\right)\cdot d\mathbf{S} + \int_{\text{Plasma}} P_{rad}\, dr^3\end{aligned} \qquad (IV.3.39)$$

For the whole plasma, the energy confinement time is

$$\tau_E = \frac{\text{Plasma Energy}}{\text{Energy Loss Rate}} \simeq \frac{\text{Plasma Energy}}{\text{Plasma Heating Rate at Equilibrium}} \qquad (IV.3.40)$$

where

$$\text{Plasma Energy} = \frac{3}{2} \int_{\text{Plasma}} n(T_e + T_i) dr^3$$

$$\text{Ion Energy Loss Rate} = \int_{\text{Surface}} (\mathbf{q}_i + 2.5 T_i \mathbf{\Gamma}_i) \cdot d\mathbf{S} + \int_{\text{Volume}} n_n n_i \langle \sigma_x v_i \rangle \frac{3}{2} T_i dr^3 \quad (\text{IV.3.41})$$

$$\text{Plasma Energy Loss} = \text{Electron Energy Loss} + \text{Ion Energy Loss}$$

# V. Conclusion

In summary, physical and technological studies and surveys considering the daily growing need of mankind to inexhaustible and clean energy, directs the researches towards nuclear fusion, where a bright future is seen for the life of the man on the earth. Fusion can be however reached only in extremely hot plasmas, which are normally confined either magnetically by strong magnetic fields, or inertially by powerful radiations of photons or energetic ions. Various plasma confinement technologies have been developed, among which tokamaks as magnetic plasma confinement machines have produced the most successful fusion experiments. At the moment, the only known promising candidate for a nuclear fusion power reactor is tokamak. The detailed theory behind the operation of magnetically confined hot plasmas was discussed in this tutorial, addressing important aspects related to the plasma equilibrium, stability, and transport.

Comparing to the nuclear fission reactions, nuclear fusion reactions enjoy an inherent safety, which is due to the fact that in case of any serious instability or runaway plasma disrupts and reactions automatically stop. In contrast, fission reactions would lead to disaster if their control is lost. From this point of view, fusion science and technology is almost entirely declassified and all its documents are openly accessible to all nations. On the other hand, it is necessary that developing countries diversify their energy resources, and assign larger budget volumes and human taskforce to investigate active areas in nuclear fusion. Since the funding needed to realize a full-size thermonuclear fusion machine is normally out of reach of developing countries, appropriate actions and decisions should be taken to minimize the technological and scientific gap between advanced and developing states in the future.

Calculations show that fission of the available uranium on earth is sufficient only for the next 300 years, while fusion of naturally abundant deuterium on the earth and oceans, should provide the necessary energy for more than a million years, or so. That is why nuclear fusion is called as the 'Tomorrow's Energy'.

# Acknowledgement

One of the authors (F. Dini) would like to acknowledge insightful discussions with Prof. Vladimir Shafranov at the Russian Research Center Kurchatov Institute, Moscow, Prof. Weston Stacey and Dr. John Mandrekas at Georgia Institute of Technology, Atlanta, and Prof. Thomas Dolan at University of Illinois at Urbana-Champaign. The authors are indebted to Mr. Mehdi Baghdadi for proofreading the manuscript and illustration of diagrams. They also wish to thank fruitful discussions with students, including Miss Shiva Shahshenas, Mr. Mohsen Mardani and Mr. Ahmad Abrishami. This work grew out of the lecture notes of an advanced graduate course on Magnetic Confinement Fusion offered at Amirkabir University of Technology by S. Khorasani.

# Correspondence

Dr. Reza Amrollahi
*Professor of Physics and Chair, Department of Physics and Nuclear Engineering*
*Amirkabir University of Technology, Tehran, Iran*
Cell:       +98-912-159-2837
Office:     +98-21-64542572    +98-21-66419506
Fax:        +98-21-6649-5519
Email:      amrollahi@aut.ac.ir    ramrollahi@yahoo.com